\documentclass[aps,prd,onecolumn,showpacs,amsmath,amssymb,floatfix,nofootinbib,11pt]{revtex4-1}
\usepackage{amsfonts,graphicx,multirow,amsmath,color,xcolor,ulem}
\usepackage[colorlinks=true,citecolor=blue,linkcolor=blue,urlcolor=blue]{hyperref}

\newcommand{\DS}[1]{/\!\!\!#1}

\allowdisplaybreaks

\begin{document}

\title{Investigating the ratio of CKM matrix elements $|V_{ub}|/|V_{cb}|$ from semileptonic decay $B_s^0\to K^-\mu^+\nu_\mu$ and kaon twist-2 distribution amplitude}

\author{Tao Zhong}
\email{zhongtao1219@sina.com}
\address{Department of Physics, Guizhou Minzu University, Guiyang 550025, P.R. China}
\author{Hai-Bing Fu}
\email{fuhb@cqu.edu.cn (Corresponding author)}
\address{Department of Physics, Guizhou Minzu University, Guiyang 550025, P.R. China}
\author{Xing-Gang Wu}
\email{wuxg@cqu.edu.cn}
\address{Department of Physics, Chongqing Key Laboratory for Strongly Coupled Physics, Chongqing University, Chongqing 401331, P.R. China}

\date{\today}

\begin{abstract}

In this paper, we calculate the ratio of Cabibbo-Kobayashi-Maskawa matrix elements, $|V_{ub}|/|V_{cb}|$, based on the semileptonic decay $B_s^0\to K^-\mu^+\nu_\mu$. Its key component, the $B_s\to K$ transition form factor $f^{B_s\to K}_+(q^2)$, is studied within the QCD light-cone sum rules approach by using a chiral correlator. The derived $f^{B_s\to K}_+(q^2)$ is dominated by the leading-twist part, and to improve its precision, we construct a new model for the kaon leading-twist distribution amplitude $\phi_{2;K}(x,\mu)$, whose parameters are fixed by using the least squares method with the help of the moments calculated by using the QCD sum rules within the background field theory. The first four moments at the initial scale $\mu_0 = 1~{\rm GeV}$ are, $\langle \xi^1\rangle _{2;K} = -0.0438^{+0.0053}_{-0.0075}$, $\langle \xi^2\rangle _{2;K} = 0.262 \pm 0.010$, $\langle \xi^3\rangle _{2;K} = -0.0210^{+0.0024}_{-0.0035}$ and $\langle \xi^4\rangle _{2;K} = 0.132 \pm 0.006$, respectively. And their corresponding Gegenbauer moments are, $a^{2;K}_1 = -0.0731^{+0.0089}_{-0.0124}$, $a^{2;K}_2 = 0.182^{+0.029}_{-0.030}$, $a^{2;K}_3 = -0.0114^{+0.0008}_{-0.0016}$ and $a^{2;K}_4 = 0.041^{-0.003}_{+0.005}$, respectively. At the large recoil region, we obtain $f^{B_s\to K} _+ (0) = 0.270^{+0.022}_{-0.030}$. By extrapolating $f^{B_s\to K}_+(q^2)$ to all the physical allowable region, we obtain a $|V_{ub}|$-independent decay width for the semileptonic decay $B_s^0\to K^-\mu^+\nu_\mu$, $5.626^{+1.271}_{-0.864} \times 10^{-12}\ {\rm GeV}$, which then leads to $|V_{ub}|/|V_{cb}| = 0.072\pm0.005$.

\end{abstract}

\maketitle

\section{Introduction}

The Cabibbo-Kobayashi-Maskawa (CKM) matrix element $|V_{ub}|$ modulates the coupling of the electroweak interaction between $u$- and $b$-quarks. The research of $|V_{ub}|$ can be performed by the weak decays of hadrons containing a $b$-quark, which occurs via the quark level transition $b \to u~ (W^\ast \to \ell\nu)$, where $\ell$ indicates a lepton and $\nu$ is for neutrino. Therefore, those decays provide a good platform to test the standard model (SM) and probe the new physics effects beyond the SM. There is a discrepancy between the measurements of $|V_{ub}|$ from exclusive decays and that from inclusive decays. So far, the data from semileptonic decay $B\to\pi\ell\nu_\ell$ dominate the world average of the exclusive $|V_{ub}|$-measurements. Therefore, it is necessary to study other exclusive processes occurring by $b\to u\ell\nu$. Especially, the LHCb collaboration reported the measurements of the branching fraction of the semileptonic decay $B_s^0\to K^-\mu^+\nu_\mu$ and of the ratio $|V_{ub}|/|V_{cb}|$~\cite{LHCb:2020ist}, which are based on the data sample from $pp$ collisions at a center-of-mass energy of $8~{\rm TeV}$ corresponding to the integrated luminosity of $2~\rm fb^{-1}$ recorded by the LHCb detector in 2012. Thus, the semileptonic decay channel $B_s^0\to K^-\mu^+\nu_\mu$ is also an significant process in dealing with $|V_{ub}|$.

The key component of $B_s^0\to K^-\mu^+\nu_\mu$ decay amplitude is the $B_s\to K$ transition form factor (TFF) $f^{B_s\to K}_+(q^2)$, which can be calculated by various approaches. Based on the light-cone sum rules (LCSRs) approach, the $f^{B_s\to K}_+(q^2)$ was studied in Ref.~\cite{Li:2001yv} by adopting the chiral correlation function (correlator) for the first time. Lately, researches about this TFF have been performed with the traditional correlator, which is arranged by the kaon's increased twist distribution amplitudes (DAs)~\cite{Khodjamirian:2003xk, Wu:2006rd, Duplancic:2008tk, Melic:2008cx, Khodjamirian:2017fxg}. Particularly, by integrating directly in the complex plane, G. Duplancic and B. Melic calculate the gluon radiative corrections to the kaon twist-2 and twist-3 DA terms, respectively~\cite{Duplancic:2008tk, Melic:2008cx}. The LCSRs calculation can also been performed in the framework of heavy quark effective field theory (HQEFT)~\cite{Wu:2006rd}. Meanwhile, the semileptonic decay $B_s\to K\ell\bar\nu_\ell$ has also been studied within the lattice QCD (LQCD)~\cite{Bouchard:2013zda, Bouchard:2014ypa, Flynn:2015mha, Bahr:2016ayy, Monahan:2018lzv} and the perturbative QCD (pQCD) factorization approach~\cite{Wang:2012ab, Meissner:2013pba, Jin:2020jtu}. Generally, the pQCD factorization approach is reliable for describing the form factors in the low $q^2$-region, the LCSR estimations for TFFs are reliable in low and intermediate $q^2$-regions, and the LQCD works well for the region near $q^2_{\rm max} = (m_{B_s} - m_K)^2 \sim 23.75 ~{\rm GeV}^2$. The predictions from those three methods can be extended from their own applicable region to all allowable $q^2$ values via appropriate extrapolations, such as the Bourrely-Caprini-Lellouch formula~\cite{Bourrely:2008za}, etc. Different methods are also complementary to each other. S. P. Jin and Z. J. Xiao present the pQCD factorization determinations of $B_s\to K$ TFFs in the low $q^2$-region firstly, then they improve their extrapolation by taking the available LQCD results at $q^2_{\rm max}$ as additional inputs~\cite{Jin:2020jtu}. In addition, there are several quark models (QMs) to be also used to study the semileptonic decay $B_s\to K\ell\bar\nu_\ell$. For example, C. Albertus studies the TFF of the semileptonic decay $\bar{B}_s\to K^+\ell^-\bar{\nu}_\ell$ within constituent quark model (CQM) and nonrelativistic quark model (NRQM), respectively, where a multiply-subtracted Omn\`{e}s dispersion relation is used to extrapolate their predictions from its applicable region near $q^2_{\rm max}$~\cite{Albertus:2014gba,Albertus:2014rna}. By using the quasipotential approach, R. Faustov and V. Galkin research the $B_s\to K$ TFFs within the framework of the QCD-motivated relativistic quark model (RQM), and they obtain the momentum dependence of TFFs in the whole $q^2$-regions without any additional extrapolations~\cite{Faustov:2013ima}. Furthermore, there are other QM researches on the TFF $f^{B\to K}_+(q^2)$ in literature, such as, the light-front covariant quark model (LFQM)~\cite{Verma:2011yw} and the light cone quark model (LCQM) within soft collinear effective theory (SCET)~\cite{Lu:2007sg}. Otherwise, the TFF $f^{B\to K}_+(q^2)$ has also been calculated in some researches on the $B/B_s$ two-body decays~\cite{Ali:2007ff, Su:2011eq, Yan:2017nlj, Yan:2019nhf, Xiao:2019mpm}. There still exist discrepancy between different theoretical groups.

Motivated by this, we will calculate the TFF $f^{B_s\to K}_+(q^2)$ within the LCSRs approach by using the chiral corrlator. The chiral correlator was first introduced to deal with the $B\to \pi$ TFF~\cite{Huang:2001xb}, where the contributions of the twist-3 DAs in $f^{B \to \pi}_+(q^2)$ vanish automatically. Since then, the chiral correlator has been widely used to study the TFFs of $B_{u,d,s,c}\to$ various pseudoscalar, vector and scalar meson semileptonic decays~\cite{Li:2001yv, Huang:2001mq, Zuo:2006dk, Huang:2008zg, Wu:2007vi, Wu:2009kq, Huang:2008sn, Sun:2010nv, Li:2012gr, Zhong:2014fma, Zhang:2017rwz, Huang:2013gra, Huang:2013yya}. By using the chiral correlator, the $B_s\to K$ TFF $f^{B\to K}_+(q^2)$ is only expressed with the kaon twist-2 and twist-4 DAs, and the contribution of the twist-2 part is dominant, which also indicates that a more precise $\phi_{2;K}(x,\mu)$ is helpful to improve the prediction of the semileptonic decay $B_s\to K\ell\bar\nu_\ell$. In view of this, the kaon leading-twist DA $\phi_{2;K}(x,\mu)$ will be another research object in this work. The meson's light-cone DAs are universal nonperturbative inputs, which enter the exclusive processes involving large momentum transfer $Q^2\gg \Lambda_{\rm QCD}^2$ and $B/D$ meson two-body decays through factorization assumption, those processes can be decomposed into the long-distance dynamics (i.e., DAs) and the perturbatively calculable hard-scattering amplitudes~\cite{Boyle:2006pw, Chetyrkin:2007vm}. The DAs are main error sources in theoretical predictions, so their precise behaviors are important~\cite{Momeni:2017moz}. Comparing with the pionic leading-twist DA, the study of kaon leading-twist DA $\phi_{2;K}(x,\mu)$ will encounter $SU_f(3)$ symmetry breaking effect originating from the $s$-quark mass effect~\cite{Chetyrkin:2007vm}. The QCD sum rules and the LQCD are the most popular methods to study $\phi_{2;K}(x,\mu)$, which usually focus on the calculation of the first two moments of $\phi_{2;K}(x,\mu)$ then whose behavior can be approximated with its truncated form of the Gegenbauer polynomial expansion series~\cite{Bali:2019dqc}. On the other hand, the $SU_f(3)$ breaking effect in the kaon leading-twst DA can be realized by the difference between the longitudinal momentum fractions of the strange and nonstrange quarks, which is proportional to the first Gegenbauer moments $a_1^{2;K}(\mu)$, and also reflected in the ratio of the pion and kaon second Gegenbauer moments, i.e., $a_2^{2;K}(\mu)/a_2^{2;\pi}(\mu)$~\cite{Chetyrkin:2007vm, Choi:2007yu}.

There are great differences in the predictions of the QCD sum rules and the LQCD on the moments of the kaon leading-twist DA in history. The earliest QCD sum rules research on $a_1^{2;K}(\mu)$ is given by Chernyak and Zhitnitsky (CZ), about $a_1^{2;K} \approx 0.1$~\cite{Chernyak:1982it,Chernyak:1983ej}. P.~Ball and M.~Boglione point out the sign error of the contribution of the perturbation term in CZ calculation, and get $a_1^{2;K}(1~{\rm GeV}) = -0.18\pm 0.09$ and $a_2^{2;K}(1~{\rm GeV}) = 0.16\pm 0.10$~\cite{Ball:2003sc}. And also the results $a_1^{2;K}(1~{\rm GeV}) = 0.050\pm 0.025$~\cite{Ball:2005vx}, $a_1^{2;K}(1~{\rm GeV}) = 0.06\pm 0.03$~\cite{Ball:2006fz}. By adopting the diagonal correlation function of local and nonlocal axial-vector currents, A. Khodjamirian obtains $a_1^{2;K}(1~{\rm GeV}) = 0.05\pm 0.02$ and $a_2^{2;K}(1~{\rm GeV}) = 0.27^{+0.37}_{-0.12}$ with QCD sum rules~\cite{Khodjamirian:2004ga}. After considering the constrains from the exact operator identities, V. Braun obtains the QCD sum rules prediction as $a_1^{2;K}(1~{\rm GeV}) = 0.10\pm 0.12$~\cite{Braun:2004vf}, which is very consistent with the sum rule result in Ref.~\cite{Chetyrkin:2007vm}. In Ref.~\cite{Chetyrkin:2007vm} the gluon radiative correction up to $\mathcal{O}(\alpha_s^2)$ is calculated, where $a_1^{2;K}(1~{\rm GeV}) = 0.10\pm 0.04$. The results of LQCD calculation are generally small, the central value of $a_1^{2;K}$ is about in $0.45\sim 0.66$ at $\mu = 2~{\rm GeV}$~\cite{Braun:2006dg, Boyle:2006pw, Arthur:2010xf, Bali:2019dqc}. In addition, especially in recent years, the kaon leading-twist DA has also been studied by other methods as a whole, such as: the light-front quark model (LFQM)~\cite{Choi:2007yu, Dhiman:2019ddr}, the light-front constituent quark model (LFCQM)~\cite{deMelo:2015yxk}, the nonlocal chiral-quark model (NLChQM) from the instanten vacuum~\cite{Nam:2006au}, the Dyson-Schwinger equation (DSE) computation~\cite{Shi:2014uwa}, the framework of the anti-de Sitter/quantum chromodynamics (AdS/QCD)~\cite{Momeni:2017moz}, by taking the infinite-momentum limit for the quasi-distribution amplitude (QDA) within NLChQM~\cite{Nam:2017gzm} and LQCD based on the large-momentum effective theory (LaMET)~\cite{Chen:2017gck, Zhang:2020gaj}. In this paper, we will study the kaon leading-twist DA $\phi_{2;K}(x,\mu)$ by combining the phenomenological light-cone harmonic oscillator (LCHO) model and the QCD sum rules in the framework of the background field theory (BFTSR). This method has been suggested to study the pionic leading-twist DA $\phi_{2;\pi}(x,\mu)$ in Ref.~\cite{Zhong:2021epq}. Following the method, a improved LCHO model is firstly introduced to achieve a better behavior of $\phi_{2;K}(x,\mu)$. New sum rules are derived to achieve the DA moments, which are adopted to fix the parameters of the LCHO model by using the least squares method.

The rest of the paper are organized as follows. In Sec.~\ref{sec:II}, the branching ratio and the TFF of $B_s^+\to K \ell^+ \nu_\ell$, the LCHO model of $K$-meson twist-2 DA, and the moments of $\phi_{2;K}(u,\mu)$ under the BFTSR are presented. In Sec.~\ref{sec:III}, we provide the numerical results and make a comparison with the experimental and other theoretical predictions. Section~\ref{sec:IV} is reserved for a summery.

\section{Calculation Technology}\label{sec:II}

\subsection{LCSR on Semileptonic Decay $B_s^0\to K^- \mu^+ \nu_\mu$}

In order to study the CKM matrix element $|V_{ub}|$ from the semileptonic decay process $B_s^0 \to K^- \mu^+\nu_\mu$, we start from the following differential decay width over the squared momentum transfer,
\begin{align}
\frac{d\Gamma}{dq^2}(B_s^0 \to K^- \mu^+ \nu_\mu) & = \frac{G_F^2 |V_{ub}|^2 } {192 \pi^3 m_{B_s^0}^3} \bigg[(m_{B_s^0}^2 + m_{K^-}^2 - q^2)^2- 4m_{B_s^0}^2m_{K^-}^2\bigg]^{3/2} |f^{B_s\to K}_+(q^2)|^2,
\label{DifferentialWidth}
\end{align}
where $G_F = 1.166\times 10^{-5} ~ {\rm GeV}^{-2}$ is the Fermi constant, $m_{B_s^0}$ and $m_{K^-}$ are $B_s^0$- and $K^-$meson masses respectively. Apart from the experimental measurement on the decay width or branching fraction, the TFF $f^{B_s\to K}_+(q^2)$ is the key component in determining the CKM matrix element $|V_{ub}|$. To derive the LCSR of the TFF, one can start with the following correlator,
\begin{align}
\Pi_\mu(p,q) &= i\int d^4x e^{iq\cdot x}\langle K(p)| T \{j_V^\mu(x), j_{B_s}^\dag (0) \} | 0 \rangle \nonumber\\
&= F\left( q^2, (p+q)^2 \right) p_\mu + \widetilde{F} \left( q^2, (p+q)^2 \right) q_\mu,
\label{LCSRcorrelator}
\end{align}
where $j_V^\mu(x) = \bar u(x)\gamma_\mu(1+\gamma_5)b(x)$. For the $B_s$-meson current $j^\dag_{B_s}(0)$, we choose the right-handed current $j^\dag_{B_s}(0) = m_b \bar{b}(0) i (1 + \gamma_5 ) s(0)$, which highlights the contribution of the kaon twist-2 DA $\phi_{2;K}(x,\mu)$ and removes the less certain twist-3 DAs' contributions. The high-twist contributions are generally power suppressed to the twist-2 one, thus the accuracy of the derived LCSR can be greatly improved in comparison to the case of conventional choice of the correlator, if we know the twist-2 DA well.

On the one hand, the $B_s\to K$ matrix elements are related to the correlator via hadronic dispersion relation in the channel of the $(\bar b\gamma_5 s)$ current with the squared four-momentum $(p+q)^2$ based on the LCSR approach. After inserting hadronic states between the two current in correlator, one then isolates the ground state of $B_s$-meson contributions in the dispersion relations, and the hadron representation of the invariant amplitude $F(q^2,(p+q)^2)$ can be read off
\begin{eqnarray}
F(q^2,(p+q)^2) = \frac{2m_{B_s}^2f_{B_s}f_+^{B_s\to K}(q^2)}{m_{B_s}^2 - (p+q)^2} + \cdots,
\label{Eq:HadronicExpression}
\end{eqnarray}
where $f_{B_s}$ is the $B_s$-meson decay constant, the ellipses indicate the contribution of heavier states, and we have implicitly expressed the $B_s\to K$ transition matrix element as
\begin{align}
&\langle K(p)|\bar u\gamma^\mu b|B_s(p+q)\rangle = 2 f^{B_s\to K}_+(q^2) p_\mu + \left( f^{B_s\to K}_+(q^2) + f^{B_s\to K}_-(q^2) \right) q_\mu.
\label{Eq:TFFmatrix}
\end{align}
In this paper, we shall focus on the semileptonic decay $B_s^0\to K^- \mu^+ \nu_\mu$, where only the TFF $f^{B_s\to K}_+(q^2)$ contributes due to negligible muon mass. Thus, the $p_\mu$-terms in correlator \eqref{LCSRcorrelator} from transition matrix element \eqref{Eq:TFFmatrix} as well as the TFF $f^{B_s\to K}_-(q^2)$ shall not be considered here. On the other hand, in the region of $q^2\leq m_b^2$ and $(p+q)^2\leq m_b^2$, which is far from the $b$-flavour threshold, the $b$-quark propagating in the correlator is highly virtual and the distances near the light-cone $x^2 = 0$ dominate. The light-cone expansion of the $b$-quark propagator have the following form
\begin{eqnarray}
\langle 0 |b_\alpha^i(x) \bar b_\beta^j(0) |0\rangle &=& - i \int\frac{d^4k}{(2\pi^4)} e^{-ik\cdot x} \bigg[ \delta^{ij} \frac{\DS k + m_b}{m_b^2 - k^2} + g_s \int_0^1 dv G^{\mu\nu a}(vx) \bigg(\frac{\lambda^a}{2}\bigg)^{ij}
\nonumber\\
&\times& \bigg( \frac{\DS k + m_b }{2(m_b^2-k^2)^2}\sigma_{\mu\nu} + \frac{1}{m_b^2-k^2}vx_\mu\gamma_\nu\bigg)
\bigg]_{\alpha\beta}.
\end{eqnarray}
In calculation of the operator product expansion (OPE), the ${\cal O}(\alpha_s)$ gluon radiative corrections to the dominant twist-2 of correlator is considered. So the OPE result for the invariant amplitude $F(q^2,(p+q)^2)$ can be represented as
\begin{eqnarray}
F(q^2,(p+q)^2) = F_0(q^2,(p+q)^2) + \frac{\alpha_s C_F}{4\pi} F_1(q^2,(p+q)^2)
\end{eqnarray}

After substituting the OPE results of $F(q^2,(p+q)^2)$ into the Eq.~\eqref{Eq:HadronicExpression}, one can introduce an effective threshold parameter $s_0^{B_s}$ such that all the continuum states and excited states are separated, whose contributions could be approximated by using the quark-hadron duality. By further making the usual Borel transformation to suppress those less certain contributions from continuum and excited states, one then obtains the required LCSR for the TFF $f^{B_s\to K}_+(q^2)$
\begin{align}
&f^{B_s\to K}_+(q^2) = \frac{e^{m_{B_s}^2/M^2}}{m_{B_s}^2 f_{B_s}}\bigg[F_0(q^2,M^2,s_0^{B_s})+\frac{\alpha_s C_F}{4\pi} F_1(q^2,M^2,s_0^{B_s})\bigg], \label{Eq:f+}
\end{align}
where $C_F=4/3$. The LO contribution is
\begin{align}
F_0(q^2,M^2,s_0^{B_s})  &= m_b(m_b + m_s) f_K \int_{u_0}^1 du e^{-(m_b^2 -\bar u q^2 + u\bar u m_K^2)/(u M^2)} \bigg\{\frac{\phi_{2;K}(u)}{u} + \frac{1}{m_b^2 - q^2 + u^2 m_K^2}
\nonumber\\
&\times \bigg\{u \psi_{4;K}(u) + \bigg(1-\frac{2u^2m_K^2}{m_b^2 - q^2 + u^2 m_K^2}\bigg)\int_0^u \,dv \psi_{4;K}(v) - \frac{u m_b^2} {4(m_b^2 - q^2 + u^2 m_K^2)}
\nonumber\\
&\times \bigg(\frac{d^2}{du^2}~-~\frac{6um_K^2}{m_b^2 - q^2 + u^2 m_K^2}\frac{d}{du}~+~ \frac{12um_K^4}{(m_b^2 - q^2 + u^2 m_K^2)^2}\bigg)~\phi_{4;K}(u)~-~ \bigg(~\frac{d}{du}
\nonumber\\
& -\frac{2um_K^2}{m_b^2 - q^2 + u^2 m_K^2}\bigg)\int_0^u ~d\alpha_1 \int_{\frac{u-\alpha_1}{1-\alpha_1}}^1 \frac{dv}{v}~ \bigg(2\Psi_{4;K}(\alpha_i) \,-\, \Phi_{4;K}(\alpha_i) + 2\tilde\Psi_{4;K}(\alpha_i)
\nonumber\\
&- \tilde\Phi_{4;K}(\alpha_i)\bigg)~-~ \frac{2um_K^2}{m_b^2 - q^2 + u^2 m_K^2}~\bigg[u\frac{d}{du} +\bigg(1-\frac{4u^2m_K^2}{m_b^2 - q^2 + u^2 m_K^2}\bigg) \bigg] \int_0^u d\alpha_1
\nonumber\\
&\times \int_{\frac{u -\alpha_1}{1-\alpha_1}}^1 ~\frac{dv}{v}~ \bigg(~\Psi_{4;K}(\alpha_i) ~+~\Phi_{4;K}(\alpha_i) ~+~ \tilde\Psi_{4;K}(\alpha_i)~+~ \tilde\Phi_{4;K}(\alpha_i)~\bigg) ~+~ 2um_K^2
\nonumber\\
& \times \frac{m_b^2-q^2-u^2m_K^2}{(m_b^2-q^2+u^2m_K^2)^2}~\bigg(\frac{d}{du} ~-~ \frac{6um_K^2}{m_b^2-q^2+u^2m_K^2}\bigg) \int_u^1 dw \int_0^w d\alpha_1~\int_{\frac{u-\alpha_1}{1-\alpha_1}}^1 ~\frac{dv}{v}
\nonumber\\
&\times \bigg(\Psi_{4;K}(\alpha_i) +\Phi_{4;K}(\alpha_i) + \tilde\Psi_{4;K}(\alpha_i)+ \tilde\Phi_{4;K}(\alpha_i)\bigg)
\bigg\}\bigg\},
\end{align}
where $m_b$ is $b$-quark mass, $f_K$ is the kaon decay constant, $\bar{u} = 1 - u$ and
\begin{align}
u_0 = \frac{1}{2m_K^2} \left[ q^2 - s_0^{B_s} + m_K^2 + \sqrt{(q^2 - s_0^{B_s} + m_K^2)^2 - 4m_K^2(q^2 - m_b^2)} \right].
\end{align}
Since the contributions from the twist-4 terms are small, to do the numerical calculation, we take the kaon twist-4 DAs, i.e., $\psi_{4;K}(u)$, $\phi_{4;K}(u)$, $\Psi_{4;K}(\alpha_i)$, $\Phi_{4;K}(\alpha_i)$, $\tilde\Psi_{4;K}(\alpha_i)$ and $\tilde\Phi_{4;K}(\alpha_i)$ as those of the corresponding ones of pion~\cite{Duplancic:2008ix} due to small $SU_f(3)$ breaking effect. As for the three particle twist-4 DAs, the momentum fractions $\alpha_2$ and $\alpha_3$ are $\alpha_2 = 1-\alpha_1-\alpha_3$ and $\alpha_3 = (u - \alpha_1)/v$. The NLO term to the twist-2 part can be expressed in the form of the dispersion relation
\begin{align}
F_1(q^2,M^2,s_0^{B_s}) &= \frac{f_K}{\pi} \int_{m_b^2}^{s_0^{B_s}} ds e^{-s/M^2}
\int_0^1~ du~ {\rm Im} T_1(q^2,s,u) ~ \phi_{2;K}(u),
\end{align}
where the expression of $T_1(q^2,M^2,s_0^{B_s})$ can be found in Ref.~\cite{Duplancic:2008ix}.

\subsection{Improved LCHO model of $\phi_{2;K}(x,\mu)$}\label{sec:II.II}
The main nonperturbative uncertainty to the above LCSR \eqref{Eq:f+} comes from the kaon twist-2 DA, which could be derived from its twist-2 wavefunction (WF). The kaon WF can be constructed via the similar way of that of pion. The pion WF has been constructed by using LCHO model~\cite{BHL,Guo:1991eb,Huang:1994dy}. More explicitly, the pion WF starts from the following $SU(6)$ instant-form in the rest frame,
\begin{eqnarray}
\Psi_{\rm CM}(\mathbf{q}^2) = A \exp\bigg(-\frac{\mathbf{q}^2}{2\beta^2}\bigg) \frac{1}{\sqrt{2}} \left( \chi_1^\uparrow \chi_2^\downarrow - \chi_1^\downarrow \chi_2^\uparrow \right),
\label{InstantFormWF}
\end{eqnarray}
where $A$ is the normalization constant, the exponential factor $\exp\left[-\mathbf{q}^2/(2\beta^2)\right]$ is from an harmonic oscillator model for the meson bound-state within the valence quark model~\cite{WF_restframe}, and the remaining part is the spin WF with the two-component Pauli spinor $\chi_i^{\uparrow,\downarrow}$. In Eq.~(\ref{InstantFormWF}), the momenta of two quarks are indicated as $q_1^\mu = (q_1^0, \mathbf{q}_1)$ and $q_2^\mu = (q_2^0, \mathbf{q}_2)$, respectively, where $\mathbf{q}_1 = -\mathbf{q}_2 = \mathbf{q}$ and $q_1^0 = q_2^0 = q^0 = \sqrt{\hat{m}^2 + \mathbf{q}}$ with the $u$- and/or $d$- constituent quark mass $\hat{m}_u = \hat{m}_d = \hat{m}$ in the rest frame. Based on the BHL description~\cite{BHL}, that is, there is a connection between the equal-times WF in the rest frame and the light-cone WF, i.e.,
\begin{eqnarray}
\Psi_{\rm CM}(\mathbf{q}^2) \longleftrightarrow \Psi_{\rm LC} \left[ \frac{\mathbf{k}_\perp^2 + \hat{m}^2}{4x(1-x)} - \hat{m}^2 \right],
\label{BHLdescription}
\end{eqnarray}
the spatial part of LCHO model can be obtained, which is proportional to
\begin{eqnarray}
\exp \left[ - \frac{\mathbf{k}_\perp^2 + \hat{m}^2}{8\beta^2 x(1-x)} \right].
\label{ExponentialFactor}
\end{eqnarray}
By further using the Wigner-Melosh rotation~\cite{Wigner:1939cj,Melosh:1974cu,Kondratyuk:1979gj}, the spin WF in the light-cone frame can be obtained from the spin part in Eq.~(\ref{InstantFormWF})\footnote{For the specific derivation details, one can refer to Eqs.~(4.1) $\sim$ (4.12) in Ref.~\cite{Huang:1994dy}. In addition, it should be noted that the spin WF (\ref{SpinWF}) includes not only the two ordinary helicity ($\lambda_1 + \lambda_2 = 0$) components in Eq.~(\ref{InstantFormWF}), but also the two higher helicity ($\lambda_1 + \lambda_2 = \pm 1$) components, which arise from the Wigner-Melosh rotation.},
\begin{eqnarray}
\frac{a_1 a_2 - \mathbf{k}^2}{\left[ (a_1^2 + \mathbf{k}^2) (a_2^2 + \mathbf{k}^2) \right]^{1/2}},
\label{SpinWF}
\end{eqnarray}
where $a_1 = x\tilde{M} + \hat{m}$ and $a_2 = (1-x)\tilde{M} + \hat{m}$ with $\tilde{M} = \sqrt{(\mathbf{k}_\perp^2 + \hat{m}^2)/[x(1-x)]}$. Then, combining Eqs.~(\ref{ExponentialFactor}) and (\ref{SpinWF}), Ref.~\cite{Huang:1994dy} suggests a LCHO model of the pionic leading-twist WF, i.e.
\begin{align}
\Psi(x,\mathbf{k}_\perp^2) &= A \frac{a_1 a_2 - \mathbf{k}^2}{\left[ (a_1^2 + \mathbf{k}^2) (a_2^2 + \mathbf{k}^2) \right]^{1/2}} \exp \left[ - \frac{1}{8\beta^2} \left( \frac{\mathbf{k}_\perp^2 + \hat{m}^2}{ x} + \frac{\mathbf{k}_\perp^2 + \hat{m}^2}{ 1-x } \right) \right].
\label{PionLCHO}
\end{align}
By returning $\hat{m}$ in Eq.~(\ref{PionLCHO}) back to $\hat{m}_u$ and $\hat{m}_d$, and replacing one of them by the constituent $s$-quark mass $\hat{m}_s$, one then obtains the LCHO model of the kaon leading-twist WF~\cite{Wu:2008yr}. In the present paper, we will build on the LCHO model of Ref.~\cite{Wu:2008yr} and suggest a way to improve it.

\begin{table*}[t]
\centering
\caption{The specific forms of the four spin-space WF $\chi_{2;K}^{\lambda_1\lambda_2}(x,\textbf{k}_\bot)$ with different $\lambda$. }
\begin{tabular}{ c  c  c }
\hline\hline
$\lambda_1\lambda_2$& ~~~~~~~~~~~~~~~~~~~~~~~~~~~~~~~~~~~~~~~$\uparrow\uparrow$
~~~~~~~~~~~~~~~~~~~~~~~~~~~~~~~~~~~~~~~&
$\uparrow\downarrow$ \\
\hline
$\chi_{2;K}^{\lambda_1\lambda_2}(x,\textbf{k}_\bot)$ & $-\dfrac{(a_1 + a_2)(k_x - ik_y)}{\left[ 2(a_1^2 + \textbf{k}_\bot^2)(a_2^2 + \textbf{k}_\bot^2) \right]^{1/2}}$ & $\dfrac{a_1a_2 - \textbf{k}_\bot^2}{\left[ 2(a_1^2 + \textbf{k}_\bot^2)(a_2^2 + \textbf{k}_\bot^2) \right]^{1/2}}$ \\
\hline\hline
$\lambda_1\lambda_2$~ & ~$\downarrow\uparrow$~ & ~$\downarrow\downarrow$~ \\
\hline
$\chi_{2;K}^{\lambda_1\lambda_2}(x,\textbf{k}_\bot)$ & $-\dfrac{a_1a_2 - \textbf{k}_\bot^2}{\left[ 2(a_1^2 + \textbf{k}_\bot^2)(a_2^2 + \textbf{k}_\bot^2) \right]^{1/2}}$ & $-\dfrac{(a_1 + a_2)(k_x + ik_y)}{\left[ 2(a_1^2 + \textbf{k}_\bot^2)(a_2^2 + \textbf{k}_\bot^2) \right]^{1/2}}$ \\
\hline\hline
\end{tabular}
\label{tFourSpinWF}
\end{table*}
Let's introduce the follow-up work via usual way. Based on the BHL description~\cite{BHL}, the LCHO model of the kaon leading-twist WF $\Psi_{2;K}(x,\textbf{k}_\bot)$ can be written as:
\begin{align}
\Psi_{2;K}(x,\textbf{k}_\bot) = \chi_{2;K}(x,\textbf{k}_\bot) \Psi^R_{2;K}(x,\textbf{k}_\bot),
\label{WF_full}
\end{align}
where $\textbf{k}_\bot$ is the kaon transverse momentum. $\chi_{2;K}(x,\textbf{k}_\bot)$ stands for the total spin-space WF that comes from the Wigner-Melosh rotation~\cite{Wigner:1939cj,Melosh:1974cu,Kondratyuk:1979gj}, and
\begin{eqnarray}
\chi_{2;K}(x,\textbf{k}_\bot) = \sum_{\lambda_1\lambda_2} \chi_{2;K}^{\lambda_1\lambda_2}(x,\textbf{k}_\bot),
\label{Chi}
\end{eqnarray}
where $\chi_{2;K}^{\lambda_1\lambda_2}(x,\textbf{k}_\bot)$ is the spin-space WF, corresponding to four different types of the helicities of the two constituent quarks, i.e., $\lambda_1\lambda_2 = (\uparrow\uparrow, \uparrow\downarrow, \downarrow\uparrow, \downarrow\downarrow)$, respectively. Their specific forms are listed in the table~\ref{tFourSpinWF}, which lead to~\cite{Wu:2008yr}
\begin{align}
\chi_{2;K}(x,\textbf{k}_\bot) = \frac{\tilde{m}}{\sqrt{\textbf{k}_\bot^2 + \tilde{m}^2}},
\label{WF_spin}
\end{align}
where $\tilde{m} = \hat{m}_q x + \hat{m}_s (1-x)$. $q$ indicates the light quark, $q=u$ is for $K^0$ and $q=d$ is for $K^+$. For the values of the constituent quark masses $\hat{m}_s$ and $\hat{m}_q$, several schemes have been adopted in literature. For example, $\hat{m}_s = 370\ {\rm MeV}$ and $\hat{m}_q = 250\ {\rm MeV}$~\cite{Choi:1997iq,Jaus:1991cy} in the invariant meson mass scheme (MS)~\cite{Jaus:1989au,Jaus:1991cy,Chung:1988mu, Choi:1997qh,Huang:1994dy,Schlumpf:1994bc,Cardarelli:1994yq, Cardarelli:1995ap,Cardarelli:1994ix}, $\hat{m}_s = 450\ {\rm MeV}$ and $\hat{m}_q = 330\ {\rm MeV}$ in the spin-averaged meson MS~\cite{Dziembowski:1986dr, Dziembowski:1987zp, Ji:1990rd, Ji:1992yf, Choi:1996mq}, $\hat{m}_s = 450\ {\rm MeV}$ and $\hat{m}_q = 300\ {\rm MeV}$ for the simplest in Refs.~\cite{Wu:2007rt, Wu:2007vi, Wu:2008yr, Wu:2009kq, Wu:2011gf}. We will analyze the behavior of the kaon leading-twist DA under different schemes in Sec.~\ref{sec:III}, and choose the resultant DA corresponding MS to further study the semileptonic decay $B_s^0\to K^-\mu^+\nu_\mu$. $\Psi^R_{2;K}(x,\textbf{k}_\bot)$ in Eq.~(\ref{WF_full}) stands for the spatial WF, and which reads\footnote{In principle, the spatial part of kaon leading-twist WF should contain a Jacobi factor~\cite{Choi:1997iq}; However, numerical prediction in Ref.~\cite{Zhong:2021epq} shows that the influence of such factor on the pionic leading-twist DA is small. Therefore, we also ignore the effect of Jacobi factor to the kaon WF/DA.}
\begin{align}
\Psi^R_{2;K}(x,\textbf{k}_\bot) &= A_{2;K} \varphi_{2;K}(x) \exp \left[ - \frac1{8\beta_{2;K}^2} \left( \frac{\textbf{k}^2_\bot + \hat{m}_s^2}{x} + \frac{\textbf{k}^2_\bot +
\hat{m}_q^2}{1-x} \right) \right],
\label{WF_spatial}
\end{align}
where $A_{2;K}$ is the normalization constant, $\beta_{2;K}$ is the harmonious parameter that dominates the WF's transverse distribution, and $\varphi_{2;K}(x)$ dominates the WF's longitudinal distribution, we take its form as
\begin{align}
\varphi_{2;K}(x) &= \left[ x(1-x) \right]^{\alpha_{2;K}}  \bigg[ 1 + \hat{B}_1^{2;K} C_1^{3/2}(2x-1)
+ \hat{B}_2^{2;K} C_2^{3/2}(2x-1) \bigg],
\label{varphi}
\end{align}
where $C_n^{3/2}(2x-1)$ is the Gegenbauer polynomial. The $\varphi_{2;K}(x)$ is constructed by applying the idea of constructing pionic longitudinal distribution function $\varphi_{2;\pi}^{\rm IV}(x)$ suggested in Ref.~\cite{Zhong:2021epq} to the present case of kaon. The factor $\left[ x(1-x) \right]^{\alpha_{2;K}}$ regulates the behavior of $\Psi_{2;K}(x,\textbf{k}_\bot)$ and $\phi_{2;K}(x,\mu)$. Considering the $SU_f(3)$ breaking effect, we add a term proportional to $C_1^{3/2}(2x-1)$. We set $\hat{B}^{2;K}_1 = 0.4\hat{B}^{2;K}_2$ so as to make the undetermined model parameters as few as possible, the factor $0.4$ is from the ratio of the first and second Gegenbauer moments, e.g., $|a^{2;K}_1/a^{2;K}_2|$. For the values of those two Gegenbauer moments, one can find in Sec.~\ref{sec:III}. The rationality of the relationship between $\hat{B}_1^{2;K}$ and $\hat{B}_2^{2;K}$ can be judged by the goodness of fit. Substituting the WF formula (\ref{WF_full}) with Eqs.~(\ref{WF_spin}), (\ref{WF_spatial}) and (\ref{varphi}) into the relationship between the kaon leading-twist DA and its WF, i.e.,
\begin{align}
\phi_{2;K}(x,\mu) = \frac{2\sqrt{6}}{f_K} \int_{| \mathbf{k}_\bot |^2 \leq \mu^2} \frac{d^2\mathbf{k}_\bot}{16\pi^3}
\Psi_{2;K}(x,\mathbf{k}_\bot),
\label{DA_WF}
\end{align}
and after integrating over the transverse momentum $\mathbf{k}_\bot$, the kaon leading-twist DA, $\phi_{2;K}(x,\mu)$, can be written as
\begin{align}
\phi_{2;K}(x,\mu) &= \frac{\sqrt{3} A_{2;K} \beta_{2;K} \tilde{m}}{2\pi^{3/2}f_K} \sqrt{x(1-x)} \varphi_{2;K}(x)
\left[
-\frac{\hat{m}_q^2x + \hat{m}_s^2(1-x) - \tilde{m}^2}{8\beta_{2;K}^2 x(1-x)} \right]
\nonumber\\
&
\times  \bigg\{ {\rm Erf} \left( \sqrt{\frac{\tilde{m}^2 + \mu^2}{8\beta_{2;K}^2x(1-x)}} \right)
-  {\rm Erf} \left( \sqrt{
\frac{\tilde{m}^2}{8\beta_{2;K}^2x(1-x)} } \right) \bigg\}.
\label{DA_model}
\end{align}
We ignore the mass difference between $u$ and $d$-quarks, the WF $\Psi_{2;K}(x,\textbf{k}_\bot)$ and the DA $\phi_{2;K}(x,\mu)$ are the same for $K^0$ and $K^+$. By replacing $x$ with $(1-x)$ in Eqs. \eqref{WF_full} and \eqref{DA_model}, one can obtain the leading-twist WF and DA of $\bar K^0$ and $K^-$.

The input parameters $A_{2;K}$, $\beta_{2;K}$, $\alpha_{2;K}$ and $\hat{B}_2^{2;K}$ satisfy the following two constraints,
\begin{itemize}
\item The normalization condition of the kaon leading-twist DA,
\begin{align}
\int^1_0 dx \phi_{2;K}(x,\mu) = 1;
\label{DA_constraint1}
\end{align}
\item The probability of finding the leading Fock-state $|\bar{s}q\rangle $ in the kaon Fock state expansion~\cite{Guo:1991eb},
\begin{align}
P_K &= \int^1_0 dx \int \frac{d^2 \textbf{k}_\bot}{16\pi^3} | \Psi_{2;K}(x,\textbf{k}_\bot) |^2.
\label{DA_constraint2}
\end{align}
\end{itemize}
The pionic leading-twist WF satisfies $P_\pi \simeq 0.2$~\cite{Zhong:2021epq}, we then adopt $P_K \simeq 0.3$ following the discussion of Ref.~\cite{Guo:1991eb}. Using the constraints (\ref{DA_constraint1}) and (\ref{DA_constraint2}), there are two free parameters left, which can be selected as $\alpha_{2;K}$ and $\hat{B}_2^{2;K}$. They are determined by adopting the least squares method to fit the moments $\langle\xi^n \rangle _{2;K}|_\mu$ of $\phi_{2;K}(x,\mu)$, defined as
\begin{align}
\langle \xi^n\rangle _{2;K} |_\mu = \int^1_0 dx (2x-1)^n \phi_{2;K}(x,\mu),
\label{moment}
\end{align}
which will be calculated in next subsection with BFTSR. In Sec.~\ref{sec:III}, we will adopt the values of the first ten moments to give strong constraint on those parameters. In the specific fitting, the undetermined model parameters $\alpha_{2;K}$ and $\hat{B}_2^{2;K}$ are regarded as the fitting parameters, i.e., $\mathbf{\theta} = (\alpha_{2;K}, \hat{B}^{2;K}_2)$. The moments $\langle\xi^n \rangle _{2;K}|_\mu$ from Eqs. \eqref{varphi}, \eqref{DA_model} and \eqref{moment} are regarded as the mean function $\mu(x_i;\mathbf{\theta})$ ($x_i \longrightarrow n$), while those moments with their errors calculated with BFTSR are regarded as the independent measurements $y_i$ with the known variance $\sigma_i$. Obviously, our goal is to obtain the best values of fitting parameters $\mathbf{\theta}$, and which can be achieved by minimizing the likelihood
function
\begin{align}
\chi^2(\mathbf{\theta}) =
\sum^{10}_{i=1} \frac{(y_i - \mu(x_i,\mathbf{\theta}))^2}{\sigma_i^2}.
\label{ls}
\end{align}
The goodness of fit is judged by the magnitude of the probability
\begin{align}
P_{\chi^2} = \int^\infty_{\chi^2} f(y;n_d) dy.
\label{px2}
\end{align}
Here $f(y;n_d)$ with the number of degrees of freedom $n_d$ is the probability density function of $\chi^2(\theta)$, and
\begin{align}
f(y;n_d) = \frac{1}{\Gamma\left(\dfrac{n_d}{2}\right) 2^{\frac{n_d}{2}}} y^{\frac{n_d}2-1} e^{-\frac y2}.
\end{align}

\begin{figure*}[t]
\centering
\includegraphics[width=0.6\textwidth]{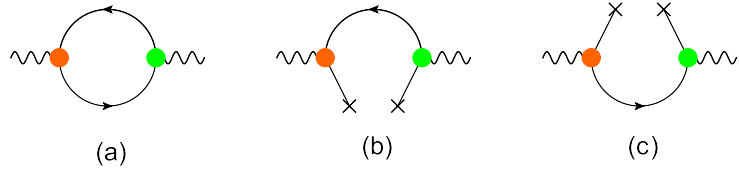}
\caption{The Feynman diagrams for Eq.\eqref{correlatorOPE}. The left big dot and the right big dot stand for the vertex operators ${z\!\!\!\slash} \gamma_5 (iz\cdot\tensor{D})^n$ and ${z\!\!\!\slash} \gamma_5$ from the currents $J_n(x)$ and $J^\dag_0(0)$, respectively. The cross symbol attached to the quark line indicates the local $s$- or $u/d$-quark background field.}
\label{ffeyn}
\end{figure*}

\subsection{The moments $\langle\xi^n \rangle _{2;K}|_\mu$ of $\phi_{2;K}(x,\mu)$ under the BFTSR approach}
To derive the sum rules for the kaon leading-twist DA moments $\langle \xi^n\rangle _{2;K}$, we introduce the following correlation function
(correlator),
\begin{align}
\Pi_{2;K} (z,q) & = i \int d^4x e^{iq\cdot x} \langle 0| T \{ J_n(x) , J^\dag_0(0) \} |0\rangle
\nonumber\\ &
= (z\cdot q)^{n+2} I_{2;K} (q^2) ,
\label{Correlator}
\end{align}
where $n=0,1,2,\cdots$ and $z^2 = 0$. The current $J_n(x) = \bar{s}(x) {z\!\!\!\slash} \gamma_5 (i z\cdot \tensor{D})^n q(x)$ with the fundamental representation of the gauge covariant derivative $D_\mu = \partial_\mu - ig_s T^A \mathcal{A}^A_\mu(x) (A = 1, \cdots, 8)$, and in which $s$ and $q$ indicate the $s$-quark and $u/d$-quark fields respectively.

In physical region, the correlator~\eqref{Correlator} can be treated by inserting a complete set of intermediate hadronic states. With $\langle 0| J_n(0) |K(q)\rangle  = i(z\cdot q)^{n+1} f_K \langle \xi^n\rangle_{2;K}$, the hadronic representation of correlator~\eqref{Correlator} reads
\begin{align}
\textrm{Im} I^{\rm had}_{2;K}(s) &= \pi \delta (s - m_K^2) f_K^2 \langle \xi^n\rangle _{2;K} + \pi {\rm Im} I^{\rm pert}_{2;K}(s) \theta
(s - s_K), \label{HadronRepresentation}
\end{align}
where the quark-hadron duality has been adopt and $m_K$, $s_K$ are the kaon mass and the continuum threshold, respectively. On the other hand, in deep Euclidean region, we apply the OPE for the correlator \eqref{Correlator} in the framework of BFT. The basic idea of BFT is that the quark and gluon fields are composed of background fields and quantum fluctuations (quantum fields) around them. By adopting the Feynman rule of BFT, that is, the quark and gluon quantum fields are contracted into the corresponding propagators, while the quark and gluon background fields combine the vacuum operators to form the vacuum matrix elements, the correlator~\eqref{Correlator} can be rewritten as
\begin{align}
\Pi_{2;K} (z,q) = i \int d^4x e^{iq\cdot x}  \Big\{&-{\rm Tr} \langle0 | S_F^s(0,x) {\DS z} \gamma_5 (iz\cdot \tensor{D})^n S^q_F(x,0) {z\!\!\!\slash} \gamma_5  | 0 \rangle \nonumber\\
&+ \langle0 | \bar{s}(x)s(0) {z\!\!\!\slash} \gamma_5 (iz\cdot \tensor{D})^n S^q_F(x,0) {z\!\!\!\slash} \gamma_5  | 0 \rangle
\nonumber\\
&+  \langle0 | S_F^s(0,x) {z\!\!\!\slash} \gamma_5 (iz\cdot \tensor{D})^n \bar{q}(0)q(x) {z\!\!\!\slash} \gamma_5  | 0 \rangle  \Big\} \nonumber\\
&+ \cdots ,
\label{correlatorOPE}
\end{align}
where $\rm Tr$ indicates trace of the $\gamma$-matrix and color matrix, $S_F^s(0,x)$ is the $s$-quark propagator from $x$ to $0$, $S^q_F(x,0)$ stands for the $u/d$-quark propagator from $0$ to $x$, ${z\!\!\!\slash} \gamma_5 (iz\cdot \tensor{D})^n$ and ${z\!\!\!\slash} \gamma_5$ are the vertex operators from currents $J_n(x)$ and $J^\dag_0(0)$, respectively. The Feynman diagrams for Eq.~\eqref{correlatorOPE} are shown in Figure~\ref{ffeyn}, in which the left big dot and the right big dot stand for the vertex operators ${z\!\!\!\slash} \gamma_5 (iz\cdot \tensor{D})^n$ and ${z\!\!\!\slash} \gamma_5$, respectively. The cross symbol attached to the quark line indicates the local $s$- or $u/d$-quark background field. Figs.~\ref{ffeyn}(a), \ref{ffeyn}(b) and \ref{ffeyn}(c) correspond to the first, second and third terms in Eq.~\eqref{correlatorOPE}, respectively. The expressions up to dimension-six of quark propagator and operator $(iz\cdot \tensor{D})^n$ have been derived in Refs.~\cite{Zhong:2014jla,Hu:2021zmy}. By substituting the formulae of $S_F^{s(q)}(0,x)$ and $(iz\cdot \tensor{D})^n$ into Eq.~\eqref{correlatorOPE}, while expanding $\bar{s}(x)$ and $q(x)$ near $0$, the long- and short-distance quark-gluon interactions can be separated with the help of the vacuum matrix element formulae~\cite{Zhong:2014jla,Zhong:2011rg}. Finally, we obtain
\begin{widetext}
\begin{align}
\hat{L}_M I^{\rm QCD}_{2;K} &= \frac{1}{\pi} \frac{1}{M^2} \int^{\infty}_{m_s^2} ds e^{-s/M^2} {\rm Im} I^{\rm pert}_{2;K}(s) + \frac{m_s \langle
\bar{s}s \rangle  + (-1)^n m_q \langle  \bar{q}q \rangle }{(M^2)^2} + \frac{\langle \alpha_sG^2\rangle }{(M^2)^2} \,\,[1+(-1)^n]
\nonumber\\[1.2ex]
&\times \frac{1}{24\pi} \frac{1 + n\theta(n-2)}{n+1} - \frac{8n+1}{18} ~\frac{m_s \langle  g_s\bar{s}\sigma TGs \rangle  + (-1)^n m_q \langle
g_s\bar{q}\sigma TGq \rangle }{(M^2)^3} + \frac{2(2n+1)}{81}
\nonumber\\[1.2ex]
&\times  \frac{\langle g_s\bar{s}s\rangle ^2 \!+\! (-1)^n \langle g_s\bar{q}q\rangle ^2}{(M^2)^3}+ \frac{\langle g_s^3fG^3\rangle }{(M^2)^3}
[1+(-1)^n] \frac{-n\theta(n-2)}{96\pi^2} + \frac{\langle g_s^2\bar{q}q\rangle ^2}{(M^2)^3} [1+(-1)^n]
\nonumber\\[1.2ex]
&\times \frac{2 + \kappa^2}{972\pi^2} ~\bigg\{ -2(51n+25) \Big( -\ln \frac{M^2}{\mu^2} \Big) + 3~(17n+35) +  \theta(n-2)~ \bigg[ ~2n\Big(
-\ln \frac{M^2}{\mu^2} \Big)
\nonumber\\[1.2ex]
&- 25(2n+1) \tilde\psi (n) + \frac{1}{n}(49n^2+100n+56) \bigg] \bigg\} + \mathcal{O}(m_s^2).
\label{OPE}
\end{align}
where $\hat{L}_M$ indicates the Borel transformation operator with the Borel parameter $M$, and we have taken $g_s^2 \sum_{\psi=u,d,s} \langle g_s\bar{\psi}\psi\rangle ^2 = (2 + \kappa^2) \langle g_s^2\bar{q}q\rangle ^2$ with $\langle  \bar{s}s \rangle /\langle  \bar{q}q \rangle  = \kappa$. In Eq.~\eqref{OPE},
\begin{align}
{\rm Im} I^{\rm pert}_{2;K}(s) = \frac{3}{8\pi(n+1)(n+3)} \left\{ \left[ 2(n+1)\frac{m_s^2}{s}\left( 1 - \frac{m_s^2}{s} \right) + 1 \right] \left( 1 -
\frac{2m_s^2}{s} \right)^{n+1} + (-1)^n \right\},
\label{pert}
\end{align}
\begin{align}
\mathcal{O}(m_s^2) &= \frac{\langle \alpha_sG^2\rangle }{M^6} ~ \frac{m_s^2}{24n\pi}~ \bigg\{ 2n ~\bigg[ -2n
\Big(-\ln\frac{M^2}{\mu^2}\Big) + n + 2 \bigg] ~+~ \theta(n-1)~ \bigg[ (-1)^n~ \bigg(   n \tilde{\tilde\psi}(n) - 2\bigg) \bigg]
\nonumber\\[1.2ex]
&+ \theta(n-2) ~\bigg[ 3n+(-1)^n\bigg(n+2-n(2n+1) \tilde{\tilde\psi}(n)\bigg) \bigg] \bigg\} + \frac{\langle g_s^3fG^3\rangle }{M^8}
\frac{m_s^2}{576\pi^2}~ \bigg\{ ~\bigg[ 8n(3n-1)
\nonumber\\[1.2ex]
&\times\left(-\ln\frac{M^2}{\mu^2}\right) - 21n^2-53n+6-10\delta^{n0} \bigg] ~+\,\, \theta(n-1) \bigg[ 4n(2n-1)
\bigg(-\ln\frac{M^2}{\mu^2}\bigg) - 4(n^2
\nonumber\\[1.2ex]
&+3n-1) \bigg]~+~ \theta(n-2) ~\bigg[ ~2n~[6n + 3(-1)^n-1] + 4(-1)^n~\bigg(1-n \tilde{\tilde\psi}(n) \bigg) \bigg] ~+~ \theta(n-3)
\nonumber\\[1.2ex]
&\times \bigg[ -19n^2 - (3+16(-1)^n)n  6(-1)^n + 2(-1)^n n(8n-1) \tilde{\tilde\psi}(n) \bigg] \bigg\} + \frac{\langle g_s^2\bar{q}q\rangle
^2}{(M^2)^4} ~m_s^2~ \frac{2 + \kappa^2}{7776\pi^2}
\nonumber\\[1.2ex]
&\times\bigg\{ ~72\delta^{n0}~ -~ 768\delta^{n1}~+~ 8 \bigg[  ~6n(1+17(-1)^n) -108n^2 - (-1)^n - 1 \bigg]
~\left(-\ln\frac{M^2}{\mu^2}\right) - 12
\nonumber\\[1.2ex]
&\times\bigg[ -63n^2 ~+~ (-193 + 34(-1)^n)n ~+ 106(-1)^n - 56 \bigg] + \theta(n-1) ~\bigg[ - 16n(2n+(-1)^n
\nonumber\\[1.2ex]
&-1) \bigg(-\ln\frac{M^2}{\mu^2}\bigg) + \frac{8}{n} \bigg( -6n^3 + (6+4(-1)^n)n^2 + 2(-1+(-1)^n)n - 23(1+(-1)^n) \bigg) \bigg]
\nonumber\\[1.2ex]
&+ \theta(n-2)~ \bigg[ -\frac{8}{n(n-1)} ~\bigg( 4n^4 + (-3+53(-1)^n)n^3 + (146-74(-1)^n)n^2 ~- 3~(49+9
\nonumber\\[1.2ex]
&\times (-1)^n)~n ~+~ 24~(1+(-1)^n)~ \bigg)
-  8n~(-50+21(-1)^n) \tilde{\tilde\psi}(n) ~+~ 92(1+(-1)^n) \tilde\psi(n) ~\bigg]
\nonumber\\[1.2ex]
&+ \theta(n-3) \bigg[ \frac{4}{n-1} \bigg( 139n^3 + 16(-10+7(-1)^n)n^2  +  (69-106(-1)^n)n - 54(-1)^n \bigg) - 8
\nonumber\\[1.2ex]
&\times\bigg(~ 56n^2(-1)^n  ~-~ 25n(-1)^n  ~+ 12(1+(-1)^n) \bigg) \tilde{\tilde\psi}(n) \bigg] \bigg\} + \frac{2n+1}{3} \frac{m_s^2}{M^2}
\frac{m_s\langle \bar{s}s\rangle }{(M^2)^2} - (-1)^n
\nonumber\\[1.2ex]
&\times \frac{m_s^2}{M^2} \frac{m_q\langle \bar{q}q\rangle }{(M^2)^2} + (-1)^n\frac{8n-3}{18} \frac{m_s^2}{M^2} \frac{m_q\langle
g_s\bar{q}\sigma TGq\rangle }{(M^2)^3} - (-1)^n\frac{2(2n+1)}{81} \frac{m_s^2}{M^2} \frac{\langle g_s\bar{q}q\rangle ^2}{(M^2)^3},
\label{MsMassCorrection}
\end{align}
\end{widetext}
where $\tilde\psi (n) = \psi\left(\frac{n+1}{2}\right) - \psi\left(\frac{n}{2}\right) + \ln 4$ and $\tilde{\tilde\psi}(n) = \psi\left(\frac{n+1}{2}\right) - \psi\left(\frac{n}{2}\right) + (-1)^n \ln 4$. In specific calculation OPE, $m_q^2 \sim 0$ have been adopted for very small $u/d$ current quark mass, while the $s$-quark mass corrections proportional to $m_s^2$ for double-gluon condensate $\langle \alpha_sG^2\rangle$, triple-gluon condensate $\langle g_s^3fG^3\rangle$, double-quark condensate $\langle \bar{s}s\rangle$ and $\langle \bar{q}q\rangle$, quark-gluon mixed condensate $\langle g_s\bar{q}\sigma TGq\rangle$ and four-quark condensate $\langle g_s\bar{q}q\rangle ^2$, i.e., $\mathcal{O}(m_s^2)$ shown in Eq.~\eqref{MsMassCorrection}, are calculated due to $m_s \sim 0.1~{\rm GeV}$. In addition, the full $s$-quark mass effect in the perterbative part is preserved (see Eq.~\eqref{pert}).

Substituting the hadronic representation \eqref{HadronRepresentation} and OPE \eqref{OPE} of the correlator \eqref{Correlator} into the dispersion relation,
\begin{eqnarray}
\frac{1}{\pi} \frac{1}{M^2} \int ds e^{-s/M^2} {\rm Im} I^{\rm had}_{2;K}(s) = \hat{L}_M I^{\rm QCD}_{2;K}(q^2),
\label{DispersionRelation}
\end{eqnarray}
the sum rules of the moments of Kaon leading-twist DA reads:
\begin{widetext}
\begin{align}
\frac{\langle \xi^n\rangle _{2;K}\langle \xi^0\rangle _{2;K} f_{K}^2}{M^2 e^{m_K^2/M^2}}
&= \frac{1}{\pi} \frac{1}{M^2} \int^{s_K}_{m_s^2} ds e^{-s/{M^2}} {\rm Im} I^{\rm pert}_{2;K}(s) \!+\! \frac{m_s \langle  \bar{s}s \rangle  + (-1)^n m_q \langle  \bar{q}q \rangle
}{(M^2)^2} + \frac{\langle \alpha_sG^2\rangle }{(M^2)^2} [1+(-1)^n] \nonumber\\[1.2ex]
&\times \frac1{24\pi}\frac{1 + n\theta(n-2)}{n+1}\!- \frac{8n+1}{18} \frac{m_s \langle  g_s\bar{s}\sigma TGs \rangle  + (-1)^n m_q \langle g_s\bar{q}\sigma TGq \rangle }{(M^2)^3} + \frac{2(2n+1)}{81} \nonumber\\[1.2ex]
& \times \frac{\langle g_s\bar{s}s\rangle ^2 \!+\! (-1)^n \langle g_s\bar{q}q\rangle ^2}{(M^2)^3}\!+\! \frac{\langle g_s^3fG^3\rangle }{(M^2)^3} [1+(-1)^n]\frac{-n\theta(n-2)}{96\pi^2} + \!\frac{\langle g_s^2\bar{q}q\rangle ^2}{\left(M^2\right)^3} [1+(-1)^n] \nonumber\\
& \times\frac{2 + \kappa^2}{972\pi^2}~ \bigg\{-2(51n+25) \left( -\ln \frac{M^2}{\mu^2} \right) ~ +  3(17n+35) + \theta(n-2) \bigg[ 2n\!\left( -\ln \frac{M^2}{\mu^2} \right)
\nonumber\\[1.2ex]
&- 25(2n+1) \tilde\psi (n) + \frac{1}{n}(49n^2+100n+56) \bigg] \bigg\} + \mathcal{O}(m_s^2).
\label{MomentsSumRules}
\end{align}
Obviously, by performing the replacement $s\to d$, $q\to u$ and $K\to \pi$, and taking $m_s^2 = 0$, the sum rules~\eqref{MomentsSumRules} with even ``$n$'' degenerates to the case of the pionic leading-twist DA, i.e., Eq. (7) in Ref.~\cite{Zhong:2021epq}. By taking $n=0$ in Eq.~\eqref{MomentsSumRules}, one can get the sum rule of zeroth moment, which reads
\begin{align}
\frac{\langle \xi^0\rangle ^2_{2;K} f_{K}^2}{M^2 e^{m_K^2/M^2}}
&=\frac{1}{8\pi^2M^2} \!\int^{s_K}_{m_s^2} \!ds e^{-s/{M^2}} \!\left\{\! \left[ \frac{2m_s^2}{s} \left( 1\! -\!\frac{m_s^2}{s} \right) \!+\! 1 \right] \left( 1 - \frac{2m_s^2}{s} \right) \!+\! 1 \right\} \!+\! \left( 1 + \frac{m_s^2}{3M^2} \right) \frac{m_s \langle  \bar{s}s \rangle }{(M^2)^2}
\nonumber\\[1.2ex]
&+ \left( 1 \!-\! \frac{m_s^2}{M^2} \right) \frac{ m_q \langle  \bar{q}q \rangle }{(M^2)^2} + \frac{1}{12\pi} \left( \!1 + 2\frac{m_s^2}{M^2} \!\right) \frac{\langle \alpha_sG^2\rangle }{(M^2)^2} - \frac{1}{18} \frac{m_s \langle  g_s\bar{s}\sigma TGs \rangle }{(M^2)^3} - \frac{1}{18} \left( 1 + 3\frac{m_s^2}{M^2} \right)
\nonumber\\[1.2ex]
&\times \frac{ m_q \langle g_s\bar{q}\sigma TGq \rangle }{(M^2)^3} - \frac{m_s^2}{144\pi^2M^2} \frac{\langle g_s^3fG^3\rangle }{(M^2)^3} + \frac{2}{81} \frac{\langle g_s\bar{s}s\rangle ^2}{(M^2)^3} + \frac{2}{81} \left( 1 - \frac{m_s^2}{M^2} \right) \frac{ \langle g_s\bar{q}q\rangle ^2}{(M^2)^3} + \frac{2 + \kappa^2}{486\pi^2}
\nonumber\\[1.2ex]
&\times \frac{\langle g_s^2\bar{q}q\rangle ^2}{(M^2)^3} \left[ -\left( 50 + \frac{m_s^2}{M^2} \right) \left( -\ln \frac{M^2}{\mu^2} \right) + 105 - 3\frac{m_s^2}{M^2} \right].
\label{Moment0SumRules}
\end{align}
\end{widetext}
Eq.~\eqref{Moment0SumRules} indicates that the zeroth moment $\langle \xi^0\rangle _{2;K}$ in Eq.~\eqref{MomentsSumRules} cannot be normalized in the whole Borel parameter region as the case of pionic leading-twist DA (see Ref.~\cite{Zhong:2021epq}). Thus as suggested in Ref.~\cite{Zhong:2021epq}, we adopt the following sum rules of the moments $\langle \xi^n\rangle _{2;K}$ to do the calculation, i.e.
\begin{eqnarray}
\langle \xi^n\rangle _{2;K} = \frac{\left( \langle \xi^n\rangle _{2;K}\langle \xi^0\rangle _{2;K} \right) |_{\rm From\ Eq.~\eqref{MomentsSumRules}}}{\sqrt{\langle \xi^0\rangle ^2_{2;K}} |_{\rm From\ Eq.~\eqref{Moment0SumRules}}},
\label{xin}
\end{eqnarray}
Meanwhile, we also assume that the zeroth moment of kaon leading-twist DA can be normalized in an appropriate Borel window in order to ensure the QCD sum rule's predictive ability for meson decay constant or determine the continuum threshold $s_K$ with Eq.~\eqref{Moment0SumRules}.

\section{numerical analysis}\label{sec:III}

\subsection{Input Parameters}
To do the numerical calculation, we adopt the latest data from Particle Data Group (PDG)~\cite{PDGnew}: $m_K = 493.677 \pm 0.013 ~\textrm{MeV}$, the current-quark-mass for the $u${\color{blue}\uwave{-}} and $s$-quark are adopted as $m_u = 2.16^{+0.49}_{-0.26} \textrm{MeV}$ and $m_s = 93^{+11}_{-5} \textrm{MeV}$ at scale $\mu = 2~\rm GeV$, respectively. The kaon decay constant is taken to be $f_K/f_\pi = 1.1932 \pm 0.0019$~\cite{FlavourLatticeAveragingGroup:2019iem} with $f_\pi = 130.2(1.2) \textrm{MeV}$~\cite{PDGnew}. The other inputs are exhibited as follows\cite{Zhong:2021epq,Zhong:2014jla,Colangelo:2000dp,Narison:2014ska,Narison:2014wqa}:
\begin{align}
\langle \bar{q}q\rangle  &= \left( -2.417_{-0.114}^{+0.227} \right) \times 10^{-2} ~{\rm GeV}^3, \nonumber\\
\langle \bar{s}s\rangle  &= \kappa \langle \bar{q}q\rangle, \nonumber\\
\langle g_s\bar{q}\sigma TGq\rangle  &= \left( - 1.934^{+0.188}_{-0.103} \right) \times 10^{-2} ~{\rm GeV}^5, \nonumber\\
\langle g_s\bar{s}\sigma TGs\rangle  &= \kappa \langle g_s\bar{q}\sigma TGq\rangle, \nonumber\\
\langle g_s\bar{q}q\rangle ^2 &= (2.082^{+0.734}_{-0.697}) \times 10^{-3} \textrm{GeV}^6, \nonumber\\
\langle g_s\bar{s}s\rangle ^2 &= \kappa^2 \langle g_s\bar{q}q\rangle ^2, \nonumber\\
\langle g_s^2\bar{q}q\rangle ^2 &= (7.420^{+2.614}_{-2.483}) \times 10^{-3} \textrm{GeV}^6, \nonumber\\
\langle \alpha_s G^2\rangle  &= 0.038 \pm 0.011 \textrm{GeV}^4, \nonumber\\
\langle g_s^3fG^3\rangle  &\simeq \frac{8}{27} (2+\kappa^2) \langle g_s^2\bar{q}q\rangle ^2, \nonumber\\
\kappa &= 0.74 \pm 0.03.
\label{inputs}
\end{align}
In those inputs, the double-gluon condensate $\langle \alpha_s G^2\rangle$, triple-gluon condensate $\langle g_s^3fG^3\rangle$ and four-quark condensate $\langle g_s^2\bar{q}q\rangle ^2$ are scale independent, while for the scale dependence of the double-quark condensates $\langle \bar{q}q\rangle$ and $\langle \bar{s}s\rangle$, quark-gluon mixed condensates $\langle g_s\bar{q}\sigma TGq\rangle$ and $\langle g_s\bar{s}\sigma TGs\rangle$, four-quark condensates $\langle g_s\bar{q}q\rangle ^2$ and $\langle g_s\bar{s}s\rangle ^2$, current quark masses $m_s$ and $m_q$, one can find in Ref.~\cite{Zhong:2021epq}. By requiring that there is reasonable Borel window to normalize $\langle \xi^0\rangle_{2;K}$ in Eq.~\eqref{Moment0SumRules}, we obtain the continuum threshold $s_K \simeq 2.5~{\rm GeV}^2$.

\subsection{Moments of the kaon leading twist DA}
Using the above inputs, one can calculate the values of the moments of the kaon leading-twist DA with the improved sum rules formula \eqref{xin}. Firstly, one need to find out the suitable Borel windows for the sum rules. Usually, the criteria are,
\begin{itemize}
\item the contributions of the continuum state and the dimension-six condensates should be as small as possible,
\item the moment values should be as stable as possible in the correspondence Borel windows.
\end{itemize}
\begin{figure}[t]
\hspace{-0.5cm}
\centering
\hspace{0.4cm}\includegraphics[width=0.48\textwidth]{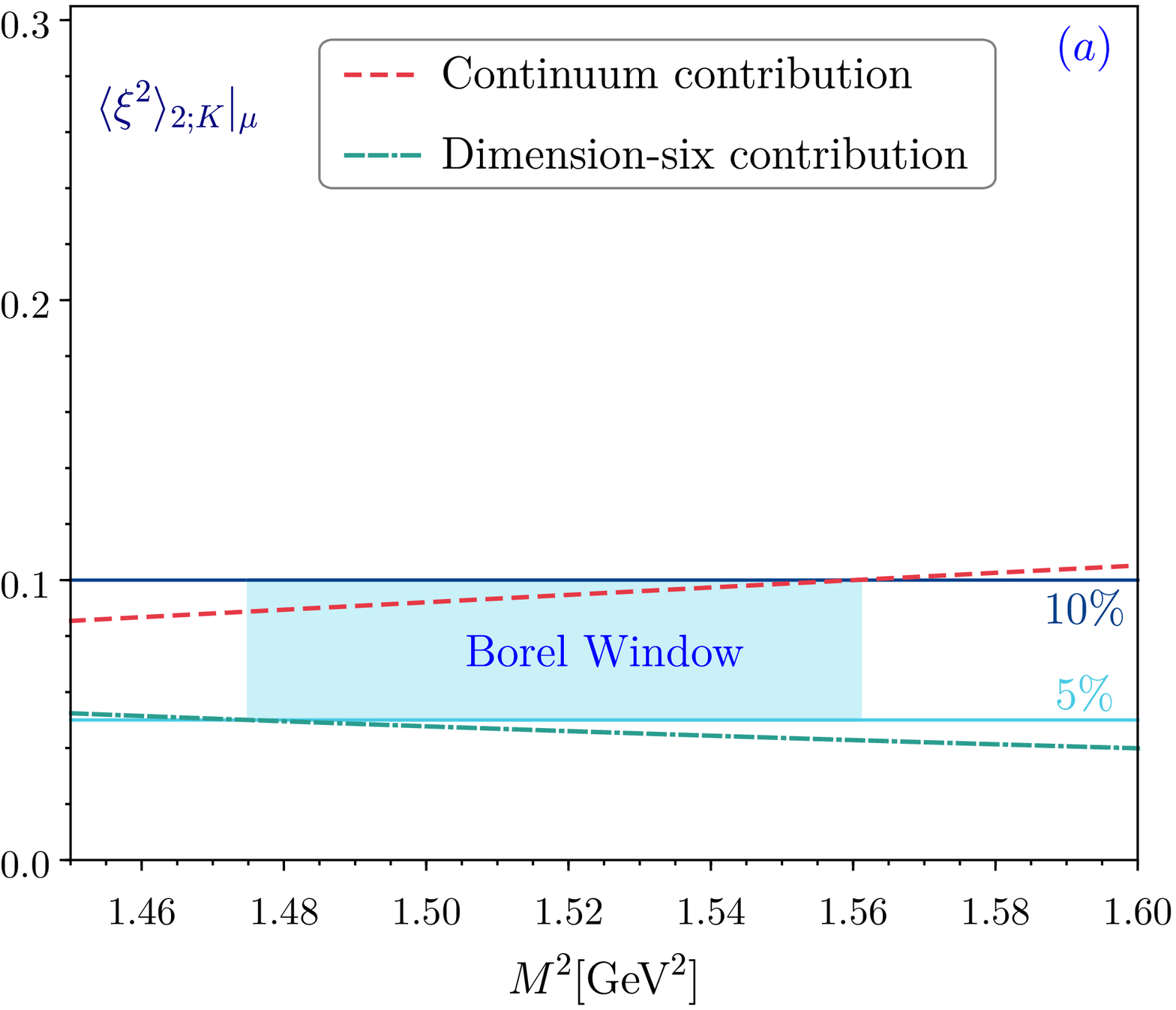}\includegraphics[width=0.48\textwidth]{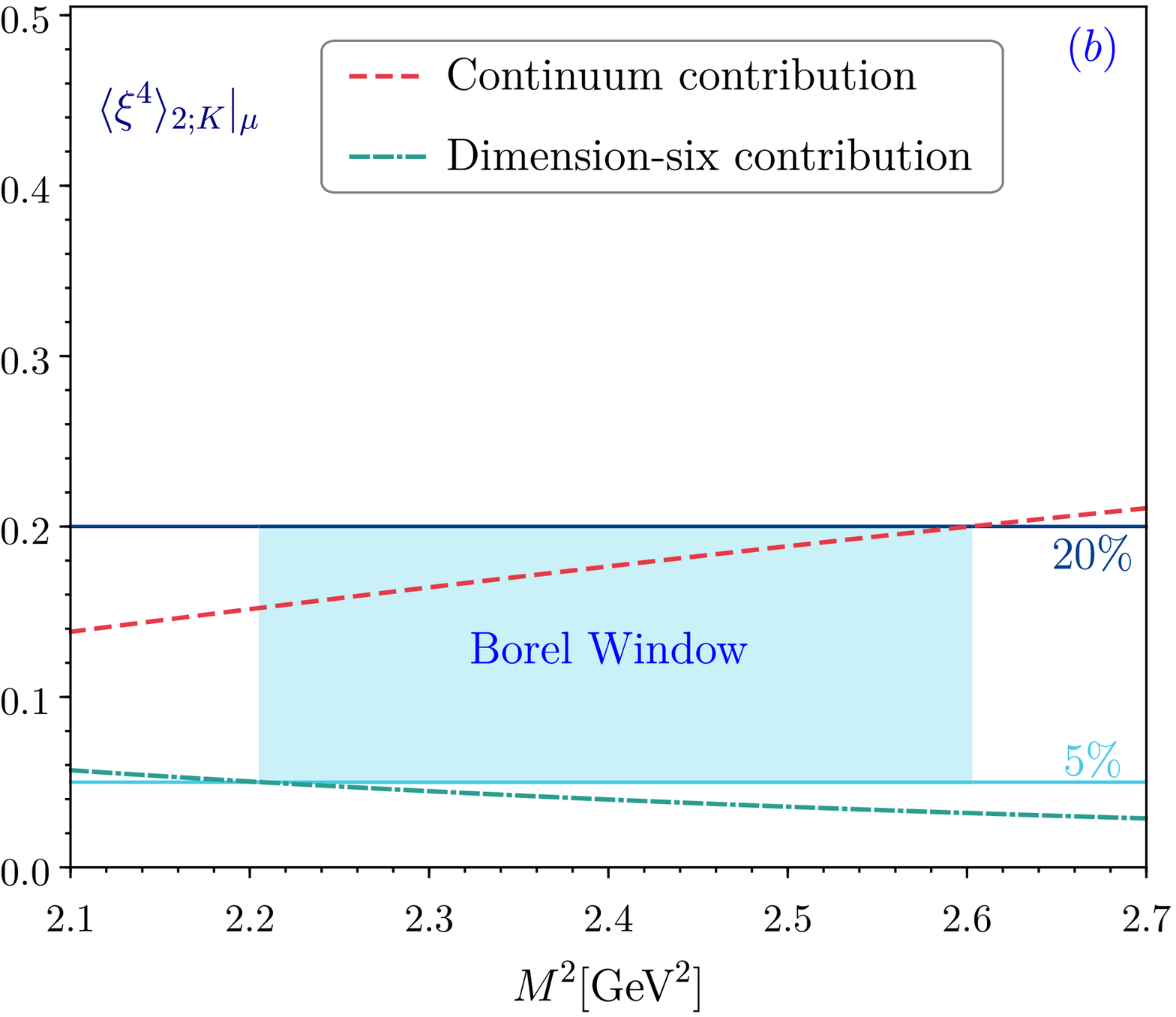}
\includegraphics[width=0.48\textwidth]{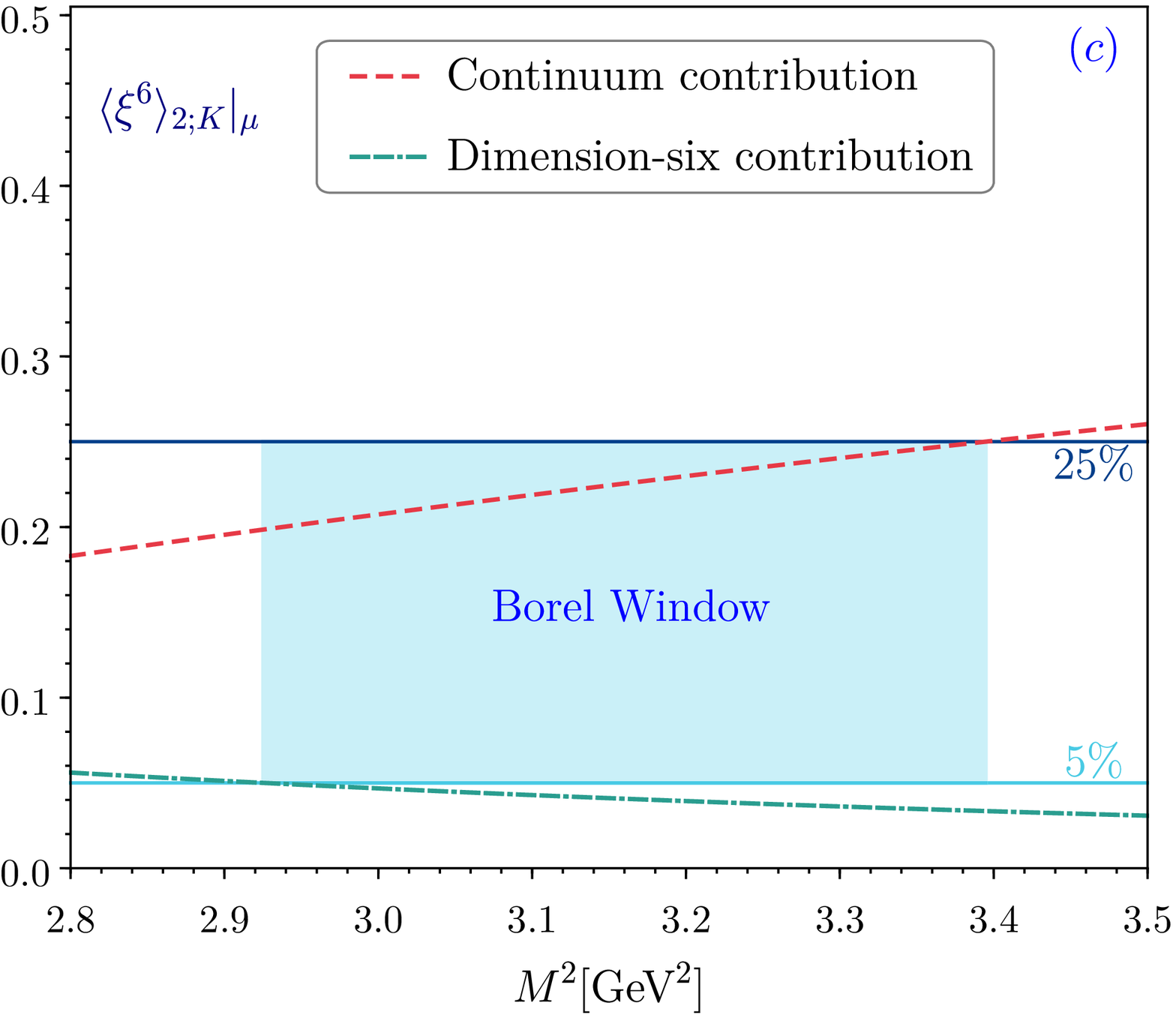}\includegraphics[width=0.48\textwidth]{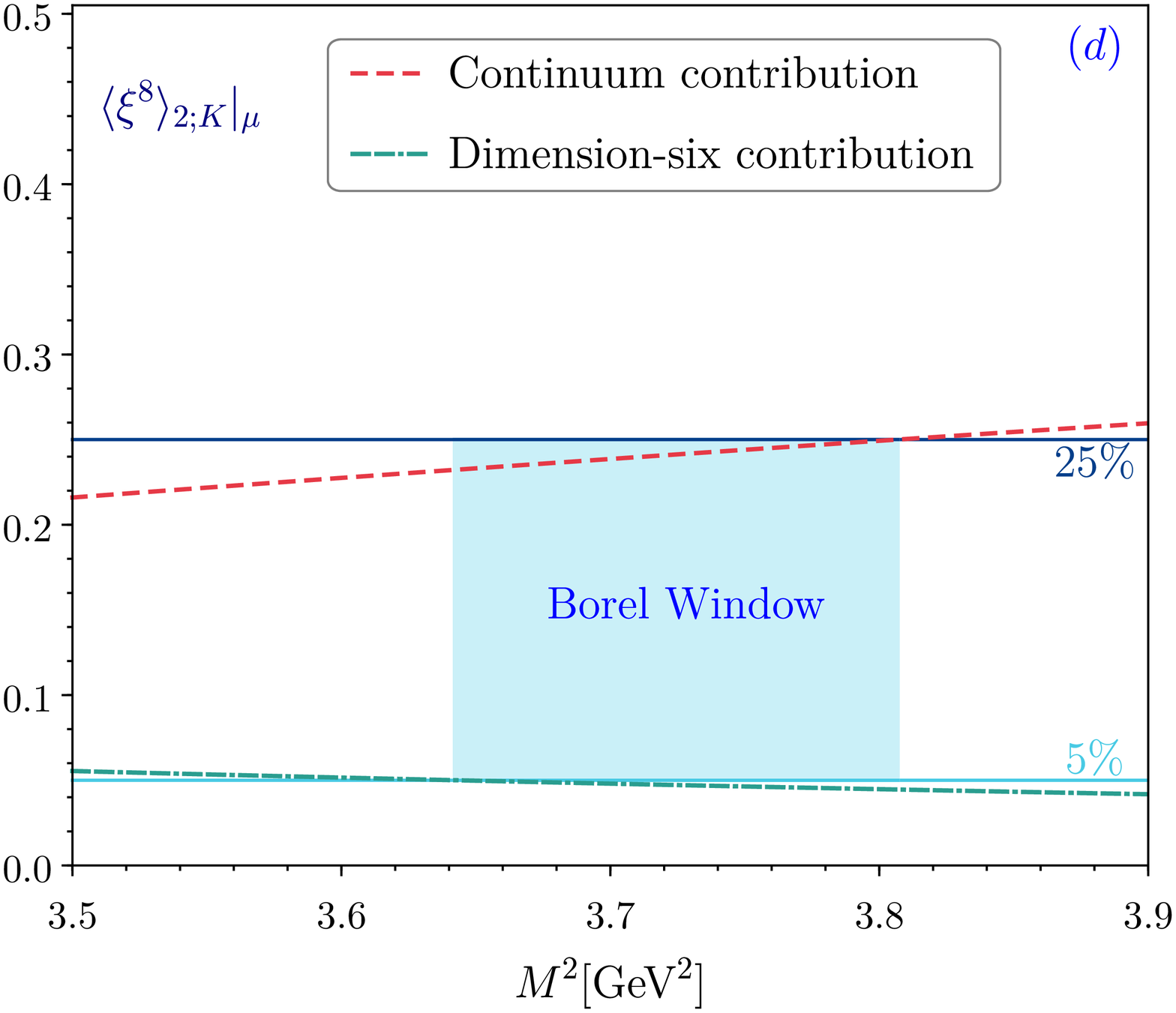}
\includegraphics[width=0.48\textwidth]{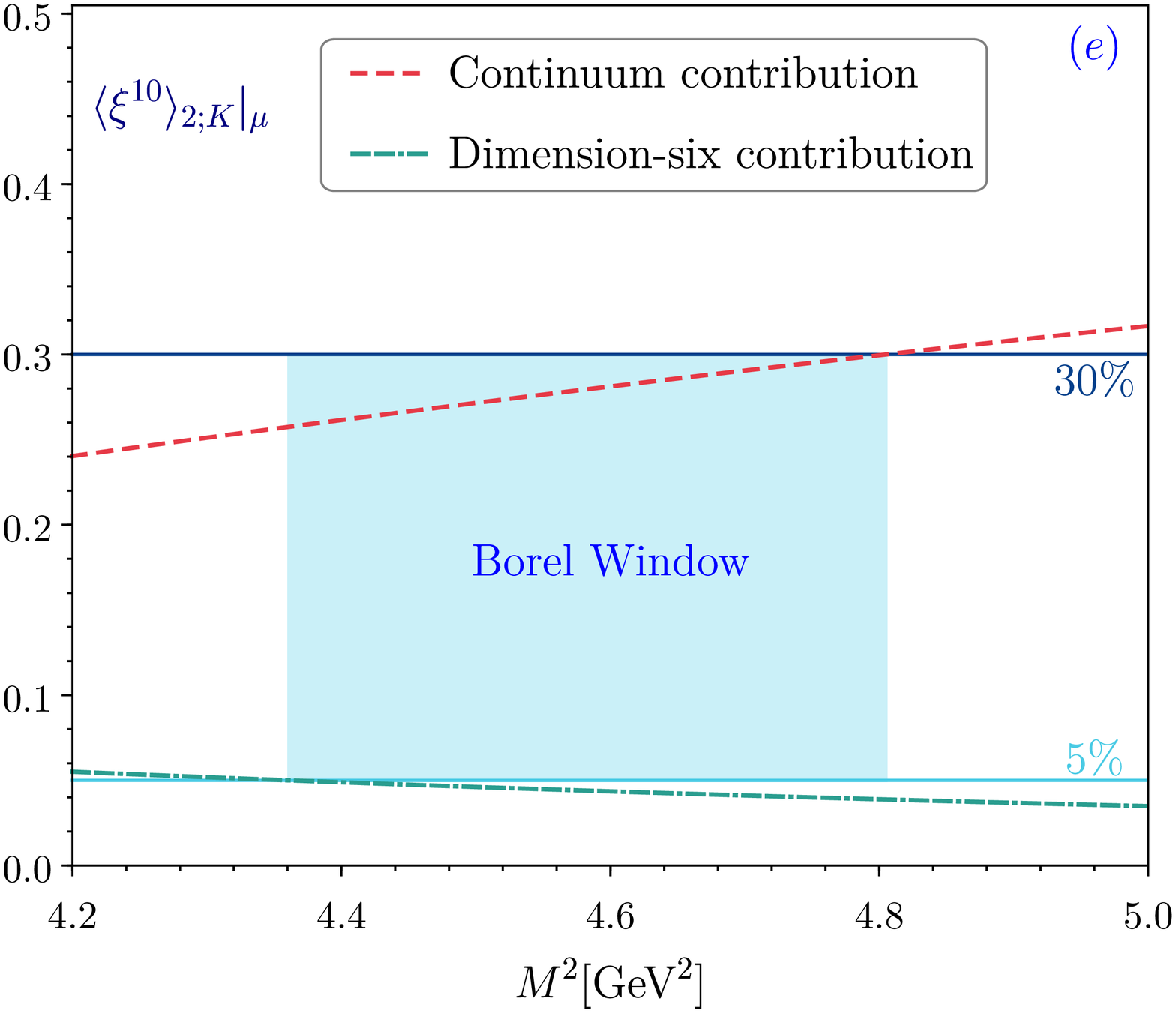}
\caption{The continuum state's contribution (dashed line) and dimension-six term's contribution (dash-dotted line) of the kaon leading-twist DA even order moments $\langle \xi^n\rangle _{2;K}$ with $n=(2,4,6,8,10)$ versus the Borel parameter $M^2$, where all input parameters are set to be their central values.}
\label{fRatioEven}
\end{figure}
\begin{figure}[t]
\hspace{-0.5cm}
\centering
\includegraphics[width=0.48\textwidth]{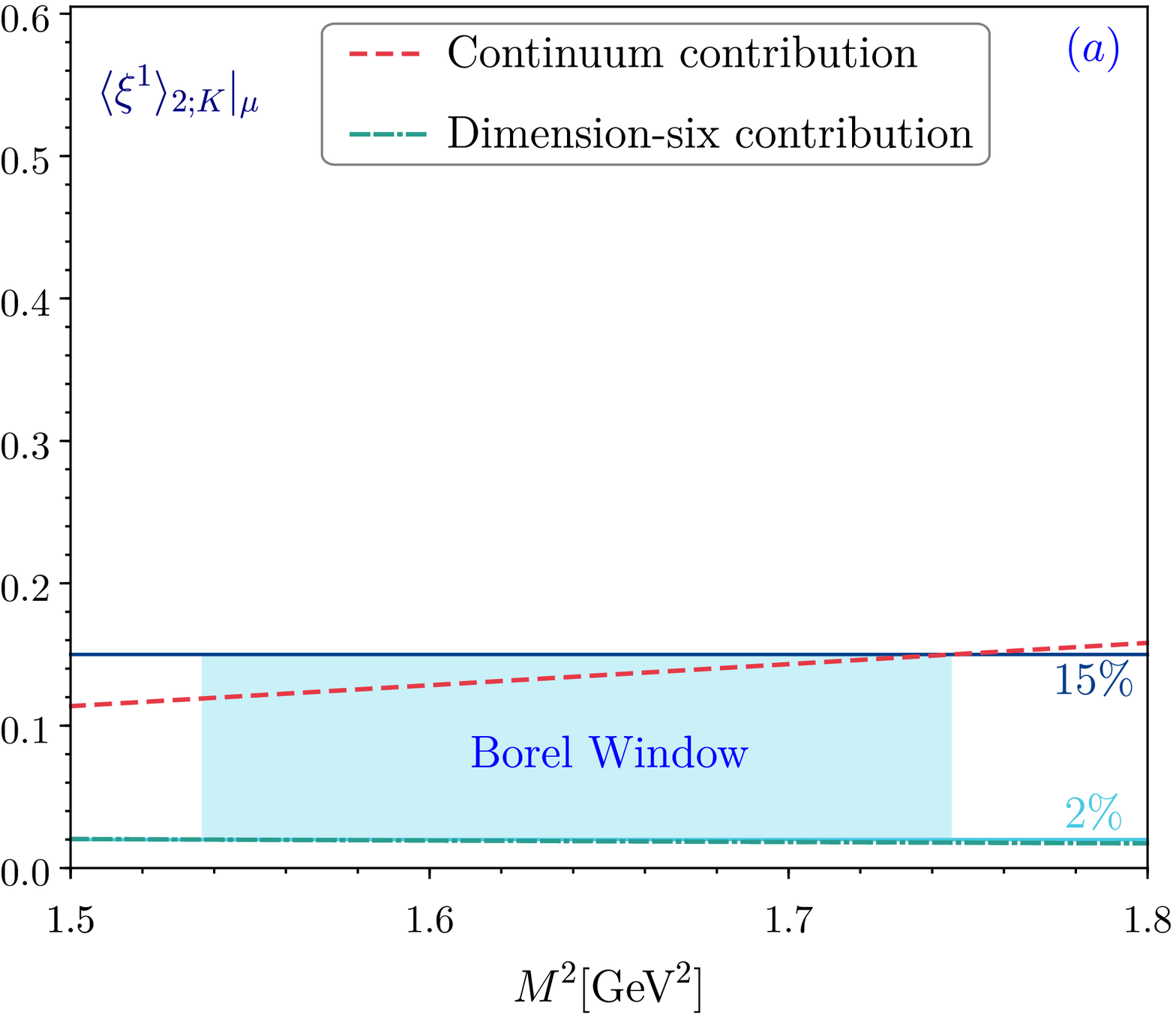}\includegraphics[width=0.472\textwidth]{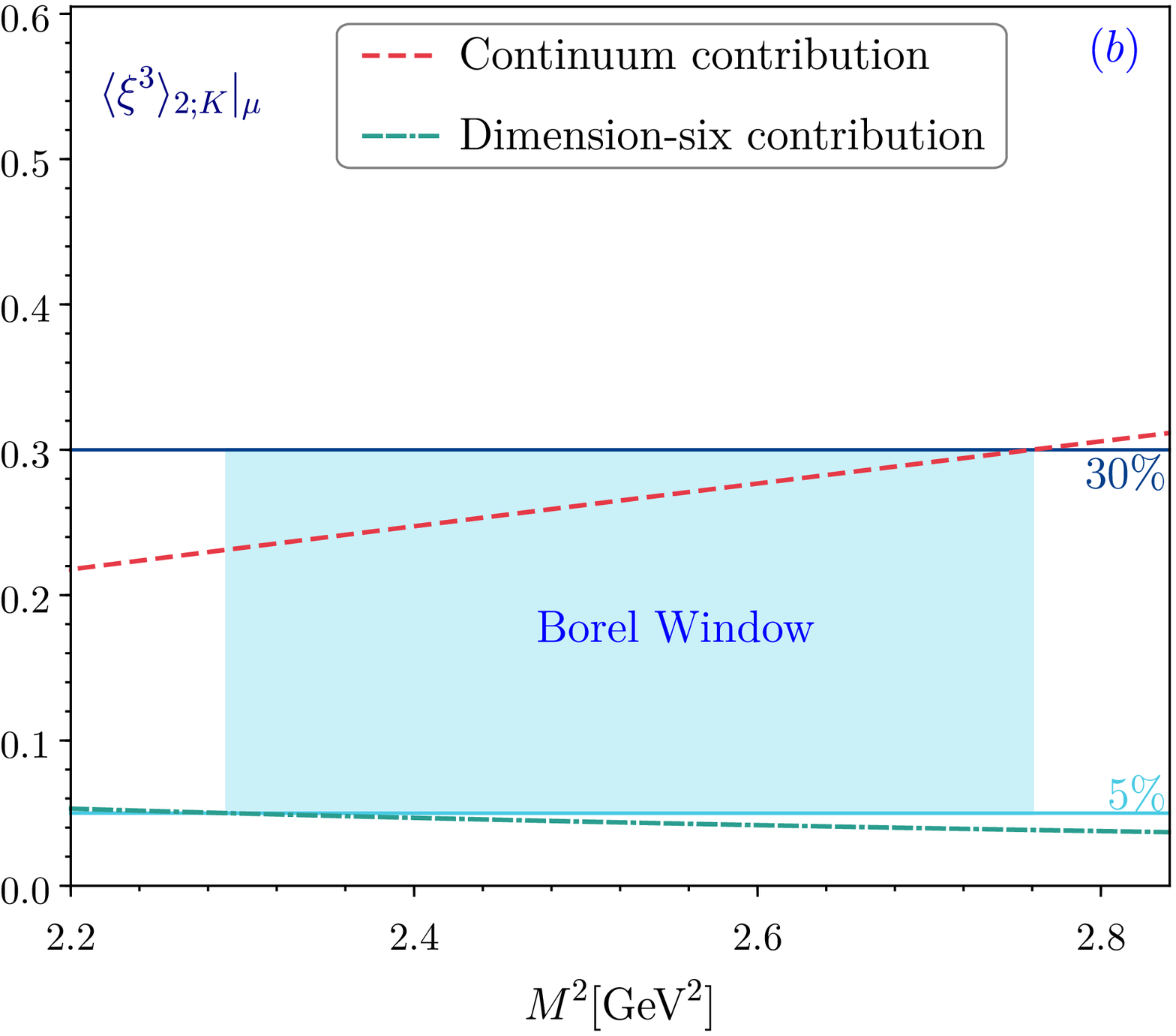}
\includegraphics[width=0.48\textwidth]{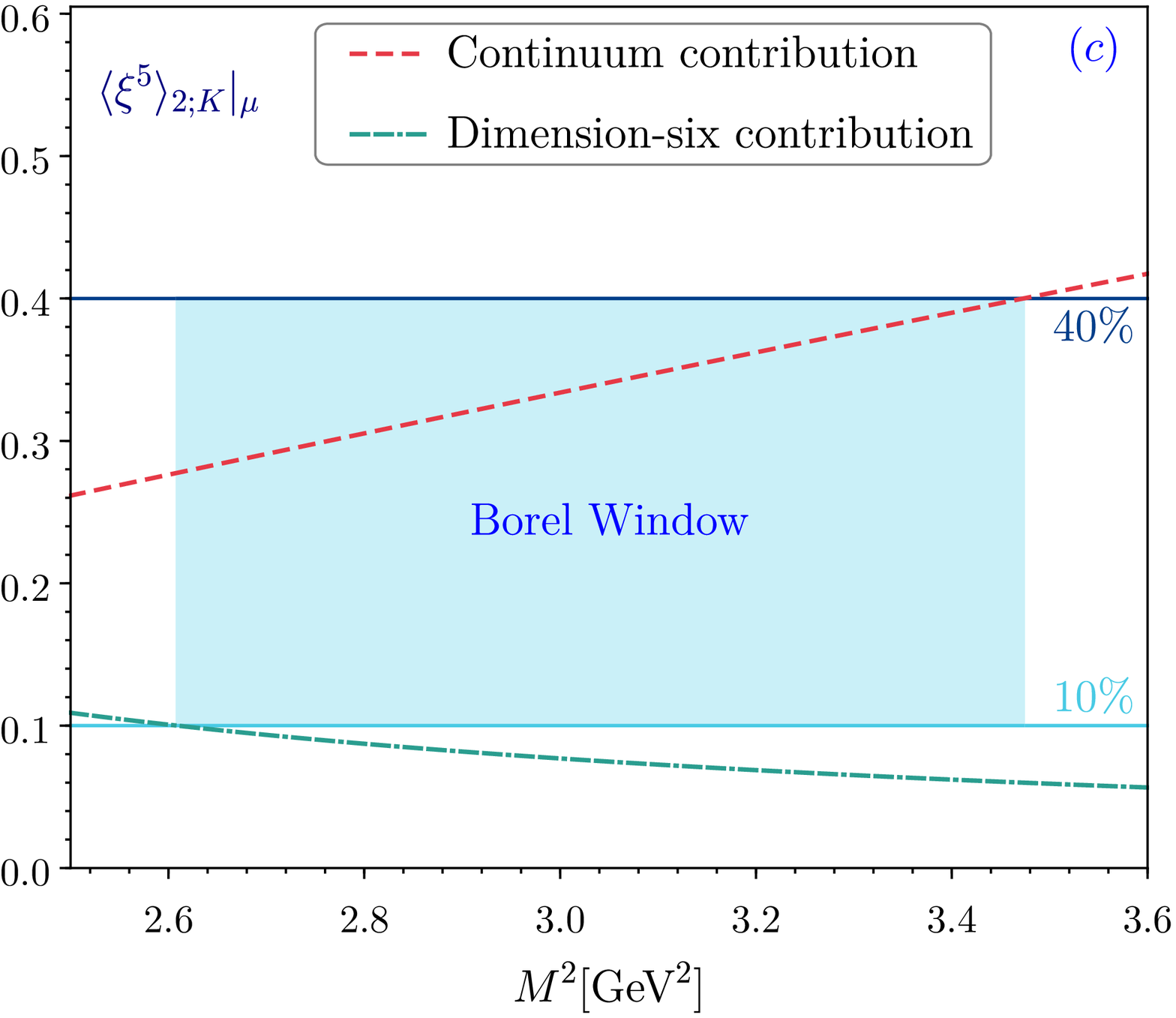}\includegraphics[width=0.48\textwidth]{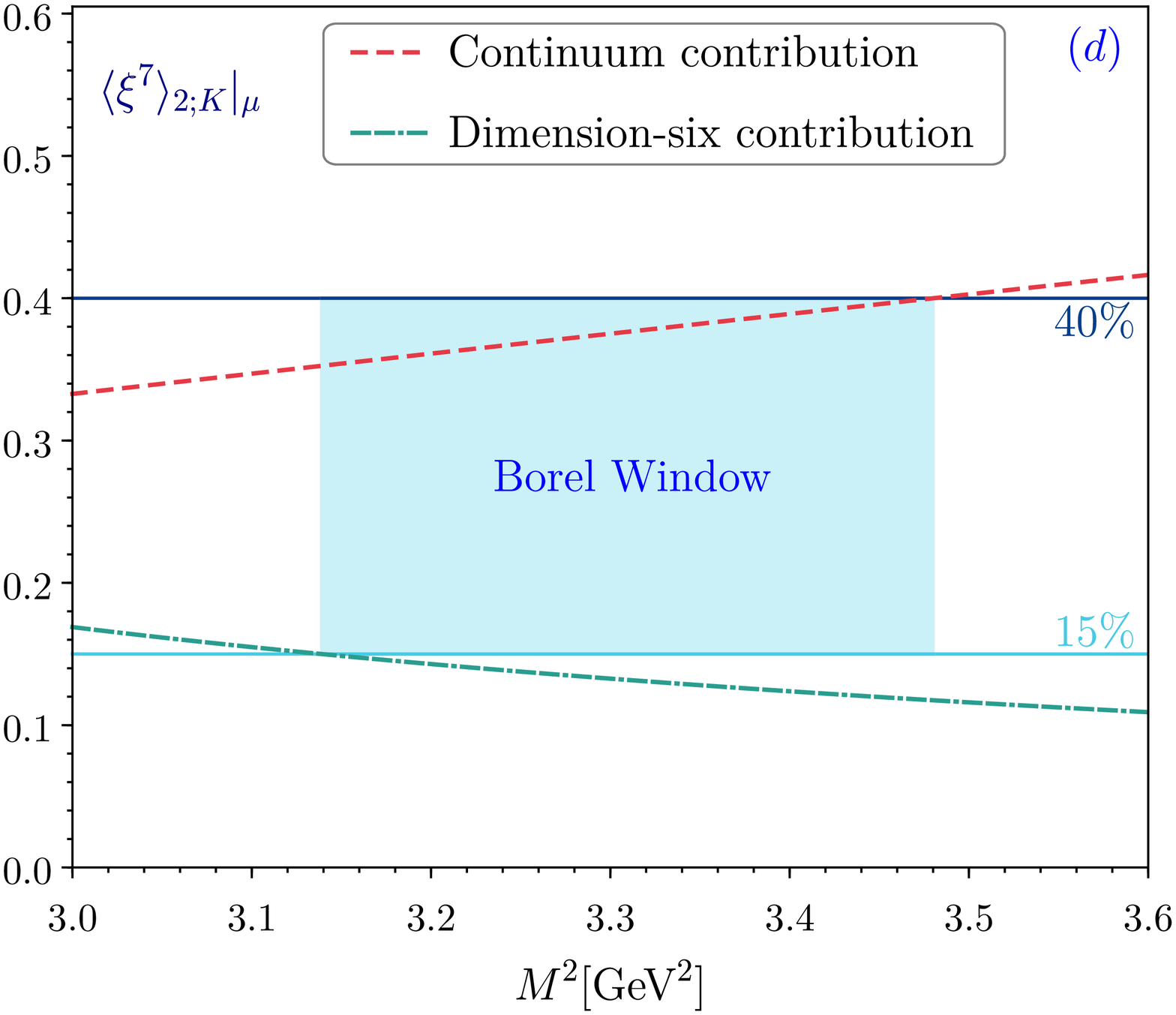}
\includegraphics[width=0.48\textwidth]{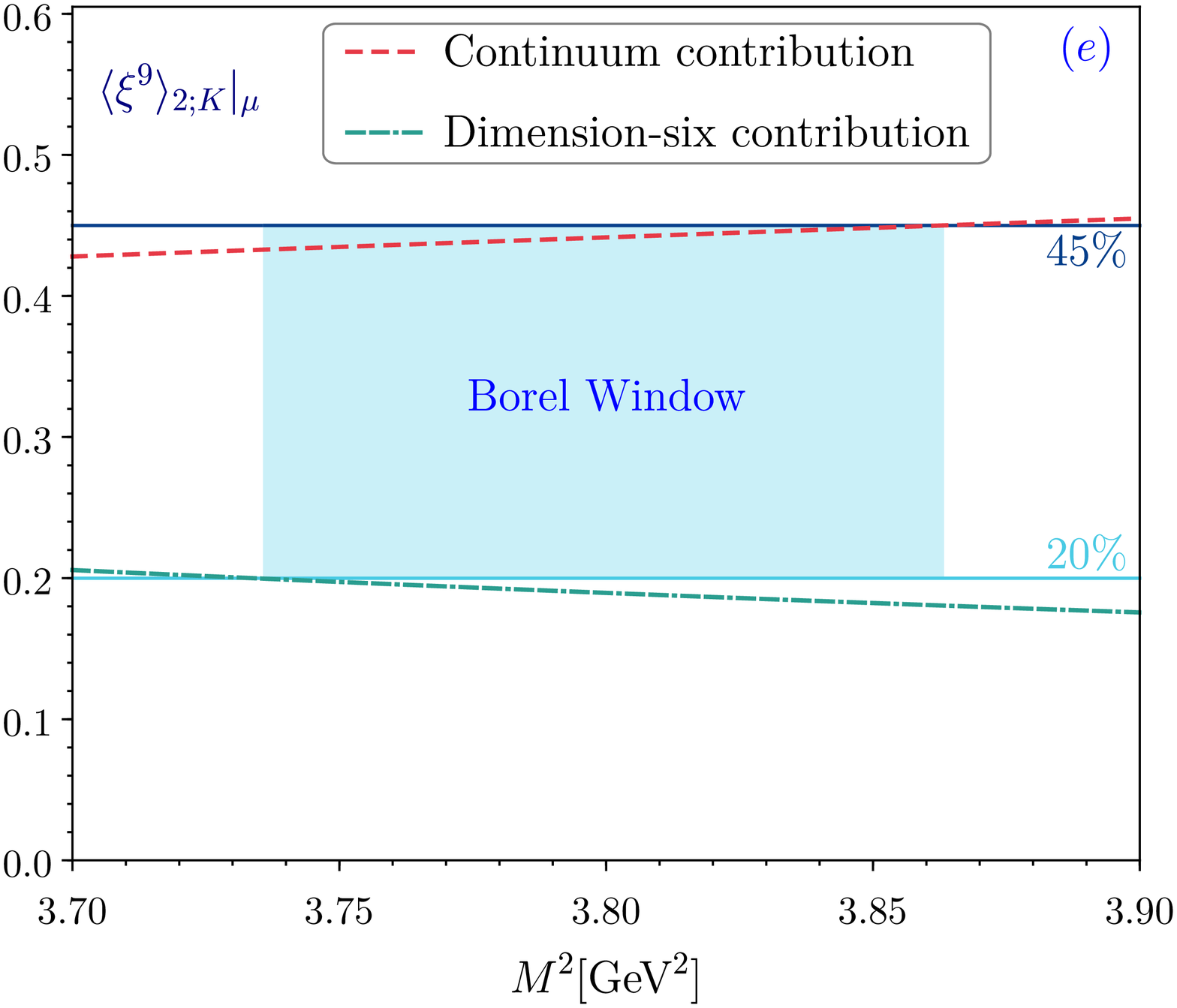}
\caption{The continuum state's contribution (dashed line) and dimension-six term's contribution (dash-dotted line) of the kaon leading-twist DA odd order moments $\langle \xi^n\rangle _{2;K}|_\mu$ with $n=(1,3,5,7,9)$ versus the Borel parameter $M^2$, where all input parameters are set to be their central values.}
\label{fRatioOdd}
\end{figure}

Specifically, we require the continuum state contributions to the even moments are not more than $10\%$, $20\%$, $25\%$, $25\%$, $30\%$ for $n = (2, 4, 6, 8, 10)$, respectively. The dimension-six term contributions are less than $5\%$ for those even moments. For the odd moments, the continuum state contributions are required to be less than $15\%$, $30\%$, $40\%$, $40\%$, $45\%$, the dimension-six term contributions are less than $2\%$, $5\%$, $10\%$, $15\%$, $20\%$ for $n=(1, 3, 5, 7, 9)$, respectively. Those criteria are shown in Figure~\ref{fRatioEven} for the even moments $n = (2,4,6,8,10)$, and in Figure~\ref{fRatioOdd} for the odd moments $n = (1,3,5,7,9)$. In those two figures, the upper solid lines indicate the criteria of the continuum state contributions, the lower solid lines indicate the criteria of the dimension-six term contributions, the dashed and dash-dotted lines stand for the continuum contributions and the dimension-six contributions, respectively. Then the corresponding Borel windows can be obtained, and which are shown with the shadow regions.

\begin{figure}[t]
\centering
\includegraphics[width=0.478\textwidth]{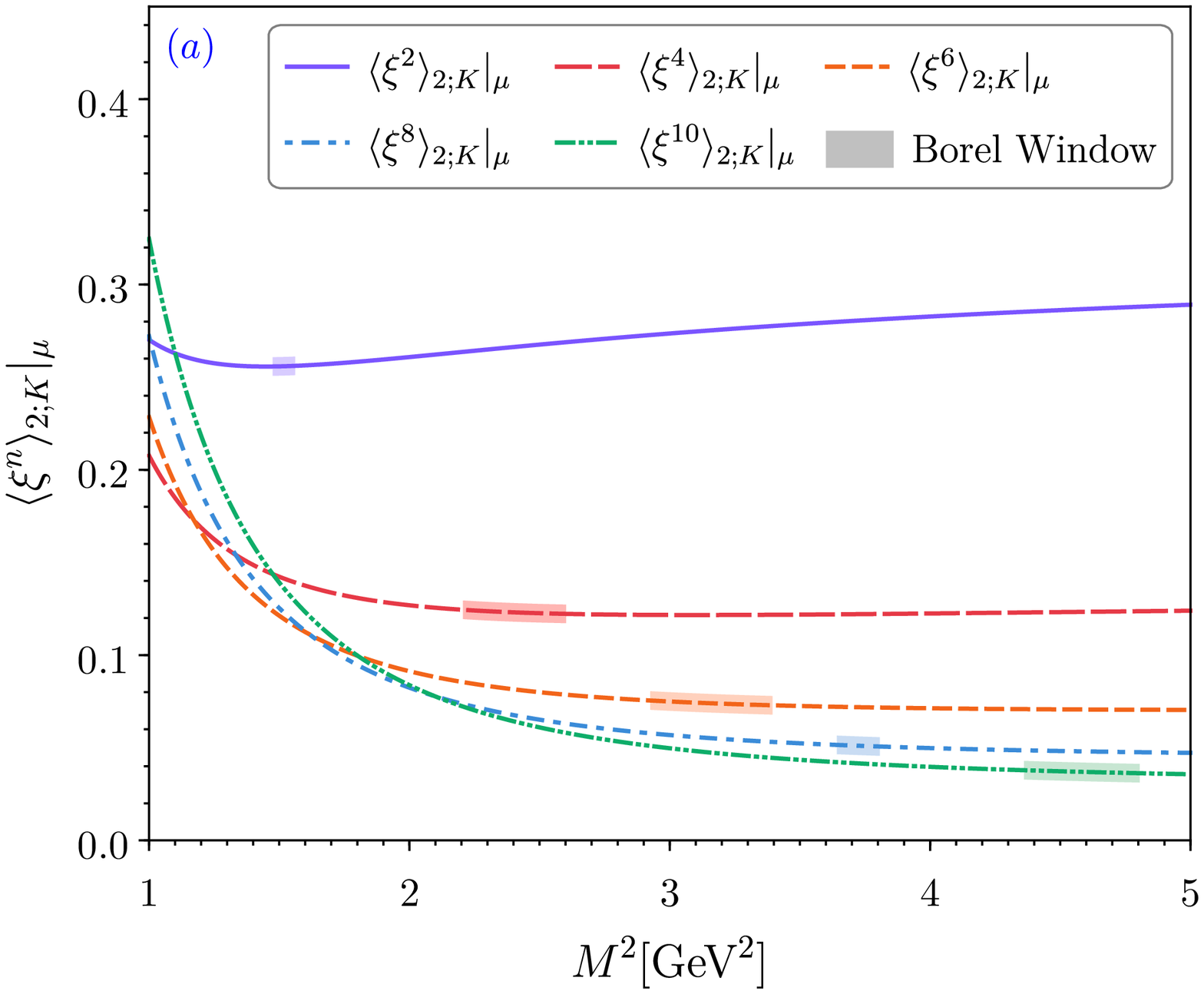}\includegraphics[width=0.5\textwidth]{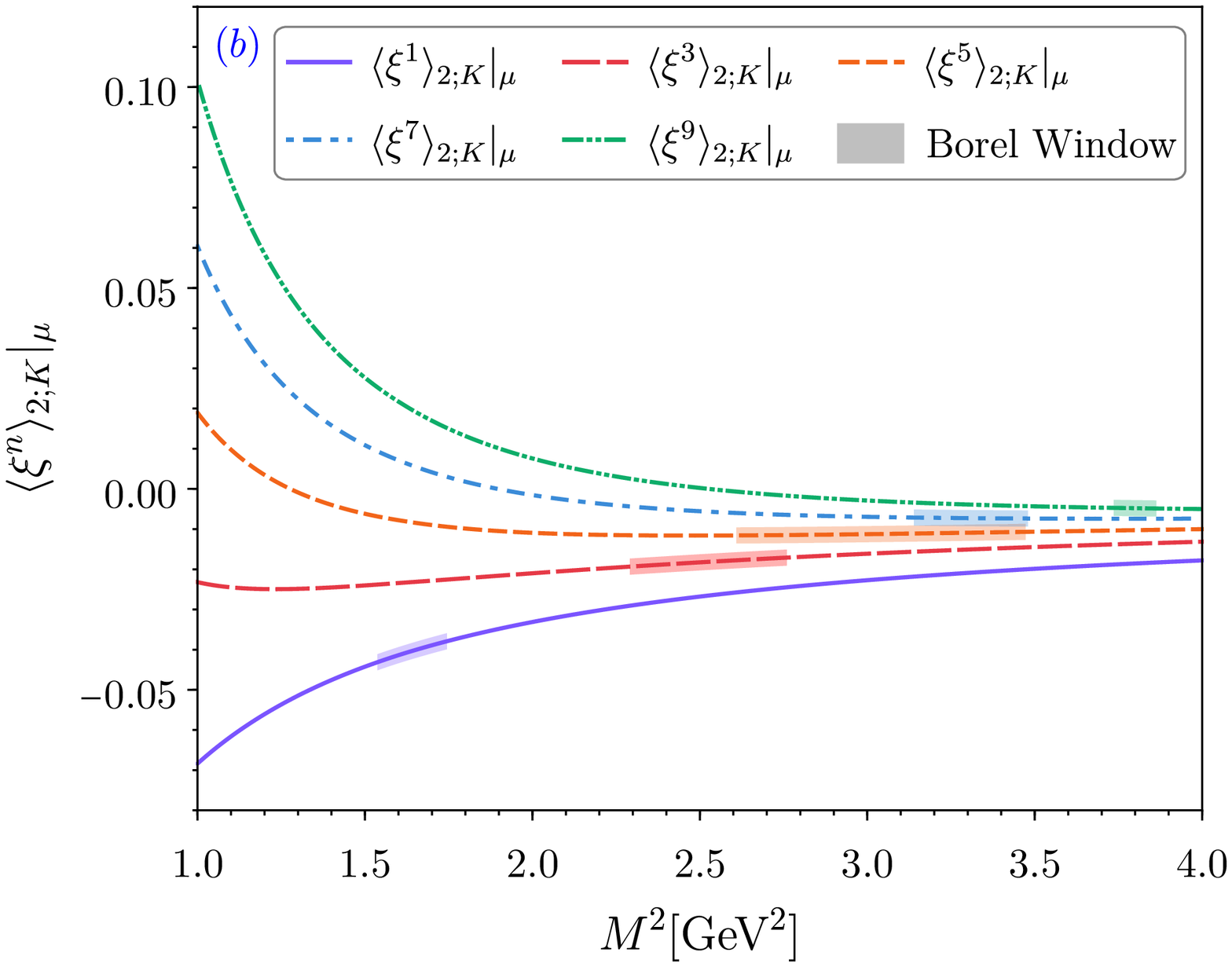}
\caption{The kaon leading-twist DA moments $\langle \xi^n\rangle _{2;K}(n=1,\cdots,10)$ versus the Borel parameter $M^2$, where all input parameters are set to be their central values. The left plan is for the even moments and the right plan is for the odd moments respectively.}
\label{fxinM2}
\end{figure}
Figure~\ref{fxinM2} shows the first ten moments of kaon leading-twist DA $\langle \xi^n\rangle _{2;K}$ versus the Borel parameter $M^2$, where all input parameters are set to be their central values. The curve segments in the shadows represent the values of $\langle \xi^n\rangle _{2;K}|_\mu$ in the Borel windows. One can find that the moments $\langle \xi^n\rangle _{2;K}|_\mu$ have good stabilities versus the Borel windows. By taking all error sources into account, and with the renormalization group equation of the moments shown in Ref.~\cite{Zhong:2021epq}, the values of $\langle \xi^n\rangle_{2;K}|_\mu$ can be obtained. At the scale $\mu_0 = 1\ {\rm GeV}$, we obtain
\begin{align}
\langle \xi^1\rangle _{2;K} |_{1 ~{\rm GeV}} &= -0.0438^{+0.0053}_{-0.0075}, \nonumber\\
\langle \xi^3\rangle _{2;K} |_{1 ~{\rm GeV}} &= -0.0210^{+0.0024}_{-0.0035}, \nonumber\\
\langle \xi^5\rangle _{2;K} |_{1 ~{\rm GeV}} &= -0.0134^{+0.0014}_{-0.0021}, \nonumber\\
\langle \xi^7\rangle _{2;K} |_{1 ~{\rm GeV}} &= -0.0087^{+0.0009}_{-0.0014}, \nonumber\\
\langle \xi^9\rangle _{2;K} |_{1 ~{\rm GeV}} &= -0.0058^{+0.0007}_{-0.0010}.\label{xinValueOdd}
\end{align}
and
\begin{align}
\hspace{-1cm}
\langle \xi^2\rangle _{2;K} |_{1 ~{\rm GeV}} &= 0.262^{+0.010}_{-0.010}, \nonumber\\
\langle \xi^4\rangle _{2;K} |_{1 ~{\rm GeV}} &= 0.132^{+0.006}_{-0.006}, \nonumber\\
\langle \xi^6\rangle _{2;K} |_{1 ~{\rm GeV}} &= 0.082^{+0.005}_{-0.005}, \nonumber\\
\langle \xi^8\rangle _{2;K} |_{1 ~{\rm GeV}} &= 0.058^{+0.004}_{-0.004}, \nonumber\\
\langle \xi^{10}\rangle _{2;K} |_{1 ~{\rm GeV}} &= 0.044^{+0.004}_{-0.004}. \label{xinValueEven}
\end{align}
\begin{table}[t]
\footnotesize
\caption{Our predictions for the first four moments $\langle \xi^n\rangle _{2;K}(n=1,2,3,4)$ and the first two Gegenbauer moments $a^{2;K}_1$ and $a^{2;K}_2$ of the kaon leading-twist DA, compared to other theoretical predictions.}\label{table:xin_value}
\begin{tabular}{l c c c c c c c}
\hline\hline
& ~$\mu{\rm [GeV]}$~ & ~$\langle\xi^1\rangle_{2;\pi}|_\mu$~ & ~$\langle\xi^2\rangle_{2;\pi}|_\mu$~ & ~$\langle\xi^3\rangle_{2;\pi}|_\mu$~ & ~$\langle\xi^4\rangle_{2;\pi}|_\mu$~ & ~$a_1^{2;K}(\mu)$~ & ~$a_2^{2;K}(\mu)$  \\ \hline
BFTSR (This Work)               & 1  & $-0.0438^{+0.0053}_{-0.0075}$ & $0.262^{+0.010}_{-0.010}$ & $-0.0210^{+0.0024}_{-0.0035}$ & $0.132^{+0.006}_{-0.006}$ & $-0.0731^{+0.0089}_{-0.0125}$ & $0.182^{+0.029}_{-0.030}$ \\
BFTSR (This Work)               & 2  & $-0.0368^{+0.0045}_{-0.0063}$ & $0.246^{+0.007}_{-0.008}$ & $-0.0173^{+0.0020}_{-0.0029}$ & $0.120^{+0.005}_{-0.005}$ & $-0.0614^{+0.0075}_{-0.0105}$ & $0.139^{+0.022}_{-0.023}$ \\
QCD SR~\cite{Ball:2003sc}       & 1  &  &  &  &  & $-0.18(9)$ & $0.16(10)$ \\
QCD SR~\cite{Chetyrkin:2007vm}  & 1  &  &  &  &  & $-0.10(4)$ &  \\
QCD SR~\cite{Chetyrkin:2007vm}  & 2  &  &  &  &  & $-0.08(4)$ &  \\
QCD SR~\cite{Khodjamirian:2004ga} & 1  &  &  &  &  & $-0.05(2)$ & $0.27^{+0.37}_{-0.12}$ \\
QCD SR~\cite{Braun:2004vf}      & 1  & $-0.06(7)$ &  &  &  & $-0.10(12)$ & \\
QCD SR~\cite{Ball:2005vx}      & 1  &  &  &  &  & $-0.050(25)$ & \\
QCD SR~\cite{Ball:2006fz}      & 1  &  &  &  &  & $-0.06(3)$ & \\
LQCD~\cite{Braun:2006dg}        & 2  & $-0.0272(5)$ & $0.260(5)$ &  &  & $-0.0453(9)(29)$ & $0.175(18)(47)$ \\
LQCD~\cite{Boyle:2006pw}        & 1  & $-0.040(4)$ &  &  &  & $-0.066(6)$ & \\
LQCD~\cite{Boyle:2006pw}        & 2  & $-0.032(3)$ &  &  &  & $-0.053(5)$ & \\
LQCD~\cite{Zhang:2020gaj}      & 2  &  & $0.198(16)$ &  &  &  & \\
LQCD~\cite{Arthur:2010xf}      & 2  & $-0.036(1)(2)$ & $0.26(1)(1)$ &  &  &  & \\
LQCD~\cite{Bali:2019dqc}      & 2  &  &  &  &  & $-0.0525^{+0.031}_{-0.033}$ & $0.106^{+0.015}_{-0.016}$ \\
AdS/QCD~\cite{Momeni:2017moz}  & 1  &  & $0.21(2)$ &  & $0.09(1)$ &  & \\
NLChQM~\cite{Nam:2017gzm}      & 1  & $-0.0277$ & $0.2043$ & $-0.0122$ & $0.0887$ &  & \\ \hline\hline
\end{tabular}
\end{table}

\begin{figure}[t]
\centering
\includegraphics[width=0.48\textwidth]{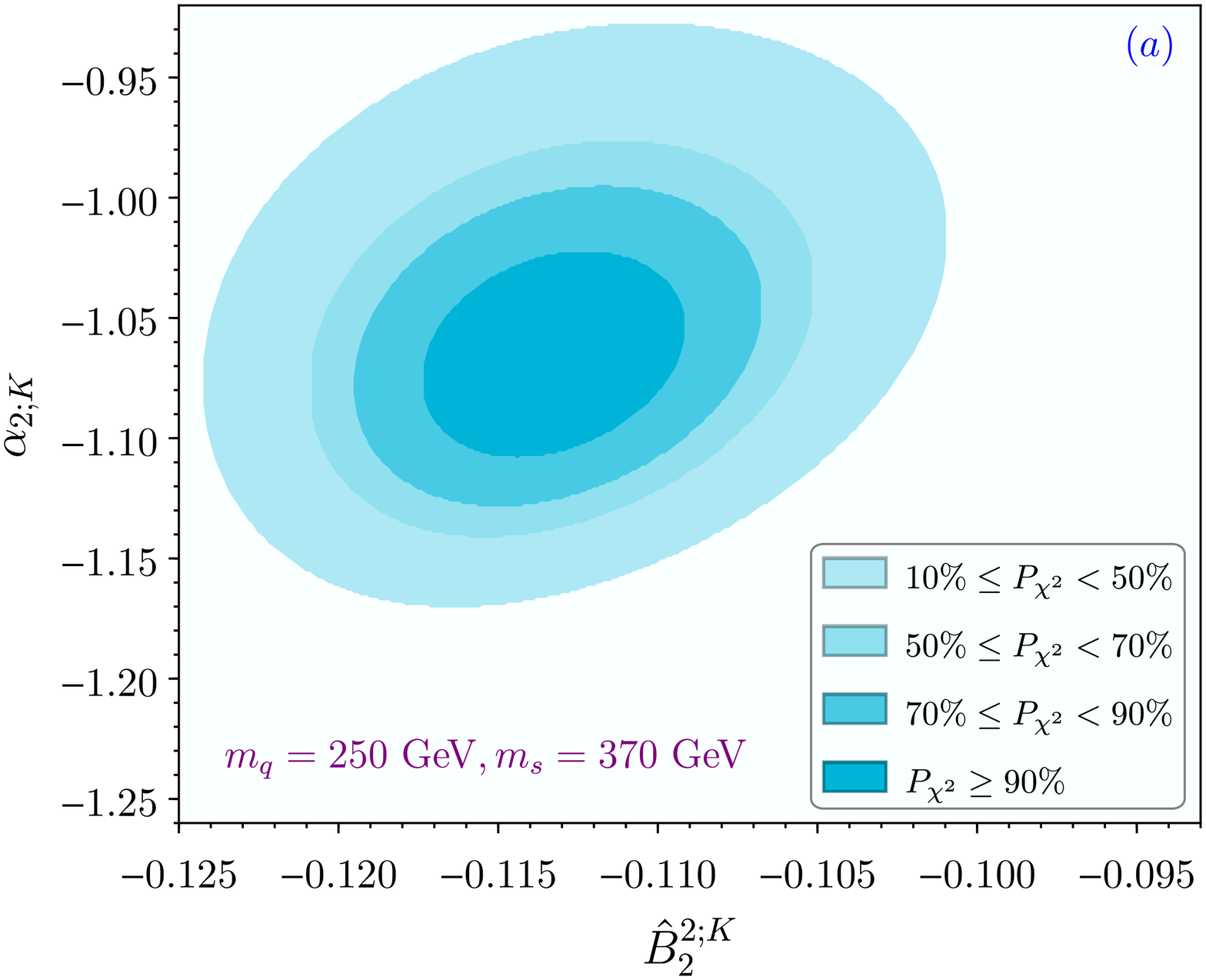}\includegraphics[width=0.50\textwidth]{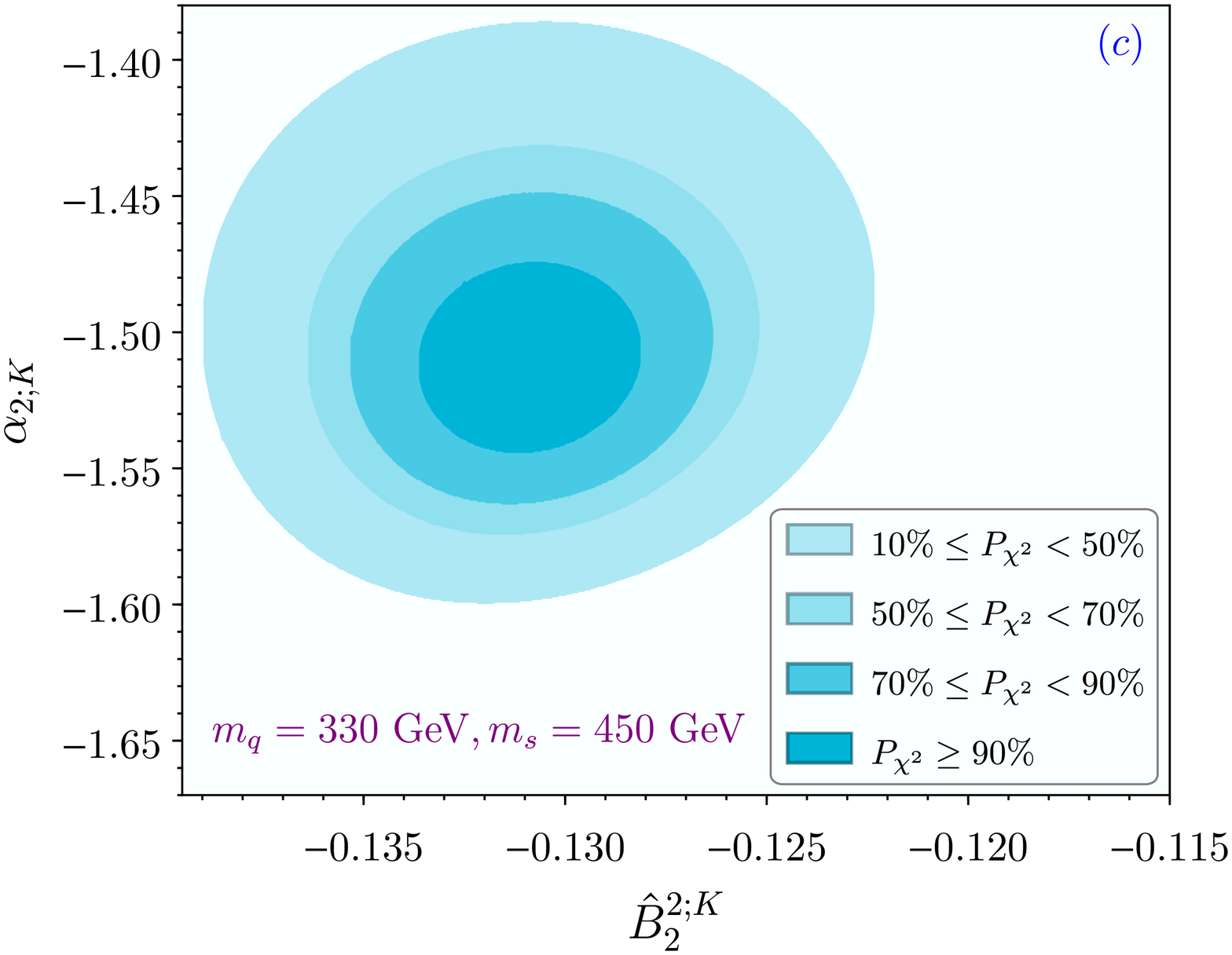}
\includegraphics[width=0.50\textwidth]{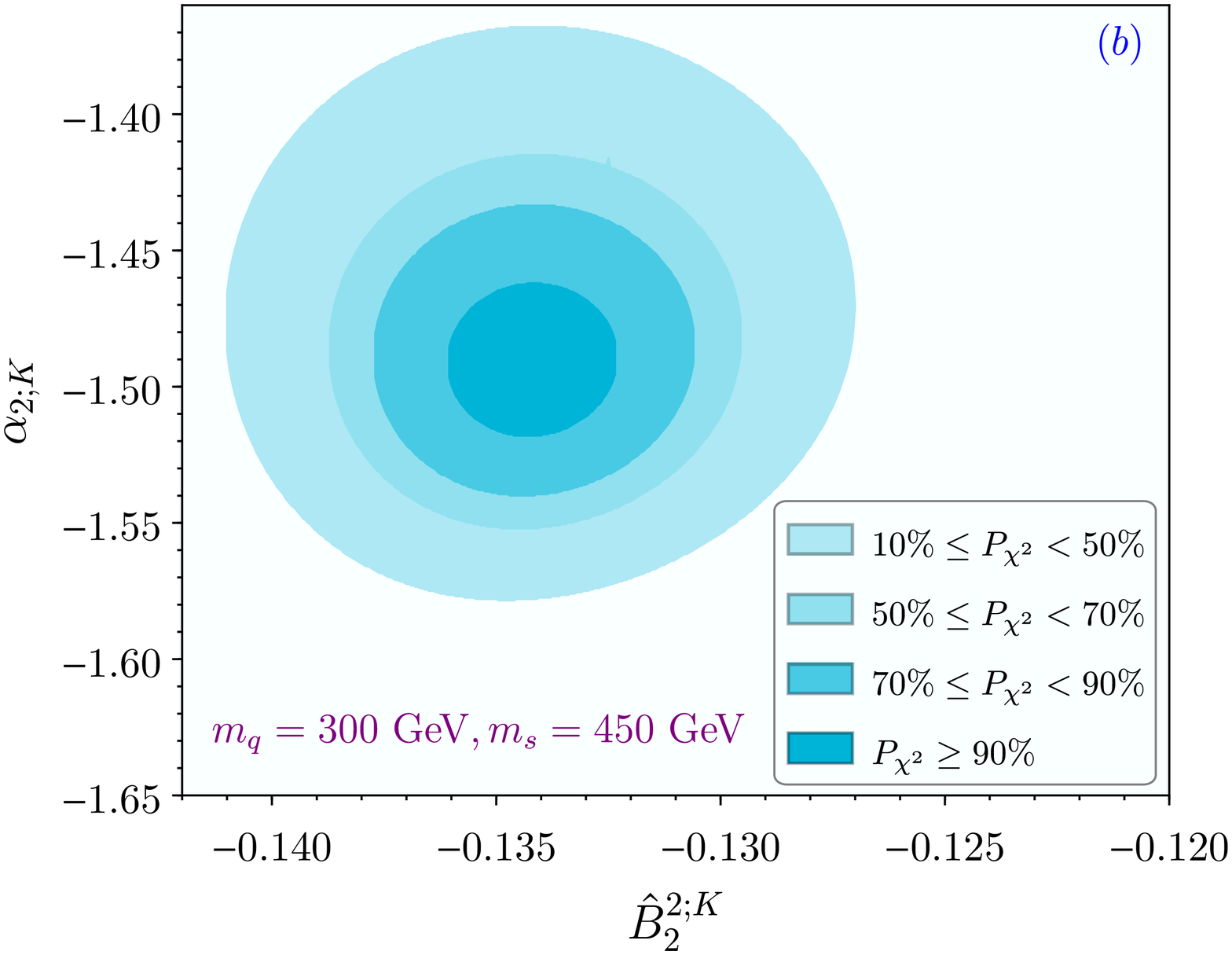}
\caption{The goodness of fit $P_{\chi^2}$ versus the fitting parameters $\theta = (\alpha_{2;K},\hat{B}_2^{2;K})$ for three different constituent quark MSs respectively.}
\label{fgof}
\end{figure}
\begin{table}[t]
\centering
\caption{Several tipycal model parameters of kaon leading-twist DA $\phi_{2;K}(x, \mu_0)$ with three different constituent quark MSs at scale $\mu_0 = 1\ ~{\rm GeV}$.}\label{table:model_parameter}
\begin{tabular}{l c c c c}
\hline\hline
$(\hat{m}_q, \hat{m}_s) = (250, 370) {\rm MeV}$ ~~ & ~~$A_{2;K}({\rm GeV}^{-1})$~~ & ~~$\alpha_{2;K}$~~ & ~~$\hat{B}^{2;K}_2$~~ & ~~$\beta_{2;K}({\rm GeV})$~~ \\ \hline
$P_{\chi^2_{\rm min}}$                          & $4.088231$  & $-1.0675$ & $-0.1134$ & $0.681348$ \\
$\phi_{2;K}^{\rm max.}(0.5,\mu_0)$       & $4.775445$  & $-0.9900$ & $-0.1158$ & $0.674196$ \\
$\phi_{2;K}^{\rm min.}(0.5,\mu_0)$       & $3.599708$  & $-1.1240$ & $-0.1104$ & $0.692533$ \\ \hline\hline
$(\hat{m}_q, \hat{m}_s) = (330, 450) {\rm MeV}$ ~~ & ~~$A_{2;K}({\rm GeV}^{-1})$~~ & ~~$\alpha_{2;K}$~~ & ~~$\hat{B}^{2;K}_2$~~ & ~~$\beta_{2;K}({\rm GeV})$~~ \\ \hline
$P_{\chi^2_{\rm min}}$                          & $2.043080$ & $-1.5115$ & $-0.13090$ & $0.640731$ \\
$\phi_{2;K}^{\rm max.}(0.5,\mu_0)$       & $2.381850$ & $-1.4405$ & $-0.13335$ & $0.632128$ \\
$\phi_{2;K}^{\rm min.}(0.5,\mu_0)$       & $1.794378$ & $-1.5640$ & $-0.12850$ & $0.652403$ \\ \hline\hline
$(\hat{m}_q, \hat{m}_s) = (300, 450) {\rm MeV}$ ~~ & ~~$A_{2;K}({\rm GeV}^{-1})$~~ & ~~$\alpha_{2;K}$~~ & ~~$\hat{B}^{2;K}_2$~~ & ~~$\beta_{2;K}({\rm GeV})$~~ \\ \hline
$P_{\chi^2_{\rm min}}$                          & $2.099306$ & $-1.4915$ & $-0.1342$ & $0.647366$ \\
$\phi_{2;K}^{\rm max.}(0.5,\mu_0)$       & $2.435475$ & $-1.4215$ & $-0.1361$ & $0.639261$ \\
$\phi_{2;K}^{\rm min.}(0.5,\mu_0)$       & $1.849371$ & $-1.5445$ & $-0.1323$ & $0.658355$ \\ \hline\hline
\end{tabular}
\end{table}
Further, the values of the corresponding Gegenbauer moments can be obtained. Whose first four Gegenbauer moments are
\begin{align}
a^{2;K}_1(1 ~{\rm GeV}) &= -0.0731^{+0.0089}_{-0.0125}, \nonumber\\
a^{2;K}_2(1 ~{\rm GeV}) &= +0.182^{+0.029}_{-0.030}, \nonumber\\
a^{2;K}_3(1 ~{\rm GeV}) &= -0.0114^{+0.0008}_{-0.0016},\nonumber\\
a^{2;K}_4(1 ~{\rm GeV}) &= +0.041^{-0.003}_{+0.005}. \label{anValues}
\end{align}
As a comparison, we also exhibit our values for the first four moments $\langle \xi^n\rangle _{2;K}$ and the first two Gegenbauer moments $a^{2;K}_1$ and $a^{2;K}_2$ at the scale $\mu = 1, 2 ~{\rm GeV}$ in Table~\ref{table:xin_value}. The values for those moments and Gegenbauer moments by various method{\color{blue}\uwave{s}} such as QCD sum rules (QCD SR)~\cite{Ball:2003sc, Chetyrkin:2007vm, Khodjamirian:2004ga, Braun:2004vf, Ball:2005vx, Ball:2006fz}, LQCD~\cite{Braun:2006dg, Boyle:2006pw,Zhang:2020gaj,Arthur:2010xf, Bali:2019dqc}, AdS/QCD~\cite{Momeni:2017moz} and NLChQM~\cite{Nam:2017gzm} are listed. Then, we can further give,
\begin{align}
\frac{\langle \xi^1\rangle _{2;K}|_{1 ~{\rm GeV}}}{\langle \xi^2\rangle _{2;K}|_{1 ~{\rm GeV}}} \simeq - 0.167, \quad\quad\quad
\frac{a^{2;K}_1(1 ~{\rm GeV})}{a^{2;K}_2(1 ~{\rm GeV})} \simeq -0.402,
\end{align}
where only the central values are adopted.

\subsection{Behavior for the kaon leading-twist DA}
Using the moments $\langle \xi^n\rangle _{2;K}$ with $ n=(1,\cdots,10)$ exhibited in Eqs.~\eqref{xinValueOdd} and \eqref{xinValueEven}, we can determine the model parameters of our LCHO model for the kaon leading-twist DA $\phi_{2;K}(x,\mu)$ by the least squares method as discussed in subsection~\ref{sec:II.II}. In calculation, the initial scale is taken to be $\mu_0 = 1\ ~{\rm GeV}$, which is consistent with the moments $\langle \xi^n\rangle _{2;K}\ (n=1,\cdots,10)$ in Eqs.~\eqref{xinValueOdd} and \eqref{xinValueEven}. Figure~\ref{fgof} shows the goodness of fit $P_{\chi^2}$ versus the fitting parameters $\theta = (\alpha_{2;K},\hat{B}_2^{2;K})$, where Fig.~\ref{fgof}(a) is for $\hat{m}_q = 250\ {\rm MeV}$ and $\hat{m}_s = 370\ {\rm MeV}$, Fig.~\ref{fgof}(b) is for $\hat{m}_q = 330\ {\rm MeV}$ and $\hat{m}_s = 450\ {\rm MeV}$, Fig.~\ref{fgof}(c) is for $\hat{m}_q = 300\ {\rm MeV}$ and $\hat{m}_s = 450\ {\rm MeV}$, respectively. Simultaneously, we can obtain optimal fitting model parameters for those three sets of constituent quark MSs, which are exhibited in Table~\ref{table:model_parameter}. The corresponding values of likelihood function and goodness of fit are: $\chi^2_{\rm min}/n_d = 2.04316/8$ and $P_{\chi^2_{\rm min}} = 0.97966$ are for $\hat{m}_q = 250\ {\rm MeV}$ and $\hat{m}_s = 370\ {\rm MeV}$, $\chi^2_{\rm min}/n_d = 2.2476/8$ and $P_{\chi^2_{\rm min}} = 0.97246$ are for $\hat{m}_q = 330\ {\rm MeV}$ and $\hat{m}_s = 450\ {\rm MeV}$, $\chi^2_{\rm min}/n_d = 2.69377/8$ are $P_{\chi^2_{\rm min}} = 0.952082$ are for $\hat{m}_q = 300\ {\rm MeV}$ and $\hat{m}_s = 450\ {\rm MeV}$, respectively. In order to analyze the uncertainty of our kaon leading-twist DA, we take the upper and lower limits of the uncertainty of $\phi_{2;K}(x,\mu_0)$ as the two curves that lead to the maximum and minimum $\phi_{2;K}(0.5,\mu_0)$ respectively in the parameter region $P_{\chi^2} \geq 50\%$, the corresponding model parameters are also exhibited in Table~\ref{table:model_parameter}.  These model parameters exhibited in Table~\ref{table:model_parameter} are corresponding to initial scale $\mu_0 = 1\ ~{\rm GeV}$, and their values at any scale $\mu$ can be obtained via the renormalization group equation~\cite{Lepage:1980fj,Huang:2013yya}.

\begin{figure}[t]
\centering
\includegraphics[width=0.50\textwidth]{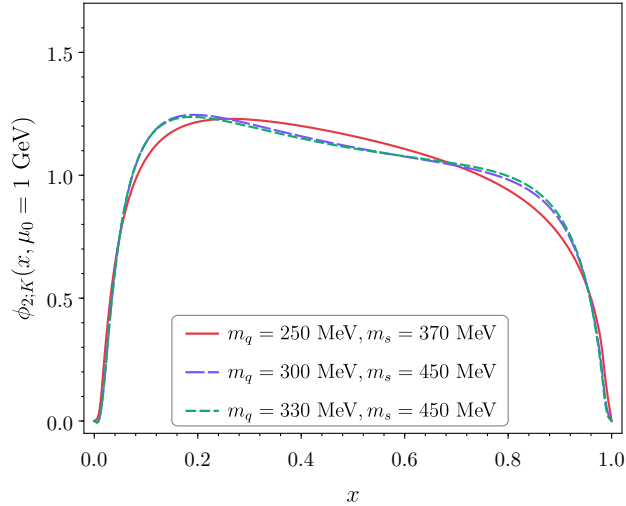}
\caption{Comparison for our DA $\phi_{2;K}(x,\mu_0=1\ ~{\rm GeV})$ with different MSs.}
\label{ourDAForDifferentMassScheme}
\end{figure}

The kaon DA $\phi_{2;K}(x,\mu_0)$ with different constituent quark MSs are shown in Figure~\ref{ourDAForDifferentMassScheme}. One can find that these cures are very close to each other. This is reasonable because they are obtained by fitting same set of data, i.e. Eqs.~\eqref{xinValueOdd} and \eqref{xinValueEven}, which in turn shows that the value of ten moments $\langle \xi^n\rangle _{2;K}$ with $n=(1,\cdots,10)$ has a strong constraint on the behavior of kaon leading-twist DA. More specifically, the curves corresponding $\hat{m}_q = 330\ {\rm MeV}$, $\hat{m}_s = 450~{\rm MeV}$ and $\hat{m}_q = 300 ~{\rm MeV}$, $\hat{m}_s = 450 ~{\rm MeV}$ almost overlap each other, and they are slightly different from the one for $\hat{m}_q = 250\ {\rm MeV}$ and $\hat{m}_s = 370~{\rm MeV}$. Obviously, this case is caused by the different constituent quark masses. Considering the goodness of fit corresponding to the optimal fitting model parameters, $P_{\chi^2_{\rm min}}$, for $\hat{m}_q = 250\ {\rm MeV}$ and $\hat{m}_s = 370\ {\rm MeV}$ is the best, we will adopt this constituent quark MS for subsequent discussion and calculation.

\begin{figure}[t]
\centering
\includegraphics[width=0.48\textwidth]{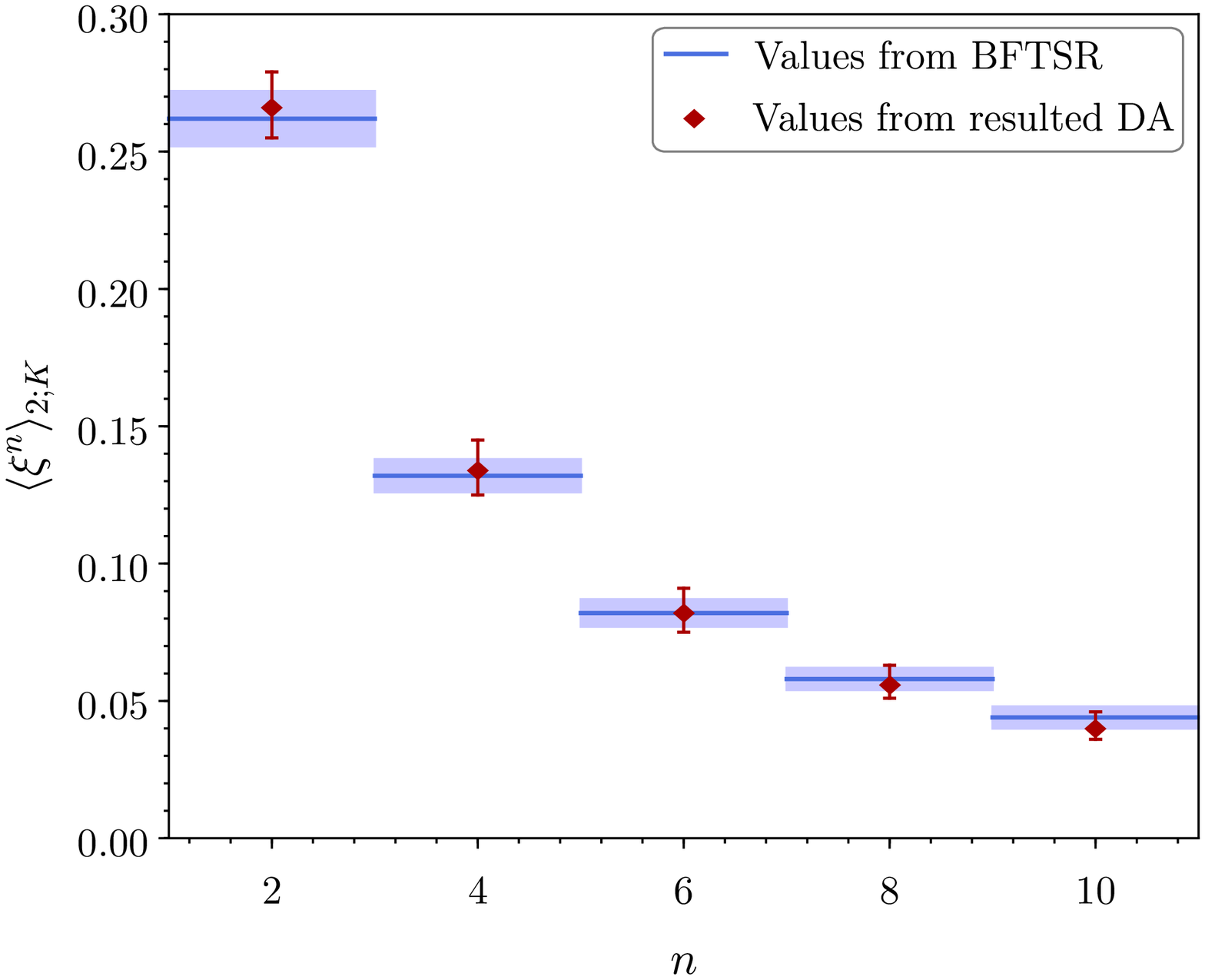}\includegraphics[width=0.495\textwidth]{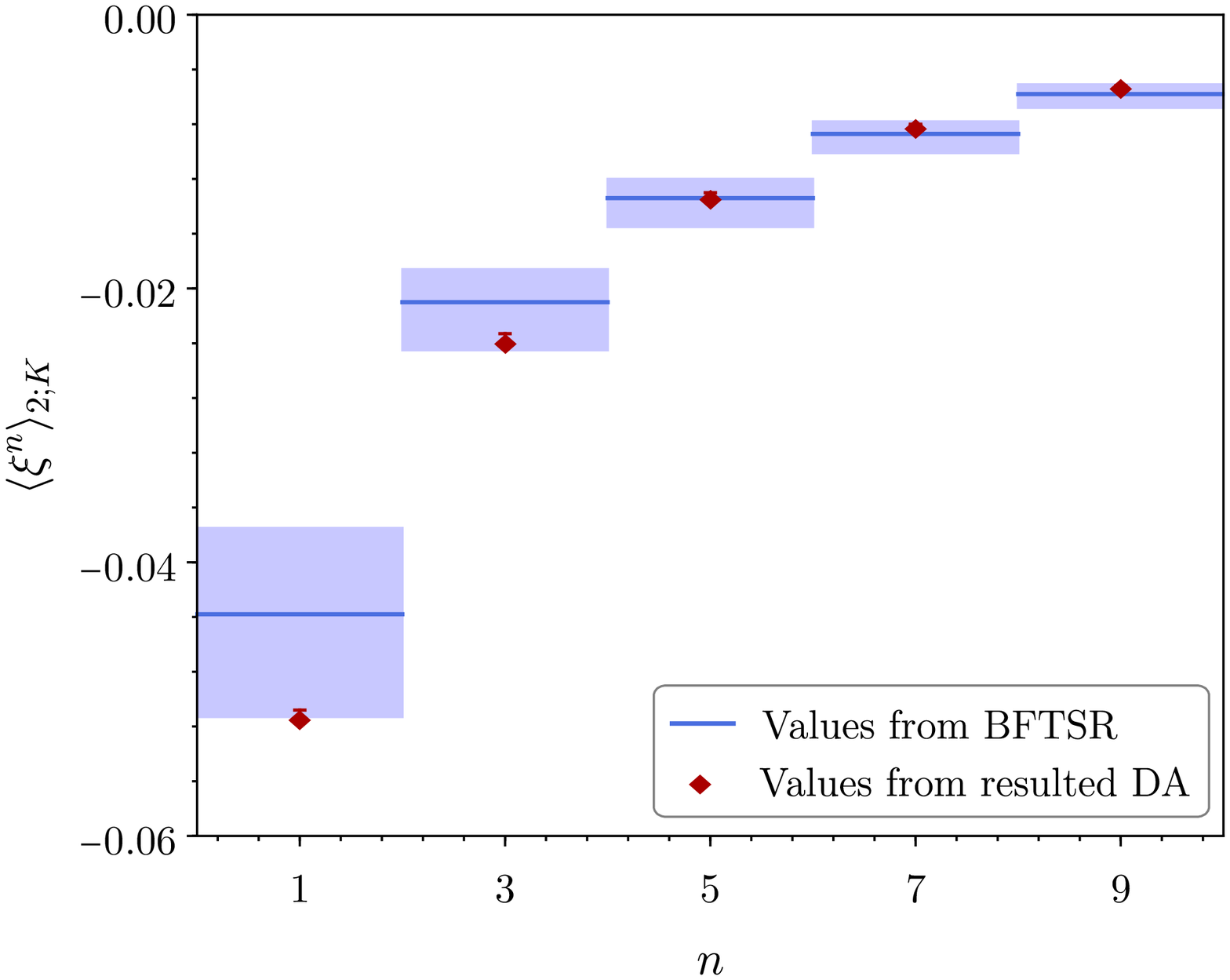}
\caption{Comparison of the moments in Eqs. (\ref{xinValueOdd}, \ref{xinValueEven}) from BFTSR and the moments in Eq. (\ref{xinValueFittingOdd}, \ref{xinValueFittingEven}) from our fitting LCHO model, where the left plan is for even moments and the right plan is for odd moments.}
\label{fCompareMoments}
\end{figure}
\begin{figure}[t]
\centering
\includegraphics[width=0.49\textwidth]{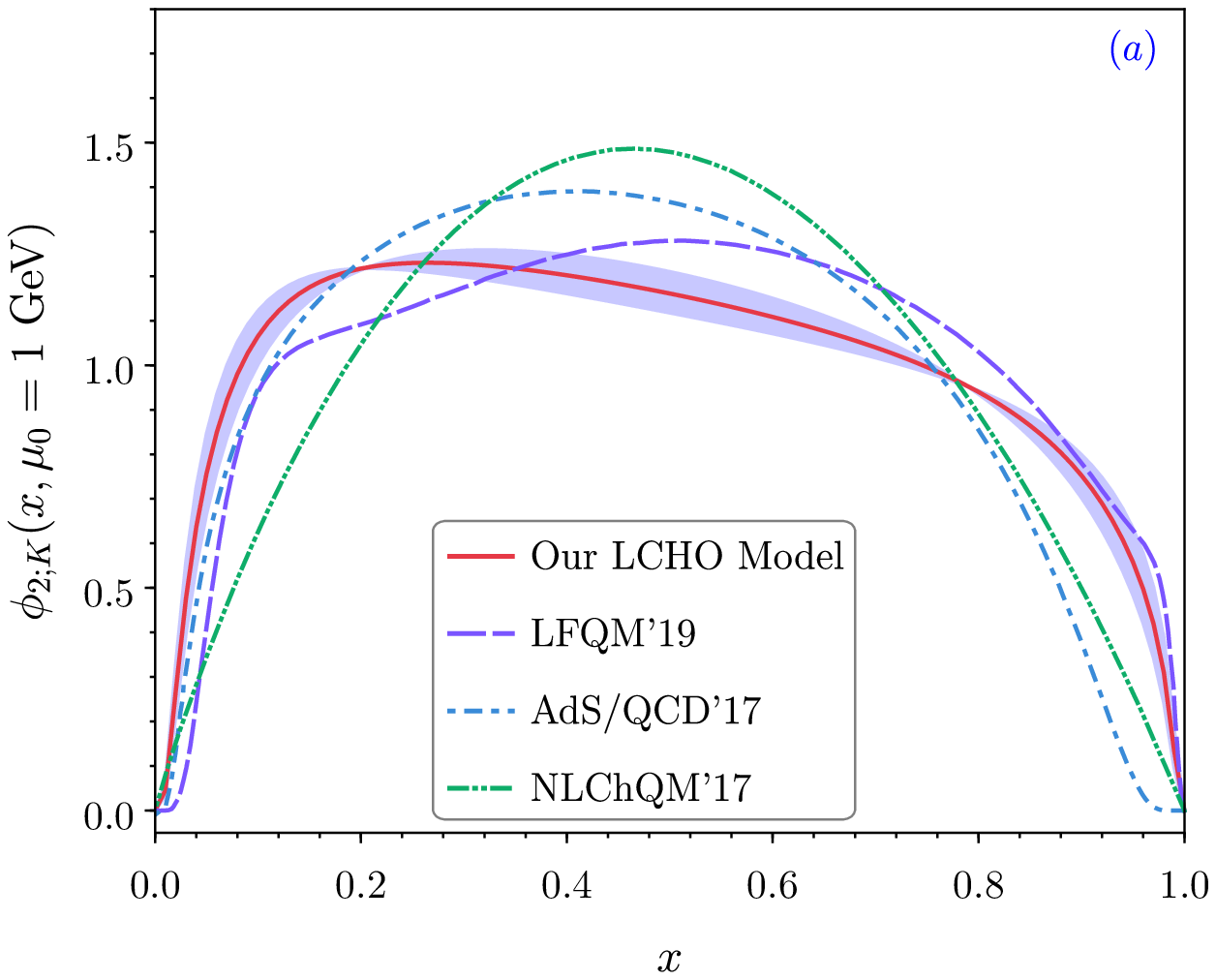}
\includegraphics[width=0.48\textwidth]{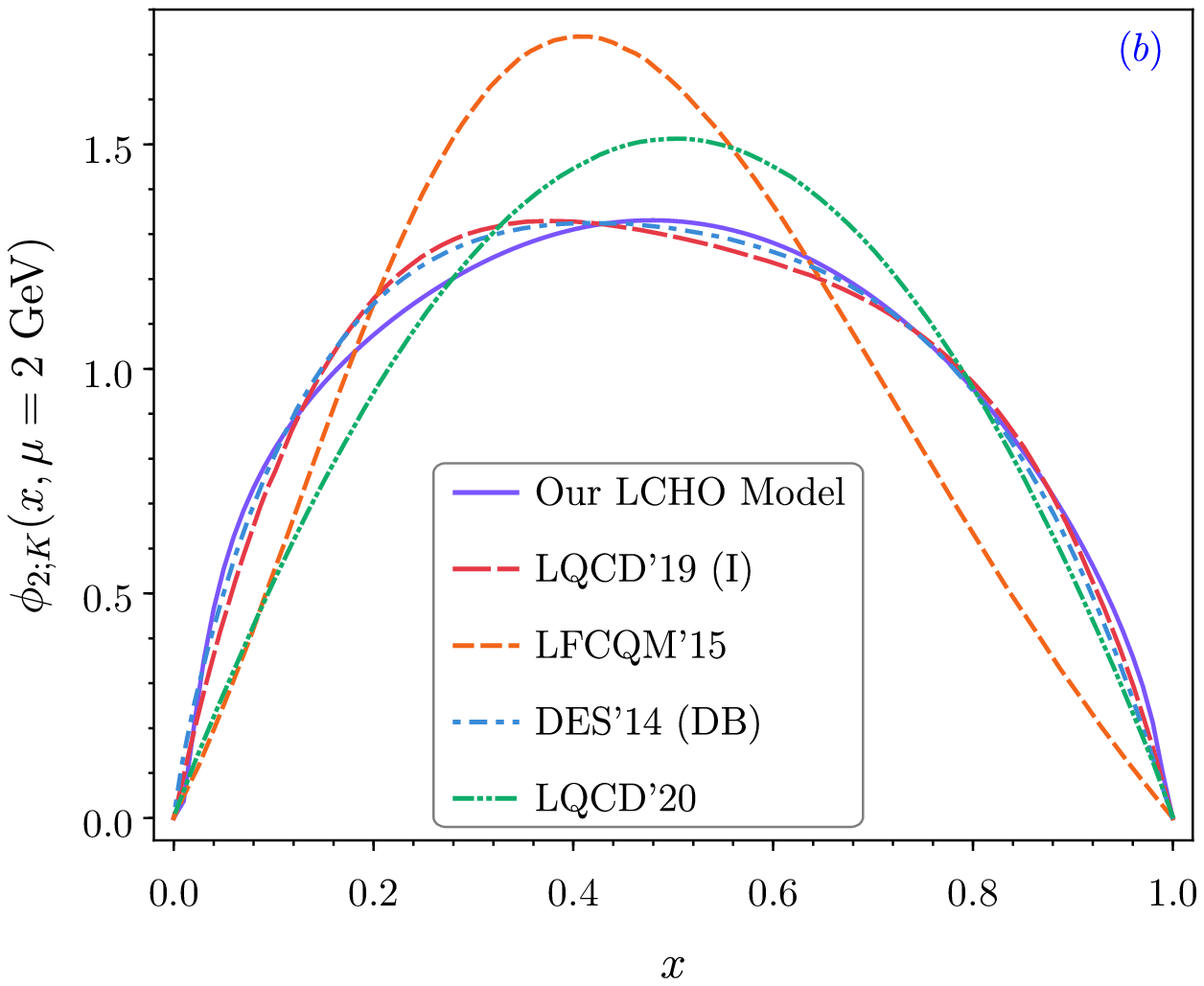}
\caption{The kaon leading-twist DA curves in this work. For comparison, the AdS/QCD model~\cite{Momeni:2017moz}, the DA by LQCD~\cite{Bali:2019dqc,Zhang:2020gaj}, by NLChQM~\cite{Nam:2017gzm}, by LFQM~\cite{Dhiman:2019ddr}, by LFCQM~\cite{deMelo:2015yxk}, by DSE~\cite{Shi:2014uwa} are presented.}
\label{fDA}
\end{figure}
In order to show the advantage of the fitting results more intuitively, we substitute the model parameters of rows $2$, $3$ and $4$ in Table~\ref{table:model_parameter} into Eqs.~(\ref{varphi}), (\ref{DA_model}) and \eqref{moment} to calculate the values of the first ten moments, which are:
\begin{align}
\langle \xi^1 \rangle_{2;K} |_{\mu_0} &= -0.0513^{+0.0005}_{-0.0001}, \nonumber\\
\langle \xi^3 \rangle_{2;K} |_{\mu_0} &= -0.0238^{+0.0005}_{-0.0002}, \nonumber\\
\langle \xi^5 \rangle_{2;K} |_{\mu_0} &= -0.0133^{+0.0003}_{-0.0001}, \nonumber\\
\langle \xi^7 \rangle_{2;K} |_{\mu_0} &= -0.0081^{+0.0001}_{-0.0000}, \nonumber\\
\langle \xi^9 \rangle_{2;K} |_{\mu_0} &= -0.0052^{+0.0000}_{-0.0001}.\label{xinValueFittingOdd}
\end{align}
and
\begin{align}
\langle \xi^2 \rangle_{2;K} |_{\mu_0} &= 0.267^{+0.012}_{-0.012}, \nonumber\\
\langle \xi^4 \rangle_{2;K} |_{\mu_0} &= 0.135^{+0.010}_{-0.010}, \nonumber\\
\langle \xi^6 \rangle_{2;K} |_{\mu_0} &= 0.083^{+0.008}_{-0.008}, \nonumber\\
\langle \xi^8 \rangle_{2;K} |_{\mu_0} &= 0.057^{+0.006}_{-0.006}, \nonumber\\
\langle \xi^{10} \rangle_{2;K} |_{\mu_0} &= 0.041^{+0.005}_{-0.005}, \label{xinValueFittingEven}
\end{align}

A comparison of the values of moments in Eqs. (\ref{xinValueOdd}), (\ref{xinValueEven}) and in Eqs. (\ref{xinValueFittingOdd}), (\ref{xinValueFittingEven}) is shown in Figure~\ref{fCompareMoments}, one can find that our fitting is good. The curves of our kaon leading-twist DA at scale $\mu_0 = 1~{\rm GeV}$ and $\mu=2~{\rm GeV}$ are shown in Figure~\ref{fDA}, where as comparison, we also present the AdS/QCD model~\cite{Momeni:2017moz}, the DA by LQCD~\cite{Bali:2019dqc,Zhang:2020gaj}, by NLChQM~\cite{Nam:2017gzm}, by LFQM~\cite{Dhiman:2019ddr}, by LFCQM~\cite{deMelo:2015yxk}, by DSE~\cite{Shi:2014uwa}. One can find that our DA is close to the LFQM and LQCD ones.

\subsection{$B_s\to K$ TFF and CKM matrix element $|V_{ub}|$ from the semileptonic decay processes $B_s\to K\ell\bar\nu_\ell$}

\begin{table}[t]
\centering
\caption{The $B_s\to K$ transition form factor at large recoil region, e.g. $f^{B_s\to K}_+(0)$. Other results from references are also listed as a comparison.}\label{Tab:TFF0}
\begin{tabular}{p{13.7cm} l}
\hline\hline
Methods  ~~~~~~~~~~~~~~~~~~~~~~~~~~~~~~~~~~~~~~~~~~~~~~~~~~~~~~~~~~~~~& $f^{B_s\to K}_+(0)$  \\ \hline
This work   & $0.270^{+0.022}_{-0.030}$\\
LQCD (FNAL/MILC)~\cite{FermilabLattice:2019ikx} & $0.135\pm0.050$\\
LCSR~\cite{Duplancic:2008tk}                    & $0.30^{+0.04}_{-0.03}$\\
LCSR~\cite{Khodjamirian:2017fxg}                & $0.336\pm0.023$\\
pQCD~\cite{Wang:2012ab}             & $0.26^{+0.04}_{-0.03}\pm0.02$ \\
RQM~\cite{Faustov:2013ima} & $0.284\pm0.014$\\
Pade approximamts~\cite{Gonzalez-Solis:2021awb} & $0.211\pm0.003$\\
\hline\hline
\end{tabular}
\end{table}

\begin{figure}[t]
\centering
\includegraphics[width=0.55\textwidth]{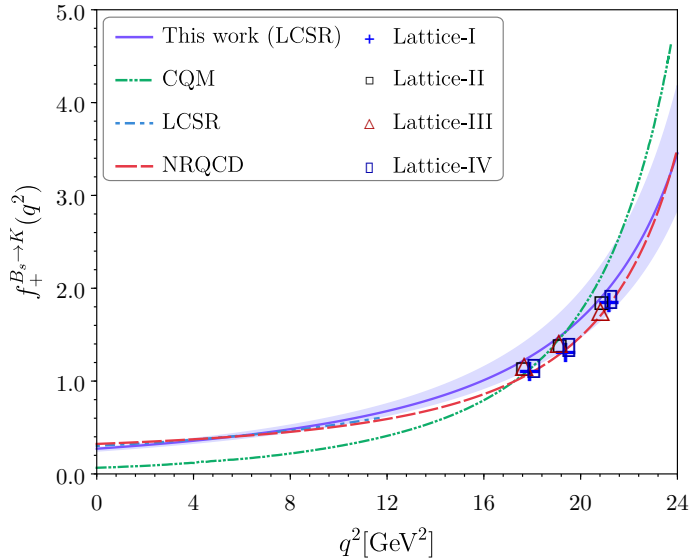}
\caption{The TFF $f^{B_s\to K}_+(q^2)$ in the whole physical region within error. The CQM~\cite{Albertus:2014gba}, LCSR~\cite{Li:2001yv}, NRQCD~\cite{Bouchard:2013zda} and LQCD results~\cite{Bouchard:2014ypa} are also present as a comparison.}
\label{Fig_TFF}
\end{figure}

Normally, in order to study the $B_s\to K$ TFF, the optimal renormalization scale for semileptonic decay $B_s^0 \to K^- \mu^+\nu_\mu$ is necessary, which is taken as $\mu_{\rm IR} = \sqrt{m_{B_s}^2 - m_b^2} \sim \sqrt{2m_b\bar\Lambda}\approx 3~{\rm GeV}$. The basic input parameters are $m_{B_s} = 5.366~{\rm GeV}$, and $f_{B_s} = 0.266\pm0.019$. By adopting the criteria of the LCSR, we determine the Borel parameter and continuum threshold are $M^2 = 20.0\pm0.5~{\rm GeV}^2$ and $s_0^{B_s} = 34.0 \pm 0.5~{\rm GeV}^2$. Then, the TFF of $B_s\to K$ at large recoil point $f_+^{B_\to K}(0)$ is shown in Table~\ref{Tab:TFF0}. The results by LQCD (FNAL/MILC)~\cite{FermilabLattice:2019ikx},  LCSR~\cite{Duplancic:2008tk, Khodjamirian:2017fxg}, pQCD~\cite{Wang:2012ab}, RQM~\cite{Faustov:2013ima} and Pade approximamts~\cite{Gonzalez-Solis:2021awb} are also shown as comparison. Our results have agreement with the perturbative QCD prediction within errors. We remind that LCSR approach for $B_s\to K$ TFF are valid up to squared momentum transfers $q^2\sim m_b^2 - 2m_b\bar\Lambda$. Meanwhile, to be on the safe side, we take the maximal allowed $q^2$ as $0\leq q^2\leq 12~{\rm GeV^2}$. To obtained the TFF at the whole physical region, e.g. $0\leq q^2 \leq (m_{B_s}^2 - m_{K}^2) = 23.74~{\rm GeV}^2$, we can use the simplified series expansion of $z$-parameterizations, which was discussed in our previous work~\cite{Hu:2021zmy}. Then, the TFF with whole physical region are shown in Figure~\ref{Fig_TFF}, which CQM~\cite{Albertus:2014gba}, LCSR~\cite{Li:2001yv}, NRQCD~\cite{Bouchard:2013zda} and LQCD results~\cite{Bouchard:2014ypa} are also present. Our predictions have agreement with the Lattice results within errors.

\begin{figure}[t]
\centering
\includegraphics[width=0.55\textwidth]{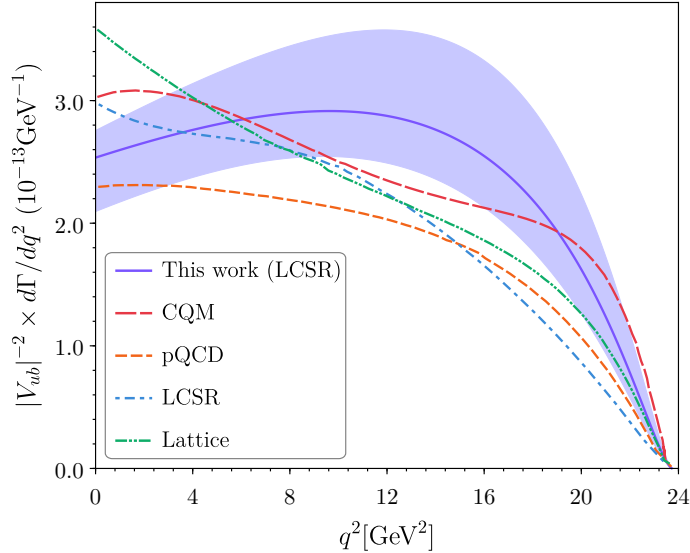}
\caption{The $|V_{ub}|$-independent differential decay width for semileptonic $B_s^0 \to K^- \mu^+\nu_\mu$. The CQM~\cite{Albertus:2014gba}, pQCD~\cite{Wang:2012ab}, LCSR~\cite{Li:2001yv}, LQCD~\cite{Bouchard:2014ypa} predictions are also present. }
\label{Fig:dGamma}
\end{figure}

With the resultant $B_s\to K$ TFF, we can get the $|V_{ub}|$-independent differential decay width of $B_s^0 \to K^- \mu^+ \nu_\mu$ shown in Figure~\ref{Fig:dGamma}. In which the CQM~\cite{Albertus:2014gba}, pQCD~\cite{Wang:2012ab}, LCSR~\cite{Li:2001yv}, LQCD~\cite{Bouchard:2014ypa} predictions are also present.
\begin{table}[t]
\centering
\caption{$|V_{ub}|^2$-independent $B_s^0 \to K^- \mu^+ \nu_\mu$ decay width in units of $10^{-12}{\rm GeV}$ from our prediction and some different methods.}\label{Tab:DecayWidth}
\begin{tabular}{l l}
\hline\hline
Methods  ~~~~~~~~~~~~~~~~~~~~~~~~~~~~~~~~~~~~~~~~~~~~~~~~~~~~~~~~~~~~~~~~~~~~~~~~& $|V_{ub}|^{-2}\times \Gamma(B_s^0 \to K^- \mu^+ \nu_\mu)[10^{-12}{\rm GeV}]$  \\ \hline
This work                                       & $5.626^{+1.271}_{-0.864}$\\
CQM~\cite{Albertus:2014gba}                     & $5.45^{+0.83}_{-0.80}$\\
LCSR+$\bar B^*$-pole~\cite{Li:2001yv}           & $4.63^{+0.97}_{-0.88}$\\
RQM~\cite{Faustov:2013ima}                      & $4.50\pm0.55$\\
LFQM~\cite{Verma:2011yw}                        & $3.17\pm 0.24$\\
pQCD~\cite{Wang:2012ab}                         & $4.2\pm2.1$ \\
\hline\hline
\end{tabular}
\end{table}
\begin{figure}[t]
\centering
\includegraphics[width=0.55\textwidth]{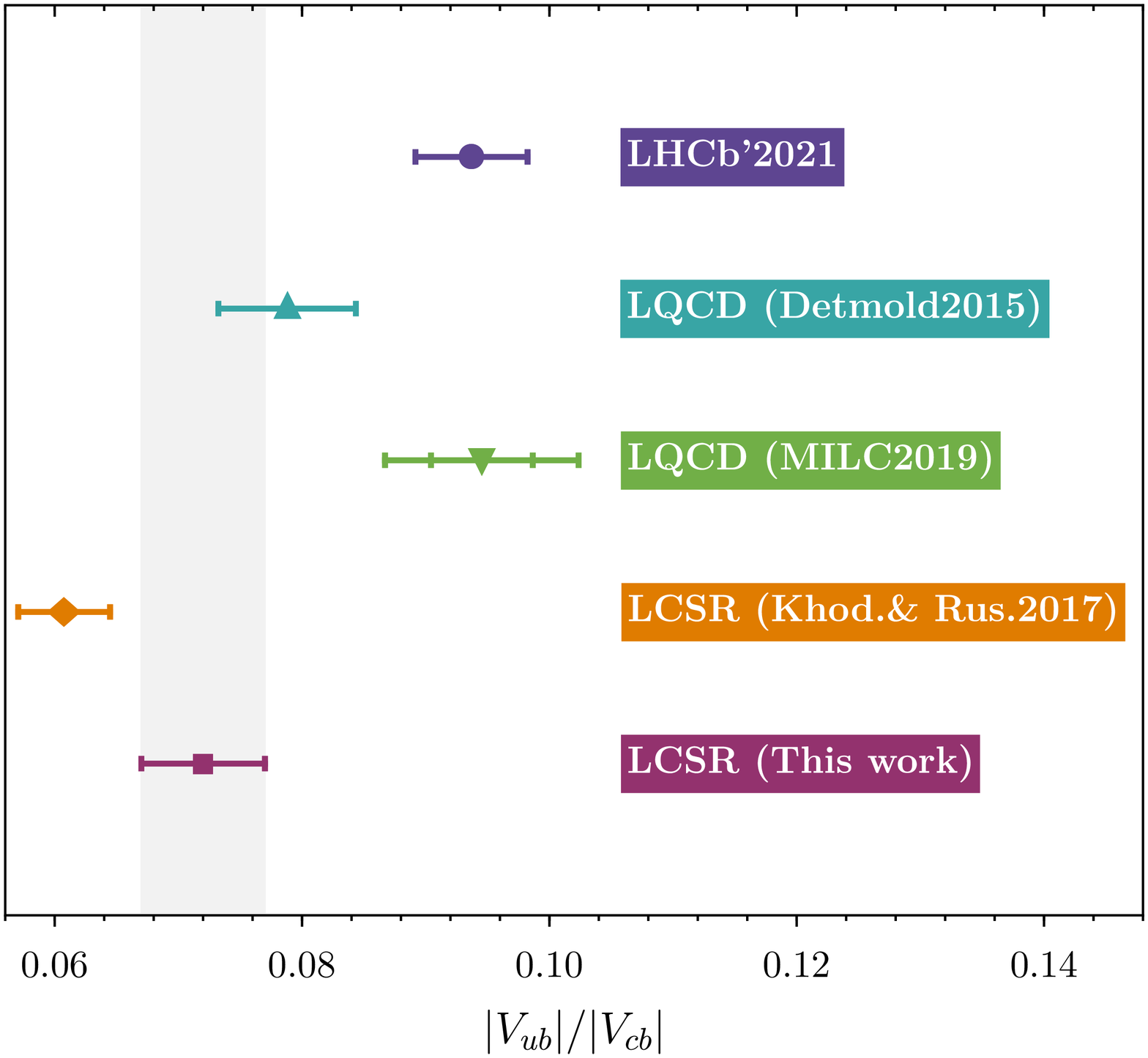}
\caption{The ratio of CKM matrix elements $|V_{ub}|/|V_{cb}|$ for our predictions with errors. The LHCb~\cite{LHCb:2020ist}, Lattice QCD~\cite{FermilabLattice:2019ikx,Detmold:2015aaa} and LCSR~\cite{Khodjamirian:2017fxg} are also present.}
\label{Fig:RatioCKM}
\end{figure}

After integrate the differential decay width with the whole physical region, we can get the $|V_{ub}|$-independent total decay width, which are listed in Table~\ref{Tab:DecayWidth}. Other theoretical results are also given. Our predictions have agreement with the CQM, LCSR, RQM and pQCD results within errors. Furthermore, to determine the CKM ratio $|V_{ub}|/|V_{cb}|$, the absolute branching fraction of $B_s^0\to K^-\mu^+\nu_\mu$ is required. Here, we take the new measurements of ${\cal B}(B_s^0\to K^-\mu^+\nu_\mu) = (1.06\pm 0.05({\rm stat})\pm 0.08({\rm syst})) \times 10^{-4}$ from the LHCb collaboration for the first time~\cite{LHCb:2020ist}. The inputs are the exclusive value of $|V_{cb}| = (39.5\pm0.9)\times 10^{-3}$~\cite{PDGnew}, the $B_s^0$-meson lifetime $\tau_{B_s} = 1.515\pm0.004 {\rm ps}$. After taking the $|V_{ub}|$-independent decay width, the obtained values are
\begin{align}
|V_{ub}|/|V_{cb}| = 0.072\pm0.005
\end{align}
In order to clearly compare the results of different groups, we depicts $|V_{ub}|/|V_{cb}| $ in Fig.~\ref{Fig:RatioCKM}. Our results have agreement with LQCD predicted by Detmold in 2015 within errors. The LHCb predictions are mainly coming from the average value $|V_{ub}|/|V_{cb}|({\rm low}) = 0.061(4)$ and $|V_{ub}|/|V_{cb}|({\rm high}) = 0.095(8)$ with the uncertainties combined. The $|V_{ub}|/|V_{cb}|$ measurement obtained with the $\Lambda_b^0$ baryon decays~\cite{LHCb:2015eia}, for which a form factor model based on a LQCD calculation was used~\cite{Detmold:2015aaa}.

\section{Summary}\label{sec:IV}

Based on the fact that the sum rule of the zeroth moment $\langle \xi^0\rangle _{2;K}$ of DA $\phi_{2;K}(x,\mu)$ can not be normalized in full Borel parameter $M^2$ regions, a more reasonable sum rules formula of the moments $\langle \xi^n\rangle _{2;K}$, i.e., Eq.~\eqref{xin}, has been adopted to do the calculation. Then more accurate values of first ten moments of DA $\phi_{2;K}(x,\mu)$ have been obtained, which are given in Eqs. (\ref{xinValueOdd}, \ref{xinValueEven}). On the other hand, we have suggested a new LCHO model for kaon leading-twist DA based on BHL prescription. By fitting those resulted first ten moments with the least squares method, the behavior of $\phi_{2;K}(x,\mu)$ has been obtained, which is consistent with that is derived by LFQM~\cite{Dhiman:2019ddr}, LQCD~\cite{Bali:2019dqc} and DSE (DB)~\cite{Shi:2014uwa}, respectively. The LCHO model for $\phi_{2;K}(x,\mu)$ is mathematical, whose rationality is judged by its goodness of fit to the moments. The results show that the goodness of fit is very close to $1$ for different constituent quark MSs. In this paper, we have used the method of fitting moments to determine the behavior of DA $\phi_{2;K}(x,\mu)$, rather than solving the constraints provided by the Gegenbauer moments. The derived results show that the goodness of fit is very close to $1$ for different constituent quark MSs. This method can further improve the accuracy of resulted $\phi_{2;K}(x,\mu)$ by improving the numerical accuracy of moments and adopting more moments.

Secondly, the TFF for $B_s\to K$ has been calculated by using the LCSR approach with chiral correlator up to NLO accuracy. The value of our prediction is shown in Table~\ref{Tab:TFF0}, which is agreement with pQCD prediction within error. After using the simplified series expansion of $z$-parameterizations, the resultant TFF for the whole physical $q^2$-region are given in Fig.~\ref{Fig_TFF}. Furthermore, the $|V_{ub}|$-independent differential decay width for $B_s^0\to K^-\mu^+\nu_\mu$, accompanied with references result are shown in Fig.~\ref{Fig:dGamma}. Meanwhile, the values of $|V_{ub}|$-independent total decay width are given in Table~\ref{Tab:DecayWidth}, which agree well with CQM, LCSR, RQM and pQCD results within errors. Finally, we have determined the ratio of CKM matrix element $|V_{ub}|/|V_{cb}| = 0.072\pm0.005$ by using the new branching fraction from LHCb collaboration and exclusive $|V_{cb}|$ value, which are shown in Fig.~\ref{Fig:RatioCKM}. Our prediction have agreement with the Lattice results within errors, which is better than the previous LCSR calculation. Yet, there still have discrepancy with the results of LHCb collaboration. We hope that this ratio will be investigated by experiments and theories in the near future.

{\bf Acknowledgments}:
This work was supported in part by the National Natural Science Foundation of China under Grant No.11765007, No.11947406, No.12147102, No.11875122, and No.12175025, the Project of Guizhou Provincial Department of Science and Technology under Grant No.KY[2019]1171 and No.ZK[2021]024, the Project of Guizhou Provincial Department of Education under Grant No.KY[2021]030 and No.KY[2021]003, the Chongqing Graduate Research and Innovation Foundation under Grant No.ydstd1912, the Fundamental Research Funds for the Central Universities under Grant No.2020CQJQY-Z003, and the Project of Guizhou Minzu University under Grant No. GZMU[2019]YB19.


\begin{thebibliography}{s2}

\bibitem{LHCb:2020ist}
  R.~Aaij \textit{et al.} [LHCb Collaboration],
  \textit{First observation of the decay $B_s^0 \to K^-\mu^+\nu_\mu$ and Measurement of $|V_{ub}|/|V_{cb}|$},
  \href{https://doi.org/10.1103/PhysRevLett.126.081804}
  {Phys. Rev. Lett. \textbf{126} (2021) 081804}.
  [\href{https://arxiv.org/abs/2012.05143}
  {arXiv:2012.05143}]

\bibitem{Li:2001yv}
  Z.~H.~Li, F.~Y.~Liang, X.~Y.~Wu and T.~Huang,
  \textit{The $B_s\to K$ form-factor in the whole kinematically accessible range},
  \href{https://doi.org/10.1103/PhysRevD.64.057901}
  {Phys.\ Rev.\ D {\bf 64} (2001) 057901}.
  [\href{https://arxiv.org/abs/hep-ph/0106186}
  {hep-ph/0106186}]

\bibitem{Khodjamirian:2003xk}
  A.~Khodjamirian, T.~Mannel and M.~Melcher,
  \textit{Flavor $SU(3)$ symmetry in charmless $B$ decays}.
  \href{https://doi.org/10.1103/PhysRevD.68.114007}
  {Phys.\ Rev.\ D {\bf 68} (2003) 114007}.
  [\href{https://arxiv.org/abs/hep-ph/0308297}
  {hep-ph/0308297}]

\bibitem{Wu:2006rd}
  Y.~L.~Wu, M.~Zhong and Y.~B.~Zuo,
  \textit{$B_{(s)}. D_{(s)} \to \pi, K, \eta, \rho, K^\ast, \omega, \phi$ transition form factors and decay rates with extraction of the CKM parameters $|V_{ub}|$, $|V_{cs}|$, $|V_{cd}|$}.
  \href{https://doi.org/10.1142/S0217751X06033209}
  {Int.\ J.\ Mod.\ Phys.\ A {\bf 21} (2006) 6125}.
  [\href{https://arxiv.org/abs/hep-ph/0604007}
  {hep-ph/0604007}]

\bibitem{Duplancic:2008tk}
  G.~Duplancic and B.~Melic,
  \textit{$B, B_s\to K$ form factors: An update of light-cone sum rule results}.
  \href{https://doi.org/10.1103/PhysRevD.78.054015}
  {Phys.\ Rev.\ D {\bf 78} (2008) 054015}.
  [\href{https://arxiv.org/abs/0805.4170}
  {arXiv:0805.4170}]

\bibitem{Melic:2008cx}
  B.~Melic,
  \textit{$B \to \pi$ and $B_s \to K$ form factors and $V_{ub}$ determination}.
  \href{https://arxiv.org/abs/0810.1144}
  {arXiv:0810.1144}.

\bibitem{Khodjamirian:2017fxg}
  A.~Khodjamirian and A.~V.~Rusov,
  \textit{$B_{s}\to K \ell \nu_\ell$ and $B_{(s)} \to \pi (K) \ell^+\ell^-$ decays at large recoil and CKM matrix elements}.
  \href{https://doi.org/10.1007/JHEP08(2017)112}
  {JHEP {\bf 1708} (2017) 112}.
  [\href{https://arxiv.org/abs/1703.04765}
  {arXiv:1703.04765}]

\bibitem{Bouchard:2013zda}
  C.~M.~Bouchard, G.~P.~Lepage, C.~J.~Monahan, H.~Na and J.~Shigemitsu,
  \textit{$B$ and $B_s$ semileptonic decay form factors with NRQCD/HISQ quarks}.
  \href{https://doi.org/10.22323/1.187.0387}
  {PoS LATTICE {\bf 2013} (2014) 387}.
  [\href{https://arxiv.org/abs/1310.3207}
  {arXiv:1310.3207}]

\bibitem{Bouchard:2014ypa}
  C.~M.~Bouchard, G.~P.~Lepage, C.~Monahan, H.~Na and J.~Shigemitsu,
  \textit{$B_s \to K \ell \nu$ form factors from lattice QCD}.
  \href{https://doi.org/10.1103/PhysRevD.90.054506}
  {Phys.\ Rev.\ D {\bf 90} (2014) 054506}.
  [\href{https://arxiv.org/abs/1406.2279}
  {arXiv:1406.2279}]

\bibitem{Flynn:2015mha}
  J.~M.~Flynn, T.~Izubuchi, T.~Kawanai, C.~Lehner, A.~Soni, R.~S.~Van de Water and O.~Witzel,
  \textit{$B \to \pi \ell \nu$ and $B_s \to K \ell \nu$ form factors and $|V_{ub}|$ from 2+1-flavor lattice QCD with domain-wall light quarks and
  relativistic heavy quarks}.
  \href{https://doi.org/10.1103/PhysRevD.91.074510}
  {Phys.\ Rev.\ D {\bf 91} (2015) 074510}.
  [\href{https://arxiv.org/abs/1501.05373}
  {arXiv:1501.05373}]

\bibitem{Bahr:2016ayy}
  F.~Bahr {\it et al.} [ALPHA Collaboration],
  \textit{Continuum limit of the leading-order HQET form factor in $B_s \to K\ell\nu$  decays}.
  \href{https://doi.org/10.1016/j.physletb.2016.03.088}
  {Phys.\ Lett.\ B {\bf 757} (2016) 473}.
  [\href{https://arxiv.org/abs/1601.04277}
  {arXiv:1601.04277}]

\bibitem{Monahan:2018lzv}
  C.~J.~Monahan, C.~M.~Bouchard, G.~P.~Lepage, H.~Na and J.~Shigemitsu,
  \textit{Form factor ratios for $B_s \rightarrow K \, \ell \, \nu$ and $B_s \rightarrow D_s \, \ell \, \nu$ semileptonic decays and $|V_{ub}/V_{cb}|$}.
  \href{https://doi.org/10.1103/PhysRevD.98.114509}
  {Phys.\ Rev.\ D {\bf 98} (2018) 114509}.
  [\href{https://arxiv.org/abs/1808.09285}
  {arXiv:1808.09285}]

\bibitem{Wang:2012ab}
  W.~F.~Wang and Z.~J.~Xiao,
  \textit{The semileptonic decays $B/B_s \to (\pi, K)(\ell^+\ell^-,\ell\nu,\nu\bar{\nu})$ in the perturbative QCD approach beyond the leading-order}.
  \href{https://doi.org/10.1103/PhysRevD.86.114025}
  {Phys.\ Rev.\ D {\bf 86} (2012) 114025}.
  [\href{https://arxiv.org/abs/1207.0265}
  {arXiv:1207.0265}]

\bibitem{Meissner:2013pba}
  U.~G.~Mei{\ss}ner and W.~Wang,
  \textit{${B_s\to K^{(*)} \ell\bar \nu}$, angular analysis, S-wave contributions and ${|V_{ub}|}$}.
  \href{https://doi.org/10.1007/JHEP01(2014)107}
  {JHEP {\bf 1401} (2014) 107}.
  [\href{https://arxiv.org/abs/1311.5420}
  {arXiv:1311.5420}]

\bibitem{Jin:2020jtu}
  S.~P.~Jin, X.~Q.~Hu and Z.~J.~Xiao,
  \textit{Study of $B_s\to K^{(*)}\ell^+ \ell^-$ decays in the PQCD factorization approach with lattice QCD input}.
  \href{https://doi.org/10.1103/PhysRevD.102.013001}
  {Phys.\ Rev.\ D {\bf 102} (2020) 013001}.
  [\href{https://arxiv.org/abs/2003.12226}
  {arXiv:2003.12226}]

\bibitem{Bourrely:2008za}
  C.~Bourrely, I.~Caprini and L.~Lellouch,
  \textit{Model-independent description of $B\to\pi l \nu$ decays and a determination of $|V_{ub}|$}.
  \href{https://doi.org/10.1103/PhysRevD.82.099902}
  {Phys.\ Rev.\ D {\bf 79} (2009) 013008}.
  \href{https://doi.org/10.1103/PhysRevD.79.013008}
  {Erratum: [Phys.\ Rev.\ D {\bf 82} (2010) 099902]}.
  [\href{https://arxiv.org/abs/0807.2722}
  {arXiv:0807.2722}]

\bibitem{Albertus:2014gba}
  C.~Albertus, E.~Hern\'{a}ndez, C.~Hidalgo-Duque and J.~Nieves,
  \textit{$\bar{B}_s\to K$ semileptonic decay from an Omn\`{e}s improved constituent quark model}.
  \href{https://doi.org/10.1016/j.physletb.2014.09.037}
  {Phys.\ Lett.\ B {\bf 738} (2014) 144}.
  [\href{https://arxiv.org/abs/1404.1001}
  {arXiv:1404.1001}]

\bibitem{Albertus:2014rna}
  C.~Albertus, C.~Hidalgo-Duque, J.~Nieves and E.~Hern\'{a}ndez,
  \textit{$\bar{B}_s\to K$ semileptonic decay from an Omn\`{e}s improved nonrelativistic quark model}.
  \href{https://doi.org/10.1088/1742-6596/556/1/012026}
  {J.\ Phys.\ Conf.\ Ser.\  {\bf 556} (2014) 012026}.
  [\href{https://arxiv.org/abs/1409.4207}
  {arXiv:1409.4207}]

\bibitem{Faustov:2013ima}
  R.~N.~Faustov and V.~O.~Galkin,
  \textit{Charmless weak $B_s$ decays in the relativistic quark model}.
  \href{https://doi.org/10.1103/PhysRevD.87.094028}
  {Phys.\ Rev.\ D {\bf 87} (2013) 094028}.
  [\href{https://arxiv.org/abs/1304.3255}
  {arXiv:1304.3255}]

\bibitem{Verma:2011yw}
  R.~C.~Verma,
  \textit{Decay constants and form factors of s-wave and p-wave mesons in the covariant light-front quark model}.
  \href{https://doi.org/10.1088/0954-3899/39/2/025005}
  {J.\ Phys.\ G {\bf 39} (2012) 025005}.
  [\href{https://arxiv.org/abs/1103.2973}
  {arXiv:1103.2973}]

\bibitem{Lu:2007sg}
  C.~D.~Lu, W.~Wang and Z.~T.~Wei,
  \textit{Heavy-to-light form factors on the light cone}.
  \href{https://doi.org/10.1103/PhysRevD.76.014013}
  {Phys.\ Rev.\ D {\bf 76} (2007) 014013}.
  [\href{https://arxiv.org/abs/hep-ph/0701265}
  {hep-ph/0701265}]

\bibitem{Ali:2007ff}
  A.~Ali, G.~Kramer, Y.~Li, C.~D.~Lu, Y.~L.~Shen, W.~Wang and Y.~M.~Wang,
  \textit{Charmless non-leptonic $B_s$ decays to $PP$, $PV$ and $VV$ final states in the pQCD approach}.
  \href{https://doi.org/10.1103/PhysRevD.76.074018}
  {Phys.\ Rev.\ D {\bf 76} (2007) 074018}.
  [\href{https://arxiv.org/abs/hep-ph/0703162}
  {hep-ph/0703162}]

\bibitem{Su:2011eq}
  F.~Su, Y.~L.~Wu, C.~Zhuang and Y.~B.~Yang,
  \textit{Charmless $B_s\to PP, PV, VV$ Decays Based on the Six-Quark Effective Hamiltonian with Strong Phase Effects II}.
  \href{https://doi.org/10.1140/epjc/s10052-012-1914-4}
  {Eur.\ Phys.\ J.\ C {\bf 72} (2012) 1914}.
  [\href{https://arxiv.org/abs/1107.0136}
  {arXiv:1107.0136}]

\bibitem{Yan:2017nlj}
  D.~C.~Yan, P.~Yang, X.~Liu and Z.~J.~Xiao,
  \textit{Anatomy of $B_s \to P V$ decays and effects of next-to-leading order contributions in the perturbative QCD factorization approach}.
  \href{https://doi.org/10.1016/j.nuclphysb.2018.04.007}
  {Nucl.\ Phys.\ B {\bf 931} (2018) 79}.
  [\href{https://arxiv.org/abs/1707.06043}
  {arXiv:1707.06043}]

\bibitem{Yan:2019nhf}
  D.~C.~Yan, X.~Liu and Z.~J.~Xiao,
  \textit{Anatomy of $B_s \to PP $ decays and effects of the next-to-leading order contributions in the perturbative QCD approach}.
  \href{https://doi.org/10.1016/j.nuclphysb.2019.114705}
  {Nucl.\ Phys.\ B {\bf 946} (2019) 114705}.
  [\href{https://arxiv.org/abs/1906.01442}
  {arXiv:1906.01442}]

\bibitem{Xiao:2019mpm}
  Z.~J.~Xiao, D.~C.~Yan and X.~Liu,
  \textit{$B_{(s)} \to \eta_c(P,V)$ decays and effects of the next-to-leading order contributions in the perturbative QCD approach}.
  \href{https://doi.org/10.1016/j.nuclphysb.2020.114954}
  {Nucl.\ Phys.\ B {\bf 953} (2020) 114954}.
  [\href{https://arxiv.org/abs/1909.10907}
  {arXiv:1909.10907}]

\bibitem{Huang:2001xb}
  T.~Huang, Z.~H.~Li and X.~Y.~Wu,
  \textit{Improved approach to the heavy to light form-factors in the light cone QCD sum rules},
  \href{https://doi.org/10.1103/PhysRevD.63.094001}
  {Phys.\ Rev.\ D {\bf 63}, 094001 (2001)}.

\bibitem{Huang:2001mq}
  T.~Huang, Z.~H.~Li and X.~Y.~Wu,
  \textit{The Heavy to light transitions in the light cone QCD sum rules},
  \href{https://arxiv.org/abs/hep-ph/0111105}
  {hep-ph/0111105}.

\bibitem{Zuo:2006dk}
  F.~Zuo, Z.~H.~Li and T.~Huang,
  \textit{Form Factor for $B\to D l \nu$ in Light-Cone Sum Rules With Chiral Current Correlator},
  \href{https://doi.org/10.1016/j.physletb.2006.07.039}
  {Phys.\ Lett.\ B {\bf 641} (2006) 177}.
  [\href{https://arxiv.org/abs/hep-ph/0606187}
  {hep-ph/0606187}]

\bibitem{Huang:2008zg}
  T.~Huang, Z.~H.~Li, X.~G.~Wu and F.~Zuo,
  \textit{Semileptonic $B(B_s, B_c)$ decays in the light-cone QCD sum rules},
  \href{https://doi.org/10.1142/S0217751X0804189X}
  {Int.\ J.\ Mod.\ Phys.\ A {\bf 23} (2008) 3237}.
  [\href{https://arxiv.org/abs/0801.0473}
  {arXiv:0801.0473}]

\bibitem{Wu:2007vi}
  X.~G.~Wu, T.~Huang and Z.~Y.~Fang,
  \textit{$SU_f(3)$-symmetry breaking effects of the $B \to K$ transition form-factor in the QCD light-cone sum rules},
  \href{https://doi.org/10.1103/PhysRevD.77.074001}
  {Phys.\ Rev.\ D {\bf 77} (2008) 074001}.
  [\href{https://arxiv.org/abs/0712.0237}
  {arXiv:0712.0237}]

\bibitem{Wu:2009kq}
  X.~G.~Wu and T.~Huang,
  \textit{Radiative Corrections on the $B \to P$ Form Factors with Chiral Current in the Light-Cone Sum Rules},
  \href{https://doi.org/10.1103/PhysRevD.79.034013}
  {Phys.\ Rev.\ D {\bf 79} (2009) 034013}.
  [\href{https://arxiv.org/abs/0901.2636}
  {arXiv:0901.2636}]

\bibitem{Huang:2008sn}
  T.~Huang, Z.~H.~Li and F.~Zuo,
  \textit{Heavy-to-light transition form factors and their relations in light-cone QCD sum rules},
  \href{https://doi.org/10.1140/epjc/s10052-008-0855-4}
  {Eur.\ Phys.\ J.\ C {\bf 60}, 63 (2009)}.
  [\href{https://arxiv.org/abs/0809.0130}
  {arXiv:0809.0130}]

\bibitem{Sun:2010nv}
  Y.~J.~Sun, Z.~H.~Li and T.~Huang,
  \textit{$B_{(s)}\to S$ transitions in the light cone sum rules with the chiral current},
  \href{https://doi.org/10.1103/PhysRevD.83.025024}
  {Phys.\ Rev.\ D {\bf 83}, 025024 (2011)}.
  [\href{https://arxiv.org/abs/1011.3901}
  {arXiv:1011.3901}]

\bibitem{Li:2012gr}
  Z.~H.~Li, N.~Zhu, X.~J.~Fan and T.~Huang,
  \textit{Form factors $f^{B\to \pi}_+(0)$ and $f^{D\to \pi}_+(0)$ in $QCD$ and determination of $|V_{ub}|$ and $|V_{cd}|$}.
  \href{https://doi.org/10.1007/JHEP05(2012)160}
  {JHEP {\bf 1205} (2012) 160}.
  [\href{https://arxiv.org/abs/1206.0091}
  {arXiv:1206.0091}]

\bibitem{Zhong:2014fma}
  T.~Zhong, X.~G.~Wu and T.~Huang,
  \textit{Heavy pseudoscalar leading-twist distribution amplitudes within QCD theory in background fields}.
  \href{https://doi.org/10.1140/epjc/s10052-015-3271-6}
  {Eur.\ Phys.\ J.\ C {\bf 75} (2015) 45}.
  [\href{https://arxiv.org/abs/1408.2297}
  {arXiv:1408.2297}]

\bibitem{Zhang:2017rwz}
  Y.~Zhang, T.~Zhong, X.~G.~Wu, K.~Li, H.~B.~Fu and T.~Huang,
  \textit{Uncertainties of the $B\to D$ transition form factor from the $D$-meson leading-twist distribution amplitude}.
  \href{https://doi.org/10.1140/epjc/s10052-018-5551-4}
  {Eur.\ Phys.\ J.\ C {\bf 78} (2018) 76}.
  [\href{https://arxiv.org/abs/1709.02226}
  {arXiv:1709.02226}]

\bibitem{Huang:2013gra}
  T.~Huang, X.~G.~Wu and T.~Zhong,
  \textit{Finding a way to determine the pion distribution amplitude from the experimental data}.
  \href{https://doi.org/10.1088/0256-307X/30/4/041201}
  {Chin.\ Phys.\ Lett.\  {\bf 30} (2013) 041201}.
  [\href{https://arxiv.org/abs/1303.2301}
  {arXiv:1303.2301}]

\bibitem{Huang:2013yya}
  T.~Huang, T.~Zhong and X.~G.~Wu,
  \textit{Determination of the pion distribution amplitude}.
  \href{https://doi.org/10.1103/PhysRevD.88.034013}
  {Phys.\ Rev.\ D {\bf 88} (2013) 034013}.
  [\href{https://arxiv.org/abs/1305.7391}
  {arXiv:1305.7391}]

\bibitem{Boyle:2006pw}
  P.~A.~Boyle {\it et al.} [UKQCD Collaboration],
  \textit{A lattice computation of the first moment of the kaon's distribution amplitude}.
  \href{https://doi.org/10.1016/j.physletb.2006.07.033}
  {Phys.\ Lett.\ B {\bf 641} (2006) 67}.
  [\href{https://arxiv.org/abs/hep-lat/0607018}
  {hep-lat/0607018}]

\bibitem{Chetyrkin:2007vm}
  K.~G.~Chetyrkin, A.~Khodjamirian and A.~A.~Pivovarov,
  \textit{Towards NNLO accuracy in the QCD sum rule for the kaon distribution amplitude}.
  \href{https://doi.org/10.1016/j.physletb.2008.02.031}
  {Phys.\ Lett.\ B {\bf 661} (2008) 250}.
  [\href{https://arxiv.org/abs/0712.2999}
  {arXiv:0712.2999}]

\bibitem{Momeni:2017moz}
  S.~Momeni and R.~Khosravi,
  \textit{Form factors and differential branching ratio of  $B \to K \mu^+ \mu^-$ in AdS/QCD}.
  \href{https://doi.org/10.1103/PhysRevD.97.056005}
  {Phys.\ Rev.\ D {\bf 97} (2018) 056005}.
  [\href{https://arxiv.org/abs/1710.10647}
  {arXiv:1710.10647}]

\bibitem{Bali:2019dqc}
  G.~S.~Bali {\it et al.} [RQCD Collaboration],
  \textit{Light-cone distribution amplitudes of pseudoscalar mesons from lattice QCD}.
  \href{https://doi.org/10.1007/JHEP08(2019)065}
  {JHEP {\bf 1908} (2019) 065}. Addendum:
  \href{https://doi.org/10.1007/JHEP11(2020)037}
  {[JHEP {\bf 2011} (2020) 037]}.
  [\href{https://arxiv.org/abs/1903.08038}
  {arXiv:1903.08038}]

\bibitem{Choi:2007yu}
  H.~M.~Choi and C.~R.~Ji,
  \textit{Distribution amplitudes and decay constants for ($\pi$, $K$, $\rho$, $K\ast$) mesons in light-front quark model}.
  \href{https://doi.org/10.1103/PhysRevD.75.034019}
  {Phys.\ Rev.\ D {\bf 75} (2007) 034019}.
  [\href{https://arxiv.org/abs/hep-ph/0701177}
  {hep-ph/0701177}]

\bibitem{Chernyak:1982it}
  V.~L.~Chernyak, A.~R.~Zhitnitsky and I.~R.~Zhitnitsky,
  \textit{Wave functions of the mesons containing $s, c, b$ quarks}.
  \href{https://lib-extopc.kek.jp/preprints/PDF/1983/8304/8304172.pdf}
  {Sov.\ J.\ Nucl.\ Phys.\  {\bf 38} (1983) 775,
  [Yad.\ Fiz.\  {\bf 38} (1983) 1277]}.

\bibitem{Chernyak:1983ej}
  V.~L.~Chernyak and A.~R.~Zhitnitsky,
  \textit{Asymptotic behavior of exclusive processes in QCD}.
  \href{https://doi.org/10.1016/0370-1573(84)90126-1}
  {Phys.\ Rept.\  {\bf 112} (1984) 173}.

\bibitem{Ball:2003sc}
  P.~Ball and M.~Boglione,
  \textit{$SU(3)$ breaking in $K$ and $K\ast$ distribution amplitudes}.
  \href{https://doi.org/10.1103/PhysRevD.68.094006}
  {Phys.\ Rev.\ D {\bf 68} (2003) 094006}.
  [\href{https://arxiv.org/abs/hep-ph/0307337}
  {hep-ph/0307337}]

\bibitem{Ball:2005vx}
  P.~Ball and R.~Zwicky,
  \textit{$SU(3)$ breaking of leading-twist $K$ and $K^\ast$ distribution amplitudes: A Reprise}.
  \href{https://doi.org/10.1016/j.physletb.2005.11.068}
  {Phys.\ Lett.\ B {\bf 633} (2006) 289}.
  [\href{https://arxiv.org/abs/hep-ph/0510338}
  {hep-ph/0510338}]

\bibitem{Ball:2006fz}
  P.~Ball and R.~Zwicky,
  \textit{Operator relations for $SU(3)$ breaking contributions to $K$ and $K\ast$ distribution amplitudes}.
  \href{https://doi.org/10.1088/1126-6708/2006/02/034}
  {JHEP {\bf 0602} (2006) 034}.
  [\href{https://arxiv.org/abs/hep-ph/0601086}
  {hep-ph/0601086}]

\bibitem{Khodjamirian:2004ga}
  A.~Khodjamirian, T.~Mannel and M.~Melcher,
  \textit{Kaon distribution amplitude from QCD sum rules}.
  \href{https://doi.org/10.1103/PhysRevD.70.094002}
  {Phys.\ Rev.\ D {\bf 70} (2004) 094002}.
  [\href{https://arxiv.org/abs/hep-ph/0407226}
  {hep-ph/0407226}]

\bibitem{Braun:2004vf}
  V.~M.~Braun and A.~Lenz,
  \textit{On the $SU(3)$ symmetry-breaking corrections to meson distribution amplitudes}.
  \href{https://doi.org/10.1103/PhysRevD.70.074020}
  {Phys.\ Rev.\ D {\bf 70} (2004) 074020}.
  [\href{https://arxiv.org/abs/hep-ph/0407282}
  {hep-ph/0407282}]

\bibitem{Braun:2006dg}
  V.~M.~Braun {\it et al.}.
  \textit{Moments of pseudoscalar meson distribution amplitudes from the lattice}.
  \href{https://doi.org/10.1103/PhysRevD.74.074501}
  {Phys.\ Rev.\ D {\bf 74} (2006) 074501}.
  [\href{https://arxiv.org/abs/hep-lat/0606012}
  {hep-lat/0606012}]

\bibitem{Arthur:2010xf}
  R.~Arthur, P.~A.~Boyle, D.~Brommel, M.~A.~Donnellan, J.~M.~Flynn, A.~Juttner, T.~D.~Rae and C.~T.~C.~Sachrajda,
  \textit{Lattice results for low moments of light meson distribution amplitudes}.
  \href{https://doi.org/10.1103/PhysRevD.83.074505}
  {Phys.\ Rev.\ D {\bf 83} (2011) 074505}.
  [\href{https://arxiv.org/abs/1011.5906}
  {arXiv:1011.5906}]

\bibitem{Dhiman:2019ddr}
  N.~Dhiman, H.~Dahiya, C.~R.~Ji and H.~M.~Choi,
  \textit{Twist-2 pseudoscalar and vector meson distribution amplitudes in light-front quark model with exponential-type confining potential}.
  \href{https://doi.org/10.1103/PhysRevD.100.014026}
  {Phys.\ Rev.\ D {\bf 100} (2019) 014026}.
  [\href{https://arxiv.org/abs/1902.09160}
  {arXiv:1902.09160}]

\bibitem{deMelo:2015yxk}
  J.~P.~B.~C.~de Melo, I.~Ahmed and K.~Tsushima,
  \textit{Parton distribution in pseudoscalar mesons with a light-front constituent quark model}.
  \href{https://doi.org/10.1063/1.4949465}
  {AIP Conf.\ Proc.\  {\bf 1735} (2016) 080012}.
  [\href{https://arxiv.org/abs/1512.07260}
  {arXiv:1512.07260}]

\bibitem{Nam:2006au}
  S.~i.~Nam, H.~C.~Kim, A.~Hosaka and M.~M.~Musakhanov,
  \textit{The leading-twist pion and kaon distribution amplitudes from the QCD instanton vacuum}.
  \href{https://doi.org/10.1103/PhysRevD.74.014019}
  {Phys.\ Rev.\ D {\bf 74} (2006) 014019}.
  [\href{https://arxiv.org/abs/hep-ph/0605259}
  {hep-ph/0605259}]

\bibitem{Shi:2014uwa}
  C.~Shi, L.~Chang, C.~D.~Roberts, S.~M.~Schmidt, P.~C.~Tandy and H.~S.~Zong,
  \textit{Flavour symmetry breaking in the kaon parton distribution amplitude}.
  \href{https://doi.org/10.1016/j.physletb.2014.07.057}
  {Phys.\ Lett.\ B {\bf 738} (2014) 512}.
  [\href{https://arxiv.org/abs/1406.3353}
  {arXiv:1406.3353}]

\bibitem{Nam:2017gzm}
  S.~i.~Nam,
  \textit{Quasi-distribution amplitudes for pion and kaon via the nonlocal chiral-quark model}.
  \href{https://doi.org/10.1142/S0217732317502182}
  {Mod.\ Phys.\ Lett.\ A {\bf 32} (2017) 1750218}.
  [\href{https://arxiv.org/abs/1704.03824}
  {arXiv:1704.03824}]

\bibitem{Chen:2017gck}
  J.~H.~Zhang {\it et al.} [LP3 Collaboration],
  \textit{Kaon distribution amplitude from lattice QCD and the flavor $SU(3)$ symmetry}.
  \href{https://doi.org/10.1016/j.nuclphysb.2018.12.020}
  {Nucl.\ Phys.\ B {\bf 939} (2019) 429}.
  [\href{https://arxiv.org/abs/1712.10025}
  {arXiv:1712.10025}]

\bibitem{Zhang:2020gaj}
  R.~Zhang, C.~Honkala, H.~W.~Lin and J.~W.~Chen,
  \textit{Pion and kaon distribution amplitudes in the continuum limit}.
  \href{https://doi.org/10.1103/PhysRevD.102.094519}
  {Phys.\ Rev.\ D {\bf 102} (2020) 094519}.
  [\href{https://arxiv.org/abs/2005.13955}
  {arXiv:2005.13955}]

\bibitem{Zhong:2021epq}
T.~Zhong, Z.~H.~Zhu, H.~B.~Fu, X.~G.~Wu and T.~Huang,
  \textit{Improved light-cone harmonic oscillator model for the pionic leading-twist distribution amplitude},
  \href{https://doi.org/10.1103/PhysRevD.104.016021}
  {Phys. Rev. D \textbf{104} (2021) 016021}.
  [\href{https://arxiv.org/abs/2102.03989}
  {arXiv:2102.03989}]

\bibitem{Duplancic:2008ix}
     G.~Duplancic, A.~Khodjamirian, T.~Mannel, B.~Melic and N.~Offen,
    \textit{Light-cone sum rules for $B\to\pi$ form factors revisited},
    \href{https://doi.org/10.1088/1126-6708/2008/04/014}
    {JHEP {\bf 0804} (2008) 014}.
    [\href{https://arxiv.org/abs/0801.1796}
    {arXiv:0801.1796}]

\bibitem{BHL} S. J. Brodsky, T. Huang, and G. P. Lepage, in \textit{Particles and Fields-2}. Proceedings of the Banff Summer Institute, Ban8; Alberta,
    1981, edited by A. Z. Capri and A. N. Kamal (Plenum, New York, 1983), p. 143; G. P. Lepage, S. J. Brodsky, T. Huang, and P. B.Mackenize,
    \textit{ibid}. , p. 83; T. Huang, in \textit{Proceedings of XXth International Conference on High Energy Physics}. Madison, Wisconsin, 1980, edited
    by L. Durand and L. G Pondrom, AIP Conf. Proc. No. 69 (AIP, New York, 1981),p. 1000.

\bibitem{Guo:1991eb}
  X.~H.~Guo and T.~Huang,
  \textit{Hadronic wave functions in $D$ and $B$ decays},
  \href{https://doi.org/10.1103/PhysRevD.43.2931}
  {Phys.\ Rev.\ D {\bf 43} (1991) 2931}.

\bibitem{Huang:1994dy}
  T.~Huang, B.~Q.~Ma and Q.~X.~Shen,
  \textit{Analysis of the pion wave function in light cone formalism},
  \href{https://doi.org/10.1103/PhysRevD.49.1490}
  {Phys. Rev. D \textbf{49}, 1490-1499 (1994)}.
  [\href{https://arxiv.org/abs/hep-ph/9402285}
  {hep-ph/9402285}]

\bibitem{WF_restframe}
    See, e.g., Elementary Particle Theory Group, Acta Phys. Sin. {\bf 25}, 415 (1976); N. Isgur, in \textit{The new aspects of subnuclear physics},
    edited by A. Zichichi (Plenum, New York, 1980), p. 107.

\bibitem{Wigner:1939cj}
  E.~P.~Wigner,
  \textit{On unitary representations of the inhomogeneous lorentz group},
  \href{https://doi.org/10.2307/1968551}
  {Annals Math. \textbf{40}, 149-204 (1939)}.

\bibitem{Melosh:1974cu}
  H.~J.~Melosh,
  \textit{Quarks: currents and constituents},
  \href{https://doi.org/10.1103/PhysRevD.9.1095}
  {Phys. Rev. D \textbf{9}, 1095 (1974)}.

\bibitem{Kondratyuk:1979gj}
  L.~A.~Kondratyuk and M.~V.~Terentev,
  \textit{The scattering problem for relativistic systems with fixed number of particles},
  Sov. J. Nucl. Phys. \textbf{31}, 561 (1980), ITEP-48-1979.

\bibitem{Wu:2008yr}
  X.~G.~Wu and T.~Huang,
  \textit{Kaon electromagnetic form-factor within the $k_T$ factorization formalism and it's light-cone wave function},
  \href{https://doi.org/10.1088/1126-6708/2008/04/043}
  {JHEP \textbf{04}, 043 (2008)}.
  [\href{https://arxiv.org/abs/0803.4229}
  {arXiv:0803.4229}]

\bibitem{Choi:1997iq}
H.~M.~Choi and C.~R.~Ji,
\textit{Mixing angles and electromagnetic properties of ground state pseudoscalar and vector meson nonets in the light cone quark model},
  \href{https://doi.org/10.1103/PhysRevD.59.074015}
  {Phys. Rev. D \textbf{59}, 074015 (1999)}.
  [\href{https://arxiv.org/abs/hep-ph/9711450}
  {hep-ph/9711450}]

\bibitem{Jaus:1991cy}
  W.~Jaus,
  \textit{Relativistic constituent quark model of electroweak properties of light mesons},
  \href{https://doi.org/10.1103/PhysRevD.44.2851}
  {Phys. Rev. D \textbf{44}, 2851-2859 (1991)}.

\bibitem{Jaus:1989au}
  W.~Jaus,
  \textit{Semileptonic decays of B and D Mesons in the light front formalism},
  \href{https://doi.org/10.1103/PhysRevD.41.3394}
  {Phys. Rev. D \textbf{41}, 3394 (1990)}.

\bibitem{Chung:1988mu}
  P.~L.~Chung, F.~Coester and W.~N.~Polyzou,
  \textit{Charge form-factors of quark model pions},
  \href{https://doi.org/10.1016/0370-2693(88)90995-1}
  {Phys. Lett. B \textbf{205}, 545-548 (1988)}.

\bibitem{Choi:1997qh}
  H.~M.~Choi and C.~R.~Ji,
  \textit{Relations among the light cone quark models with the invariant meson mass scheme and the model prediction of $\eta-\eta'$ mixing angle},
  \href{https://doi.org/10.1103/PhysRevD.56.6010}
  {Phys. Rev. D \textbf{56}, 6010-6013 (1997)}.

\bibitem{Schlumpf:1994bc}
  F.~Schlumpf,
  \textit{Charge form-factors of pseudoscalar mesons},
  \href{https://doi.org/10.1103/PhysRevD.50.6895}
  {Phys. Rev. D \textbf{50}, 6895-6898 (1994)}.
  [\href{https://arxiv.org/abs/hep-ph/9406267}
  {hep-ph/9406267}]

\bibitem{Cardarelli:1994yq}
F.~Cardarelli, I.~L.~Grach, I.~M.~Narodetsky, G.~Salme and S.~Simula,
\textit{Electromagnetic form-factors of the rho meson in a light front constituent quark model},
\href{https://doi.org/10.1016/0370-2693(95)00230-I}
{Phys. Lett. B \textbf{349}, 393-399 (1995)}.
[\href{https://arxiv.org/abs/hep-ph/9502360}
{hep-ph/9502360}]

\bibitem{Cardarelli:1995ap}
F.~Cardarelli, I.~L.~Grach, I.~Narodetsky, G.~Salme and S.~Simula,
\textit{Radiative $\pi \rho$ and $\pi \omega$ transition form-factors in a light front constituent quark model},
\href{https://doi.org/10.1016/0370-2693(95)01058-X}
{Phys. Lett. B \textbf{359}, 1-7 (1995)}.
[\href{https://arxiv.org/abs/nucl-th/9509004}
{nucl-th/9509004}]

\bibitem{Cardarelli:1994ix}
F.~Cardarelli, I.~L.~Grach, I.~M.~Narodetsky, E.~Pace, G.~Salme and S.~Simula,
\textit{Hard constituent quarks and electroweak properties of pseudoscalar mesons},
\href{https://doi.org/10.1016/0370-2693(94)90849-4}
{Phys. Lett. B \textbf{332}, 1-7 (1994)}.
[\href{https://arxiv.org/abs/nucl-th/9405014}
{nucl-th/9405014}]

\bibitem{Dziembowski:1986dr}
Z.~Dziembowski and L.~Mankiewicz,
\textit{Light meson distribution amplitude: a simple relativistic model},
\href{https://doi.org/10.1103/PhysRevLett.58.2175}
{Phys. Rev. Lett. \textbf{58}, 2175 (1987)}.

\bibitem{Dziembowski:1987zp}
Z.~Dziembowski,
\textit{Relativistic model of nucleon and pion structure: static properties and electromagnetic soft form-factors},
\href{https://doi.org/10.1103/PhysRevD.37.778}
{Phys. Rev. D \textbf{37}, 778 (1988)}.

\bibitem{Ji:1990rd}
C.~R.~Ji and S.~R.~Cotanch,
\textit{Simple relativistic quark model analysis of flavored pseudoscalar mesons},
\href{https://doi.org/10.1103/PhysRevD.41.2319}
{Phys. Rev. D \textbf{41}, 2319-2322 (1990)}.

\bibitem{Ji:1992yf}
C.~R.~Ji, P.~L.~Chung and S.~R.~Cotanch,
\textit{Light cone quark model axial vector meson wave function},
\href{https://doi.org/10.1103/PhysRevD.45.4214}
{Phys. Rev. D \textbf{45}, 4214-4220 (1992)}.

\bibitem{Choi:1996mq}
H.~M.~Choi and C.~R.~Ji,
\textit{Light cone quark model predictions for radiative meson decays},
\href{https://doi.org/10.1016/S0375-9474(97)00052-3}
{Nucl. Phys. A \textbf{618}, 291-316 (1997)}.

\bibitem{Wu:2007rt}
X.~G.~Wu, T.~Huang and Z.~Y.~Fang,
\textit{$B\to K$ transition form-factor up to $\mathcal{O}(1/m_b^2)$ within the $k_T$ factorization approach},
\href{https://doi.org/10.1140/epjc/s10052-007-0421-5}
{Eur. Phys. J. C \textbf{52}, 561-570 (2007)}.
[\href{https://arxiv.org/abs/0707.2504}
{arXiv:0707.2504}]

%

\bibitem{Wu:2011gf}
X.~G.~Wu and T.~Huang,
\textit{Constraints on the Light Pseudoscalar Meson Distribution Amplitudes from Their Meson-Photon Transition Form Factors},
\href{https://doi.org/10.1103/PhysRevD.84.074011}
{Phys. Rev. D \textbf{84} (2011) 074011}.
[\href{https://arxiv.org/abs/1106.4365}
{arXiv:1106.4365}]


\bibitem{Zhong:2014jla}
  T.~Zhong, X.~G.~Wu, Z.~G.~Wang, T.~Huang, H.~B.~Fu and H.~Y.~Han,
  \textit{Revisiting the Pion Leading-Twist Distribution Amplitude within the QCD Background Field Theory},
  \href{https://doi.org/10.1103/PhysRevD.90.016004}
  {Phys. Rev. D \textbf{90}, 016004 (2014)}.
  [\href{https://arxiv.org/abs/1405.0774}
  {arXiv:1405.0774}]

\bibitem{Hu:2021zmy}
  D.~D.~Hu, H.~B.~Fu, T.~Zhong, L.~Zeng, W.~Cheng and X.~G.~Wu,
  \textit{$\eta$-meson leading-twist distribution amplitude within QCD sum rule approach and its application to the semi-leptonic decay $ D_s^+ \to\eta{\ell}^{+} \nu_{\ell}$},
  \href{https://arxiv.org/abs/2102.05293}
  {arXiv:2102.05293}.

\bibitem{Zhong:2011rg}
  T.~Zhong, X.~G.~Wu, H.~Y.~Han, Q.~L.~Liao, H.~B.~Fu and Z.~Y.~Fang,
  \textit{Revisiting the Twist-3 Distribution Amplitudes of $K$ Meson within the QCD Background Field Approach},
  \href{https://doi.org/10.1088/0253-6102/58/2/16}
  {Commun. Theor. Phys. \textbf{58}, 261-270 (2012)}.
  [\href{https://arxiv.org/abs/1109.3127}
  {arXiv:1109.3127}]

\bibitem{PDGnew}
  P. A. Zyla et al. (Particle Data Group),
  \textit{Review of Particle Physics},
  \href{https://pdg.lbl.gov/2020/listings/contents_listings.html}
  {Prog. Theor. Exp. Phys. 2020 (2020) 083C01}.

\bibitem{FlavourLatticeAveragingGroup:2019iem}
  S.~Aoki \textit{et al.} [Flavour Lattice Averaging Group],
  \textit{FLAG Review 2019: Flavour Lattice Averaging Group (FLAG)}
  \href{https://doi.org/10.1140/epjc/s10052-019-7354-7}
  {Eur. Phys. J. C \textbf{80}, 113 (2020)}.
  [\href{https://arxiv.org/abs/1902.08191}
  {arXiv:1902.08191}]

\bibitem{Narison:2014ska}
  S.~Narison,
  \textit{Improved $f_{D^\ast_{(s)}}, f_{B^\ast_{(s)}}$ and $f_{B_{c}}$ from QCD Laplace sum rules},
  \href{https://doi.org/10.1142/S0217751X1550116X}
  {Int.\ J.\ Mod.\ Phys.\ A {\bf 30} (2015) 1550116}.
  [\href{https://arxiv.org/abs/1404.6642}
  {arXiv:1404.6642}]


\bibitem{Colangelo:2000dp}
  P.~Colangelo and A.~Khodjamirian,
  \textit{QCD sum rules, a modern perspective},
  \href{https://doi.org/10.1142/9789812810458\_0033}
  {hep-ph/0010175}.

\bibitem{Narison:2014wqa}
  S.~Narison,
  \textit{Mini-review on QCD spectral sum rules},
  \href{https://doi.org/10.1016/j.nuclphysbps.2015.01.041}
  {Nucl.\ Part.\ Phys.\ Proc.\  {\bf 258-259} (2015) 189}.
  [\href{https://arxiv.org/abs/1409.8148}
  {arXiv:1409.8148}]

\bibitem{Lepage:1980fj}
  G.~P.~Lepage and S.~J.~Brodsky,
  \textit{Exclusive Processes in Perturbative Quantum Chromodynamics},
  \href{https://doi.org/10.1103/PhysRevD.22.2157}
  {Phys. Rev. D \textbf{22}, 2157 (1980)}.

\bibitem{FermilabLattice:2019ikx}
    A.~Bazavov \textit{et al.} [Fermilab Lattice and MILC],
    \textit{$B_s\to K\ell\nu$ decay from lattice QCD},
    \href{https://doi.org/10.1103/PhysRevD.100.034501}
    {Phys. Rev. D \textbf{100} (2019) 034501}.
    [\href{https://arxiv.org/abs/1901.02561}
    {arXiv:1901.02561}]

\bibitem{Gonzalez-Solis:2021awb}
    S.~Gonz\`alez-Sol\'\i{}s, P.~Masjuan and C.~Rojas,
    \textit{Pad\'e approximants to $B\to\pi\ell\nu_{\ell}$ and $B_{s}\to K\ell\nu_{\ell}$ and determination of $|V_{ub}|$},
    \href{https://arxiv.org/abs/2110.06153}
    {arXiv:2110.06153}.

\bibitem{Detmold:2015aaa}
    W.~Detmold, C.~Lehner and S.~Meinel,
    \textit{$\Lambda_b \to p \ell^- \bar{\nu}_\ell$ and $\Lambda_b \to \Lambda_c \ell^- \bar{\nu}_\ell$ form factors from lattice QCD with relativistic heavy quarks},
    \href{https://doi.org/10.1103/PhysRevD.92.034503}
    {Phys. Rev. D \textbf{92} (2015) 034503}.
    [\href{https://arxiv.org/abs/1503.01421}
    {arXiv:1503.01421}]

\bibitem{LHCb:2015eia}
R.~Aaij \textit{et al.} [LHCb],
    \textit{Determination of the quark coupling strength $|V_{ub}|$ using baryonic decays},
    \href{https://doi.org/10.1038/nphys3415}
    {Nature Phys. \textbf{11} (2015), 743-747}.
    [\href{https://arxiv.org/abs/1504.01568}
    {arXiv:1504.01568}]

\end{thebibliography}
\end{document}